\documentclass[twoside,12pt]{article}
\usepackage{epsfig}

\newcommand{\be}{\begin{equation}}
\newcommand{\ee}{\end{equation}}
\newcommand{\bea}{\begin{eqnarray}}
\newcommand{\eea}{\end{eqnarray}}

\usepackage{amsmath}
\usepackage{amsfonts}
\usepackage{amssymb}
\usepackage{amsxtra}
\usepackage{supertabular}
\begin{document}
\title{ \vspace{1cm} 

How Clifford algebra helps understand second quantized quarks and leptons 
and corresponding vector and scalar boson fields, {\it opening a new step beyond 
the standard model}\\ 
Nucl. Phys. B 994 (2023) 116326
}

\author{N.S.\ Manko\v c Bor\v stnik$^{1}$ 
\\
$^1$Department of Physics, University of Ljubljana\\
SI-1000 Ljubljana, Slovenia } 
\maketitle
 



\begin{abstract}
This article presents the description of the internal spaces of fermion and boson fields
in $d$-dimensional spaces, with the odd and even ``basis vectors'' which are the 
superposition of odd and even products of operators $\gamma^a$. 
While the Clifford odd ``basis vectors'' manifest properties of fermion fields, appearing 
in families, 
the Clifford even ``basis vectors'' demonstrate properties of the corresponding gauge fields. 
In $d\ge (13+1)$ the corresponding creation operators manifest in $d=(3+1)$ the 
properties of all the observed quarks and leptons, with the families included, and of their 
gauge  boson fields, with the scalar fields included,  making several predictions. 
The properties of the creation and annihilation operators for fermion and boson fields
are illustrated on the case $d=(5+1)$, when $SO(5,1)$ demonstrates the symmetry 
of $SU(3)\times U(1)$.
\end{abstract}
\section {Introduction}
\label{introduction}
\vspace{2mm}

The {\it standard model} (corrected with the right-handed neutrinos) has
been experimentally confirmed without raising any severe doubts so far on its
assumptions, which, however, remain unexplained.

The {\it standard model} assumptions have several explanations in the literature, 
mostly with several new, not explained assumptions. The most popular are
the grand unifying theories~(\cite{Geor,FritzMin,PatiSal,GeorGlas,Cho} 
and many others). 

\vspace{3mm}

In a long series of works~(\cite{norma93,pikan2006,
n2014matterantimatter,JMP2013},and the references there in) 
the author has found, together with the collaborators~(\cite{nh02,pikan2006,
nd2017,nh2018,
2020PartIPartII,nh2021RPPNP} and the references therein), 
the phenomenological success with the model named the
{\it spin-charge-family} theory with the properties: \\
{\bf a.} The internal space of fermions are described by the ``basis vectors'' which are
superposition of odd products of anti-commuting objects (operators)~\footnote{
According to Eq.~(\ref{gammatildeantiher0}) $\{\gamma^a,\gamma^b\}_{+}=2\eta^{ab}$ are 
anticommuting unless $a=b$.}
$\gamma^a$ (in the sense $\{\gamma^{a},\gamma^{b}\}_{+}= 2 \eta^{ab}$),
Sect.~\ref{grassmannandclifford}, in
$d=(13+1)$-dimensional space~\cite{
2020PartIPartII,nh2021RPPNP}. 
Correspondingly the ``basis vectors'' of one Lorentz irreducible representation in internal
space of fermions, together with their Hermitian conjugated partners, anti-commute,
fulfilling (on the vacuum state) all the requirements for the second quantized fermion 
fields~(\cite{
nh02,
nh2021RPPNP} and references therein). \\
{\bf a.i.} 
The second kind of anti-commuting objects, 
$\tilde{\gamma}^a$, Sect.~\ref{grassmannandclifford}, equip each
irreducible representation of odd ``basis vectors'' with the family quantum number~%
\cite{2020PartIPartII,nh02}.\\
{\bf a.ii.} Creation operators for single fermion states --- which are tensor products, 
$*_{T}$, of a finite number of odd ``basis vectors'' appearing in $2^{\frac{d}{2}-1}$ 
families, each family with $2^{\frac{d}{2}-1}$ members, and the (continuously) 
infinite momentum/coordinate basis applying on the vacuum state~\cite{2020PartIPartII,
nh2021RPPNP} --- inherit anti-commutativity of ``basis vectors''. Creation operators
and their Hermitian conjugated partners correspondingly anti-commute. \\
{\bf a.iii.} 
The Hilbert space of second quantized fermion field is represented by the tensor
products, $*_{T_{H}}$, of all possible numbers of creation operators, from zero to
infinity~\cite{nh2021RPPNP}, applying on a vacuum state.\\
{\bf a.iv.} Spins from higher dimensions, $d>(3+1)$, described by the eigenvectors 
of the superposition of the Cartan subalgebra $S^{ab}$, Table~\ref{Table so13+1.},
manifest in $d=(3+1)$ all the charges of the {\it standard model} quarks and
leptons and antiquarks and antileptons.\\ 
{\bf b.} 
In a simple starting action, Eq.~(\ref{wholeaction}), massless fermions carry only 
spins and interact with only gravity --- with the vielbeins and the two kinds of spin 
connection fields (the gauge fields of momenta, of 
$S^{ab}=\frac{i}{4}(\gamma^a \gamma^b- \gamma^b \gamma^a)$ and of 
$\tilde{S}^{ab}=\frac{1}{4} (\tilde{\gamma}^a \tilde{\gamma}^b -
\tilde{\gamma}^b \tilde{\gamma}^a)$, respectively~%
\footnote{
If no fermions are present, the two kinds of spin connection fields are
uniquely expressible by the vielbeins.}). The starting action includes
only even products of $\gamma^a$'s and $\tilde{\gamma^a}$'s~(\cite{nh2021RPPNP} and
references therein). \\
{\bf b.i.} Gravity --- the gauge fields of $S^{ab}$, ($(a,b)=(5,6,....,d)$), with the space
index $m=(0,1,2,3)$ --- manifest as the {\it standard model} vector gauge
fields~\cite{nd2017}, with the ordinary gravity included ($(a,b)=(0,1,2,3)$). \\
{\bf b.ii.} The scalar gauge fields of $\tilde{S}^{ab}$, and of some of the
superposition of $S^{ab}$, with the space index $s=(7,8)$ manifest as the scalar Higgs 
and Yukawa couplings~\cite{JMP2013,nh2021RPPNP,NHD},
determining mass matrices (of particular symmetry) and correspondingly the masses
of quarks and leptons 
and of the weak boson fields after (some of) the scalar fields with the space index
$(7,8)$ gain constant values. \\
{\bf b.iii.} The scalar gauge fields of $\tilde{S}^{ab}$ and of
$S^{ab}$ with the space index $s=(9,10,...,14)$ and $(a,b)=(5,6,....,d)$ offer the
explanation for the observed matter/antimatter asymmetry~\cite{n2014matterantimatter,JMP2013,
nh2018,nh2021RPPNP} in the universe.\\
{\bf c.} The theory predicts at low energy two groups with four families. To the lower 
group of four families the so far  observed three belong~\cite{mdn2006,
gmdn2007,gmdn2008,gn2013,gn2014}, and the stable of the upper four families, 
the fifth family of (heavy) quarks and leptons, offers the explanation for the 
appearance of dark matter.
Due to the heavy masses of the fifth family quarks, the nuclear interaction among 
hadrons of the fifth family members is very different than the ones so far observed~\cite{gn2009,nm2015}. \\
{\bf d.} The theory offers a new understanding of the second quantized fermion 
fields, as mentioned in  {\bf a.} and it is explained in Refs.~\cite{2020PartIPartII,NHD,nh2021RPPNP}, it also enables a new understanding of the second 
quantization  of boson fields which is the main topics of this article~\cite{n2021SQ,%
n2022epjc}, both in even  dimensional spaces. \\
 {\bf d.i.} The Clifford odd ``basis vectors'' appear in $2^{\frac{d}{2}-1}$ families, 
each family having  $2^{\frac{d}{2}-1}$ members. Their Hermitian conjugated 
partners appear in a  separate group, Sect.~\ref{creationannihilation}.\\
{\bf d.ii.} The Clifford even ``basis vectors'' appear in two  groups, each with  
$2^{\frac{d}{2}-1}$ $\times 2^{\frac{d}{2}-1}$ members with their Hermitian 
conjugated  partners within the same group. One group of the Clifford even ``basis 
vectors'' transform, when applying algebraically on the Clifford odd ``basis vector'', 
this Clifford odd ``basis vector'' into other members of the same family. The other 
group of the Clifford even ``basis vectors'' transform, when being applied algebraically 
by the Clifford odd ``basis vector'', this Clifford odd ``basis vector'' into the same 
member of another family; in agreement with the action, 
Eq.~(\ref{wholeaction}).\\ 
{\bf d.iii.} In odd dimensional spaces, $d=(2n+1)$, the properties of Clifford odd 
and Clifford even ``basis vectors'' differ essentially from their properties in even 
dimensional spaces, resembling the ghosts needed to make the contributions of 
the Feynman diagrams finite~\cite{n2023MDPI}.

\vspace{3mm}

The theory seems very promising to offer a new insight into the second quantization of 
fermion and boson fields and to show the next step beyond the {\it standard model}.

The more work is put into the theory, the more phenomena the theory can
explain.

Other references used a different approach by trying to make the next step with Clifford 
algebra to the second quantized fermion, which might also be a boson 
field~\cite{MPavsic,MP2017}. 

Let us present  a simple starting action of the {\it spin-charge-family} 
theory~(\cite{nh2021RPPNP} and the references therein)  for massless fermions 
and anti-fermions which interact with massless gravitational fields only; with 
vielbeins (the gauge fields of momenta) and the two kinds of spin connection 
fields, the gauge fields of the two kinds of the Lorentz transformations in the 
internal space of fermions, of $S^{ab}$ and $\tilde{S}^{ab}$, 
in  $d=2(2n+1)$-dimensional space 
\begin{eqnarray}
{\cal A}\,  &=& \int \; d^dx \; E\;\frac{1}{2}\, (\bar{\psi} \, \gamma^a p_{0a} \psi) 
+ h.c. +
\nonumber\\  
               & & \int \; d^dx \; E\; (\alpha \,R + \tilde{\alpha} \, \tilde{R})\,,
\nonumber\\
           p_{0\alpha} &=&  p_{\alpha}  - \frac{1}{2}  S^{ab} \omega_{ab \alpha} - 
                    \frac{1}{2}  \tilde{S}^{ab}   \tilde{\omega}_{ab \alpha} \,,
                    \nonumber\\  
           p_{0a } &=& f^{\alpha}{}_a p_{0\alpha} + \frac{1}{2E}\, \{ p_{\alpha},
E f^{\alpha}{}_a\}_- \,,\nonumber\\                  
R &=&  \frac{1}{2} \, \{ f^{\alpha [ a} f^{\beta b ]} \;(\omega_{a b \alpha, \beta} 
- \omega_{c a \alpha}\,\omega^{c}{}_{b \beta}) \} + h.c. \,, \nonumber \\
\tilde{R}  &=&  \frac{1}{2} \, \{ f^{\alpha [ a} f^{\beta b ]} 
\;(\tilde{\omega}_{a b \alpha,\beta} - \tilde{\omega}_{c a \alpha} \,
\tilde{\omega}^{c}{}_{b \beta})\} + h.c.\,.               
\label{wholeaction}
\end{eqnarray}
Here~\footnote{$f^{\alpha}{}_{a}$ are inverted vielbeins to 
$e^{a}{}_{\alpha}$ with the properties $e^a{}_{\alpha} f^{\alpha}{\!}_b = 
\delta^a{\!}_b,\; e^a{\!}_{\alpha} f^{\beta}{\!}_a = \delta^{\beta}_{\alpha} $, 
$ E = \det(e^a{\!}_{\alpha}) $.
Latin indices  
$a,b,..,m,n,..,s,t,..$ denote a tangent space (a flat index),
while Greek indices $\alpha, \beta,..,\mu, \nu,.. \sigma,\tau, ..$ denote an Einstein 
index (a curved index). Letters  from the beginning of both the alphabets
indicate a general index ($a,b,c,..$   and $\alpha, \beta, \gamma,.. $ ), 
from the middle of both the alphabets   
the observed dimensions $0,1,2,3$ ($m,n,..$ and $\mu,\nu,..$), indexes from 
the bottom of the alphabets
indicate the compactified dimensions ($s,t,..$ and $\sigma,\tau,..$). 
We assume the signature $\eta^{ab} =
diag\{1,-1,-1,\cdots,-1\}$.} 
$f^{\alpha [a} f^{\beta b]}= f^{\alpha a} f^{\beta b} - f^{\alpha b} f^{\beta a}$.
The vielbeins, $f^a_{\alpha}$, and the two kinds of the spin connection fields, 
$\omega_{ab \alpha}$ (the gauge fields of $S^{ab}$) and $\tilde{\omega}_{ab \alpha}$  
(the gauge fields of $\tilde{S}^{ab}$), manifest in $d=(3+1)$ as the known vector 
gauge fields and the scalar gauge fields taking care of masses of quarks and leptons and 
antiquarks and antileptons and of the weak boson fields~\cite{nd2017,%
n2014matterantimatter,JMP2013,%
nh2018}~\footnote{
Since the multiplication with either $\gamma^a$'s or $\tilde{\gamma}^a$'s  changes 
the Clifford odd ``basis vectors'' into the Clifford even  objects, 
and even ``basis vectors'' commute, the action for fermions can not include odd 
numbers of $\gamma^a$'s or $\tilde{\gamma}^a$'s, what the simple starting action 
of Eq.~(\ref{wholeaction}) does not. In the starting action $\gamma^a$'s and 
$\tilde{\gamma}^a$'s appear as $\gamma^0 \gamma^a \hat{p}_{0a}$  or as 
$\gamma^0 \gamma^c \, S^{ab}\omega_{abc}$  and  as 
$\gamma^0 \gamma^c \,\tilde{S}^{ab}\tilde{\omega}_{abc} $.}.

The action, Eq.~(\ref{wholeaction}), assumes two kinds of the spin connection gauge 
fields, due to two kinds of the operators: $\gamma^a$ and $\tilde{\gamma}^{a}$.
Let be pointed out that the description of the internal space of bosons with the
Clifford even ``basis vectors'' offers as well two kinds of the Clifford even ``basis 
vectors'', as presented in {\bf d.ii.}.

\vspace{3mm}

In Sect.~\ref{creationannihilation} the Grassmann and the Clifford algebras are explained,
Subsect.{\ref{grassmannandclifford}, 
and creation and annihilation operators described as  tensor products of the 
``basis vectors'' offering an explanation of the internal spaces of fermion (by the Clifford 
odd algebra) and boson (by the Clifford even algebra) fields and the basis in ordinary
 space. 

In Subsect.~\ref{basisvectors}, the ``basis vectors''  are introduced and their properties 
presented in even and odd-dimensional spaces, Subsects.~\ref{deven}, Subsect.~\ref{dodd}, 
respectively.

In Subsect.~\ref{cliffordoddevenbasis5+1}, the properties of the Clifford odd and even
``basis vectors''  are demonstrated in the toy model in $d=(5+1)$.

In Subsect.~\ref{secondquantizedfermionsbosonsdeven}, the properties of the creation and 
annihilation operators for the second quantized fermion and boson fields in even dimensional
spaces are described.

Sect.~\ref{conclusions} presents what the reader could learn new from
 this article.
 
  In App.~\ref{OpenQuestionsSM}, the answers of the {\it spin-charge-family} theory to 
  some of the open questions of the {\it standard model} are discussed.
 

 In App.~\ref{A}, some useful formulas and relations are presented.

 In App.~\ref{13+1representation} one irreducible representation (one family) of
 $SO(13,1)$, group,  analysed with respect to $SO(3,1)$, $SU(2)_{I}$,  $SU(2)_{II}$,
  $SU(3)$, and $U(1)$, representing ``basis vectors'' of  quarks and leptons and antiquarks 
  and antilepons is discussed.
  

App.~\ref{basis3+1}, suggested by the referee,  illustrates on the simplest case
$ d=(3+1)$ (and $d=(1+1)$; which offers only one ``family'' of fermions, $d=(3+1)$ 
has two families) the properties of the  Clifford odd and Clifford even``basis vectors'' 
describing the internal 
spaces of fermion and boson fields, explaining in a pedagogical way in details their 
construction, manifestation of anti-commutativity (in the fermion case) and commutativity
(in the boson case) of the tensor product of the ``basis vectors'' and the basis in ordinary 
space-time.

The referee suggested also several footnotes.

%
\section{Creation and annihilation operators for fermions and bosons in even and 
odd dimensional spaces}
\label{creationannihilation}
\vspace{2mm}

Refs.~\cite{norma93,nh02,2020PartIPartII,n2014matterantimatter,nh2021RPPNP} describe 
the internal space of fermion fields by the superposition of odd products of $\gamma^a$ 
in even dimensional spaces ($d=2(2n+1)$, or $d=4n$).
In any even dimensional space there appear $2^{\frac{d}{2}-1}$ members of each
irreducible representation of $S^{ab}$, each irreducible representation representing one of
$2^{\frac{d}{2}-1}$ families, carrying quantum numbers determined by $\tilde{S}^{ab}$.
Their Hermitian conjugated partners appear in a separate group (not reachable by either
$S^{ab}$ or $\tilde{S}^{ab}$). Since the tensor products, $*_{T}$, of these Clifford
odd ``basis vectors'' and basis in ordinary momentum or coordinate space, applying on
the vacuum state, fulfil the second quantization postulates for 
fermions~\cite{Dirac,BetheJackiw,Weinberg}, it is obvious that the $2^{\frac{d}{2}-1}$
$\times \, \,2^{\frac{d}{2}-1}$ anti-commuting Clifford odd ``basis vectors'', together with 
their Hermitian conjugated partners, transferring their anti-commutativity to creation and 
annihilation operators, explain the second quantization postulates of Dirac for fermions 
and their families~\cite{2020PartIPartII}.\\

\vspace{2mm}

There are, however, the same number of the Clifford even ``basis vectors'', which obviously
commute, transferring their commutativity to tensor products, $*_{T}$, of the Clifford even
``basis vectors'' and basis in ordinary momentum or coordinate space. 

\vspace{2mm}

We shall see in what follows that the Clifford even ``basis vectors'' appear in two
groups, each with $2^{\frac{d}{2}-1}$ $\times \, \,2^{\frac{d}{2}-1}$ members.
The members of each group have their Hermitian conjugated partners within the
same group.
As we shall see, one group transforms a particular family member of a Clifford odd
``basis vector'' into other members of the same family, keeping the family quantum
number unchanged. The second group transforms a particular family member of a
Clifford odd ``basis vector'' into the same member of another
family~\cite{n2022epjc}. We shall see that the Clifford even ``basis vectors'' of
each of the two groups has, in even dimensional spaces, the properties of the
gauge boson fields of the corresponding Clifford odd ``basis vectors'': One group
with respect to $S^{ab}$, the other with respect to $\tilde{S}^{ab}$.
\vspace{2mm}

The properties of the Clifford odd and the Clifford even ``basis vectors'' in odd 
dimensional spaces, $d=(2n +1)$, differ essentially from their properties in even 
dimensional spaces, as we shall review Ref.~\cite{n2023MDPI} in 
Subsect.~\ref{dodd}. 
Although anti-commuting, the Clifford odd ``basis vectors'' manifest properties
of the Clifford even ``basis vectors'' in even dimensional spaces. And the
Clifford even ``basis vectors'', although commuting, manifest properties of the 
Clifford odd ``basis vectors'' in even dimensional spaces. 

\vspace{2mm}

\subsection{Grassmann and Clifford algebras}
\label{grassmannandclifford}

\vspace{2mm}

This part is a short overview of several references,
cited in Ref.~(\cite{nh2021RPPNP}, Subsects. 3.2,3.3),  also appearing in 
Ref.~\cite{nIARD2022,2020PartIPartII,n2023MDPI}.

The internal spaces of anti-commuting or commuting second quantized fields can be 
described by using either the Grassmann or the Clifford algebras~\cite{
norma93,nh2021RPPNP} 
 
In Grassmann $d$-dimensional space there are $d$ anti-commuting (operators) 
$\theta^{a}$,
 and $d$ anti-commuting operators which are derivatives with respect to $\theta^{a}$,
$\frac{\partial}{\partial \theta_{a}}$, 
%
\begin{eqnarray}
\label{thetaderanti0}
\{\theta^{a}, \theta^{b}\}_{+}=0\,, \, && \,
\{\frac{\partial}{\partial \theta_{a}}, \frac{\partial}{\partial \theta_{b}}\}_{+} =0\,,
\nonumber\\
\{\theta_{a},\frac{\partial}{\partial \theta_{b}}\}_{+} &=&\delta_{ab}\,, 
\;(a,b)=(0,1,2,3,5,\cdots,d)\,. 
\end{eqnarray}
Making a choice~\cite{nh2018} 
\begin{eqnarray}
(\theta^{a})^{\dagger} &=& \eta^{a a} \frac{\partial}{\partial \theta_{a}}\,,\quad
{\rm leads  \, to} \quad
(\frac{\partial}{\partial \theta_{a}})^{\dagger}= \eta^{a a} \theta^{a}\,,
\label{thetaderher0}
\end{eqnarray}
with $\eta^{a b}=diag\{1,-1,-1,\cdots,-1\}$.

$ \theta^{a}$ and $ \frac{\partial}{\partial \theta_{a}}$ are, up to the sign, Hermitian 
conjugated to each other. The identity is the self adjoint member of the algebra.
The choice for the following complex properties of $\theta^a$ 
 \begin{small}
\begin{eqnarray}
\label{complextheta}
\{\theta^a\}^* &=&  (\theta^0, \theta^1, - \theta^2, \theta^3, - \theta^5,
\theta^6,...,- \theta^{d-1}, \theta^d)\,, 
\end{eqnarray}
\end{small}
 correspondingly requires $\;\;\, $ 
$\{\frac{\partial}{\partial \theta_{a}}\}^* = (\frac{\partial}{\partial \theta_{0}},
\frac{\partial}{\partial \theta_{1}}, - \frac{\partial}{\partial \theta_{2}},
\frac{\partial}{\partial \theta_{3}}, - \frac{\partial}{\partial \theta_{5}}, 
\frac{\partial}{\partial \theta_{6}},..., - \frac{\partial}{\partial \theta_{d-1}}, 
\frac{\partial}{\partial \theta_{d}})\,. $
%

There are $2^d$ superposition of products of  $\theta^{a}$, 
the Hermitian conjugated partners of which are the corresponding superposition of products 
of $\frac{\partial}{\partial \theta_{a}}$.

There exist two kinds of the Clifford algebra elements (operators), $\gamma^{a}$ and 
$\tilde{\gamma}^{a}$, expressible with $\theta^{a}$'s and their conjugate momenta 
$p^{\theta a}= i \,\frac{\partial}{\partial \theta_{a}}$ ~\cite{norma93}, 
Eqs.~(\ref{thetaderanti0}, \ref{thetaderher0}), 
\begin{eqnarray}
\label{clifftheta1}
\gamma^{a} &=& (\theta^{a} + \frac{\partial}{\partial \theta_{a}})\,, \quad 
\tilde{\gamma}^{a} =i \,(\theta^{a} - \frac{\partial}{\partial \theta_{a}})\,,\nonumber\\
\theta^{a} &=&\frac{1}{2} \,(\gamma^{a} - i \tilde{\gamma}^{a})\,, \quad 
\frac{\partial}{\partial \theta_{a}}= \frac{1}{2} \,(\gamma^{a} + i \tilde{\gamma}^{a})\,,
\nonumber\\
\end{eqnarray}
offering together  $2\cdot 2^d$  operators: $2^d$ are superposition of products of 
$\gamma^{a}$  and  $2^d$  of $\tilde{\gamma}^{a}$.
It is easy to prove if taking into account Eqs.~(\ref{thetaderher0}, \ref{clifftheta1}),
 that they form two anti-commuting Clifford subalgebras, 
$\{\gamma^{a}, \tilde{\gamma}^{b}\}_{+} =0$, Refs.~(\cite{nh2021RPPNP} and 
references therein)
\begin{eqnarray}
\label{gammatildeantiher0}
\{\gamma^{a}, \gamma^{b}\}_{+}&=&2 \eta^{a b}= \{\tilde{\gamma}^{a}, 
\tilde{\gamma}^{b}\}_{+}\,, \nonumber\\
\{\gamma^{a}, \tilde{\gamma}^{b}\}_{+}&=&0\,,\quad
 (a,b)=(0,1,2,3,5,\cdots,d)\,, \nonumber\\
(\gamma^{a})^{\dagger} &=& \eta^{aa}\, \gamma^{a}\, , \quad 
(\tilde{\gamma}^{a})^{\dagger} =  \eta^{a a}\, \tilde{\gamma}^{a}\,.
\end{eqnarray}
%
%
%
While the Grassmann algebra offers the description of the ``anti-commuting integer spin
second quantized fields'' and of the ``commuting integer spin second quantized 
fields''~\cite{2020PartIPartII,nh2021RPPNP}, the Clifford algebras which are 
superposition of odd products of either $\gamma^a$'s or $\tilde{\gamma}^a$'s offer 
the description of the second quantized half integer spin fermion fields, which from 
the point of the subgroups of the $SO(d-1,1)$ group manifest spins and charges of 
fermions and antifermions in the fundamental representations of the group and 
subgroups, Table~\ref{Table so13+1.}.

\noindent
The superposition of even products of either $\gamma^a$'s or $\tilde{\gamma}^a$'s 
offer the description of the commuting second quantized boson fields with integer spins 
(as we can see in~\cite{n2021SQ,n2022epjc} and shall see in this contribution) which from the 
point of the subgroups of the $SO(d-1,1)$ group manifest spins and charges in the 
adjoint representations of the group and subgroups.

The following {\it postulate}, which determines how does  $\tilde{\gamma}^{a}$ 
operate on $\gamma^a$,  reduces the two Clifford subalgebras, $\gamma^a$ and 
$\tilde{\gamma}^a$, to one, to the one described by 
$\gamma^a$~\cite{nh02,norma93,JMP2013,nh2018}
\begin{eqnarray}
\{\tilde{\gamma}^a B &=&(-)^B\, i \, B \gamma^a\}\, |\psi_{oc}>\,,
\label{tildegammareduced0}
\end{eqnarray}
with $(-)^B = -1$, if $B$ is (a function of) odd products of $\gamma^a$'s,  otherwise 
$(-)^B = 1$~\cite{nh02}, the vacuum state $|\psi_{oc}>$ is defined in 
Eq.~(\ref{vaccliffodd}) of Subsect.~\ref{basisvectors}.
%

%

\vspace{2mm}

After the postulate of Eq.~(\ref{tildegammareduced0}) it follows:\\
{\bf a.} The Clifford subalgebra described by $\tilde{\gamma}^{a}$'s looses its meaning 
for the description of the internal space of quantum fields.\\
{\bf b.} The ``basis vectors'' which are  superposition of  odd or even products of 
$\gamma^a$'s obey the postulates for the second quantized fields for fermions or 
bosons, respectively, Sect.\ref{basisvectors}.\\
{\bf c.} It can be proven that the relations presented in Eq.~(\ref{gammatildeantiher0})
remain valid also after the postulate of Eq.~(\ref{tildegammareduced0}). The proof is
presented in Ref.~(\cite{nh2021RPPNP}, App.~I, Statement~3a).\\
{\bf d.} Each irreducible representation of the Clifford odd ``basis vectors''  described by
$\gamma^{a}$'s are equipped by the quantum numbers of  the Cartan subalgebra 
members of $\tilde{S}^{ab}$, chosen in Eq.~(\ref{cartangrasscliff}), 
 as follows
\begin{small}
\begin{eqnarray}
&&S^{03}, S^{12}, S^{56}, \cdots, S^{d-1 \;d}\,, \nonumber\\
&&\tilde{S}^{03}, \tilde{S}^{12}, \tilde{S}^{56}, \cdots,  \tilde{S}^{d-1\; d}\,, 
\nonumber\\
&&{\cal {\bf S}}^{ab} = S^{ab} +\tilde{S}^{ab}=
 i \, (\theta^{a} \frac{\partial}{\partial \theta_{b}} - 
 \theta^{b} \frac{\partial}{\partial \theta_{a}})\,.
\label{cartangrasscliff}
\end{eqnarray}
\end{small}
%
%
After the  postulate of Eq.~(\ref{tildegammareduced0}) no vector space of 
$\tilde{\gamma}^{a}$'s needs to be taken into account for the description of 
the internal space of either fermions or bosons, in agreement with the observed 
properties of fermions and bosons. Also the Grassmann algebra is reduced to only
one of the Clifford subalgebras. 
The operator $\tilde{\gamma}^a$ will from now on be used to describe the 
properties of fermion ``basis vectors'', determining by $\tilde{S}^{ab}=
\frac{i}{4}(\tilde{\gamma}^a \tilde{\gamma}^b - \tilde{\gamma}^b \tilde{\gamma}^a)$
the ``family'' quantum numbers of the irreducible representations of the Lorentz 
group in internal space of fermions, $S^{ab}$, and the properties  of bosons 
``basis vectors'' determined by ${\cal S}^{ab}= S^{ab} + \tilde{S}^{ab}$. 
We shall see that while the fermion ``basis vectors'' appear in ``families'', the 
boson ``basis vectors'' have no ``families'' and manifest properties of the 
gauge fields of the corresponding  fermion fields.  In App.~\ref{basis3+1} the
case of $d=(3+1)$ is discussed.

%

$\tilde{\gamma}^a$'s  equip each irreducible representation of the Lorentz group 
(with the infinitesimal generators $S^{ab}=\frac{i}{4} \{\gamma^a, \,
\gamma^b\}_{-}$) when applying on  the Clifford odd ``basis vectors'' (which are 
superposition of odd products of $\gamma^{a's}$) with the family quantum 
numbers (determined by $\tilde{S}^{ab}=\frac{i}{4} \{\tilde{\gamma}^a, 
\,\tilde{\gamma}^b\}_{-}$). 

Correspondingly the Clifford odd ``basis vectors''  (they are the superposition of odd 
products of $\gamma^{a}$'s) form $2^{\frac{d}{2}-1}$ families, with the 
quantum number $f$, each family  has $2^{\frac{d}{2}-1}$ members, $m$. 
They offer the description of the second quantized fermion fields.

The Clifford even ``basis vectors'' (they are the superposition of even products of
$\gamma^a$'s) have no families, as we shall see in what follows, but they do
carry both quantum numbers, $f$ and $m$,
offering the description of the second quantized boson fields as the gauge
fields of the second quantized fermion fields. The generators of the Lorentz
transformations in the internal space of the Clifford even ``basis vectors'' are
${\bf {\cal S}}^{ab}= S^{ab} + \tilde{S}^{ab}$.

Properties of the Clifford odd and the Clifford even ``basis vectors'' are discussed
in the following subsection.

\vspace{1mm}   
\subsection{``Basis vectors''  of fermions and bosons in even and odd dimensional spaces}
\label{basisvectors} 

\vspace{2mm}

This subsection is a short overview of similar sections of several articles of the author, 
like~\cite{nIARD2022,n2022epjc,n2023MDPI,2020PartIPartII}.

After the reduction of the two Clifford subalgebras to only one, 
Eq.~(\ref{tildegammareduced0}), we only need to define  ``basis vectors''  for the case 
that the internal space of second quantized fields is described by superposition of odd or
 even products $\gamma^{a}$'s~\footnote{
In Ref.~\cite{nh2021RPPNP}, the reader can find in 
Subsects.~(3.2.1 and 3.2.2) definitions for the ``basis vectors'' for the Grassmann and 
the two Clifford subalgebras, which are products of nilpotents and projectors chosen 
to be the eigenvectors of the corresponding Cartan subalgebra members of the Lorentz 
algebras presented in Eq.~(\ref{cartangrasscliff}).}.

Let us use the technique which makes ``basis vectors''   products of nilpotents and 
projectors~\cite{norma93,nh02} which are eigenvectors of the 
(chosen) Cartan subalgebra members, Eq.~(\ref{cartangrasscliff}), of the Lorentz 
algebra in the space of $\gamma^{a}$'s, either  in the case of the Clifford odd 
or in the case of the Clifford even products of  $\gamma^{a}$'s. \\
There  are  in  even-dimensional spaces $\frac{d}{2}$ members of the Cartan subalgebra, 
Eq.~(\ref{cartangrasscliff}). In odd-dimensional spaces there are $\frac{d-1}{2}$ members 
of the Cartan subalgebra.

One finds in even dimensional spaces for any of the $\frac{d}{2}$ Cartan subalgebra 
member, $S^{ab}$ 
applying on a nilpotent  $\stackrel{ab}{(k)}$ or on projector
$\stackrel{ab}{[k]}$
\begin{small}
\begin{eqnarray}
\label{nilproj}
\stackrel{ab}{(k)}:&=&\frac{1}{2}(\gamma^a + 
\frac{\eta^{aa}}{ik} \gamma^b)\,, \;\;\; (\stackrel{ab}{(k)})^2=0\, , \nonumber \\
\stackrel{ab}{[k]}:&=&
\frac{1}{2}(1+ \frac{i}{k} \gamma^a \gamma^b)\,, \;\;\;(\stackrel{ab}{[k]})^2=
\stackrel{ab}{[k]},
\end{eqnarray}
\end{small}
the relations
\begin{small}
\begin{eqnarray}
\label{signature0}
S^{ab} \,\stackrel{ab}{(k)} = \frac{k}{2}  \,\stackrel{ab}{(k)}\,,\quad && \quad
\tilde{S}^{ab}\,\stackrel{ab}{(k)} = \frac{k}{2}  \,\stackrel{ab}{(k)}\,,\nonumber\\
S^{ab}\,\stackrel{ab}{[k]} =  \frac{k}{2}  \,\stackrel{ab}{[k]}\,,\quad && \quad 
\tilde{S}^{ab} \,\stackrel{ab}{[k]} = - \frac{k}{2}  \,\,\stackrel{ab}{[k]}\,,
\end{eqnarray}
\end{small}
with  $k^2=\eta^{aa} \eta^{bb}$~\footnote{Let us prove one of the relations in 
Eq.~(\ref{signature0}): $S^{ab}\, \stackrel{ab}{(k)}= \frac{i}{2} \gamma^a 
\gamma^b \frac{1}{2} (\gamma^a +\frac{\eta^{aa}}{ik} \gamma^b)=
\frac{1}{2^2}\{ -i (\gamma^a)^2 \gamma^b + i (\gamma^b)^2 \gamma^a 
\frac{\eta^{aa}}{ik}\}= \frac{1}{2} \frac{\eta^{aa}\eta^{bb}}{k}
\frac{1}{2} \{\gamma^a + \frac{k^2}{\eta^{bb} ik}\gamma^b\}$. For 
$k^2 = \eta^{aa} \eta^{bb}$ the first relation follows.}, 
demonstrating that the eigenvalues of 
$S^{ab}$ on nilpotents and projectors expressed with $\gamma^a$ differ from the 
eigenvalues of $\tilde{S}^{ab}$ on  nilpotents and projectors expressed with 
$\gamma^a$, so that $\tilde{S}^{ab}$ can be used to equip each irreducible 
representation of $S^{ab}$ with the ''family'' quantum number.~\footnote{
 The reader can find the proof of Eq.~(\ref{signature0})  also in Ref.~\cite{nh2021RPPNP}, 
 App.~(I).}

We define in even $d$ the ``basis vectors'' as algebraic, $*_A$, products of nilpotents
and projectors so that each product is an eigenvector of all $\frac{d}{2}$ Cartan 
subalgebra members, Eq.(\ref{cartangrasscliff}). 
%
Fermion ``basis vectors'' are (algebraic, $*_A$,) products of  odd number of
nilpotents; each of them is the eigenvector of one of the Cartan subalgebra members,
and the rest of the projectors; again is each projector the eigenvector of one of the
Cartan subalgebra members. The boson ``basis vectors'' are (algebraic, $*_{A}$) 
products of an even number of nilpotents and the rest of the projectors. 
(In App.~\ref{basis3+1}, the reader can find concrete examples.)  

It follows that the Clifford odd ``basis vectors'', which are the superposition of odd 
products of $\gamma^{a}$, must include an odd number of nilpotents, at least one, 
while the superposition of an even products of $\gamma^{a}$, that is Clifford even 
``basis vectors'', must include an even number of nilpotents or only projectors. 

We shall see that the Clifford odd ``basis vectors'' have properties appropriate to 
describe the internal space of the second quantized fermion fields while the Clifford 
even ``basis vectors'' have properties appropriate to describe the internal space of 
the second quantized boson fields. 

\vspace{2mm}

Taking into account Eq.~(\ref{gammatildeantiher0}) one finds
\begin{small}
\begin{eqnarray}
\label{usefulrel}
 \gamma^a \stackrel{ab}{(k)}&=& \eta^{aa}\stackrel{ab}{[-k]},\; \quad
 \gamma^b \stackrel{ab}{(k)}= -ik \stackrel{ab}{[-k]}, \; \quad 
\gamma^a \stackrel{ab}{[k]}= \stackrel{ab}{(-k)},\;\quad \;\;
 \gamma^b \stackrel{ab}{[k]}= -ik \eta^{aa} \stackrel{ab}{(-k)}\,,\nonumber\\
 \tilde{\gamma^a} \stackrel{ab}{(k)} &=& - i\eta^{aa}\stackrel{ab}{[k]},\quad
 \tilde{\gamma^b} \stackrel{ab}{(k)} =  - k \stackrel{ab}{[k]}, \;\qquad  \,
 \tilde{\gamma^a} \stackrel{ab}{[k]} =  \;\;i\stackrel{ab}{(k)},\; \quad
 \tilde{\gamma^b} \stackrel{ab}{[k]} =  -k \eta^{aa} \stackrel{ab}{(k)}\,, 
 \nonumber\\
\stackrel{ab}{(k)}^{\dagger} &=& \eta^{aa}\stackrel{ab}{(-k)}\,,\quad 
(\stackrel{ab}{(k)})^2 =0\,, \quad \stackrel{ab}{(k)}\stackrel{ab}{(-k)}
=\eta^{aa}\stackrel{ab}{[k]}\,,\nonumber\\
\stackrel{ab}{[k]}^{\dagger} &=& \,\stackrel{ab}{[k]}\,, \quad \quad \quad \quad
(\stackrel{ab}{[k]})^2 = \stackrel{ab}{[k]}\,, 
\quad \stackrel{ab}{[k]}\stackrel{ab}{[-k]}=0\,.
\end{eqnarray}
\end{small}
More relations are presented in App.~\ref{A}. 

 The relations in Eq.~(\ref{usefulrel}) demonstrate that the properties of ``basis vectors''
which include an odd number of nilpotents, differ essentially from the
``basis vectors'', which include an even number of nilpotents.

One namely recognizes:\\
{\bf i.} Since the Hermitian conjugated partner of a nilpotent
$\stackrel{ab}{(k)}^{\dagger}$ is $\eta^{aa}\stackrel{ab}{(-k)}$ and since
neither $S^{ab}$ nor $\tilde{S}^{ab}$ nor both can transform odd products of
nilpotents to belong to one of the $2^{\frac{d}{2}-1}$ members of one of
$2^{\frac{d}{2}-1}$ irreducible representations (families),
the Hermitian conjugated partners of the Clifford odd ``basis vectors'' must belong
to a different group of $2^{\frac{d}{2}-1}$ members of $2^{\frac{d}{2}-1}$
families.\\
Since $S^{ac}$ transforms $\stackrel{ab}{(k)} *_A \stackrel{cd}{(k')}$ into
$\stackrel{ab}{[-k]} *_A \stackrel{cd}{[-k']}$, while $\tilde{S}^{ac}$ transforms
$\stackrel{ab}{(k)} *_A \stackrel{cd}{(k')}$ into $\stackrel{ab}{[k]} *_A $
$ \stackrel{cd}{[k']}$
%
it is obvious that the Hermitian conjugated partners of the Clifford even ``basis
vectors'' must belong to the same group of $2^{\frac{d}{2}-1}\times $
$2^{\frac{d}{2}-1}$ members. Projectors are self-adjoint. \\
{\bf ii.} Since odd products of $\gamma^{a}$ anti-commute with another group
of odd products of $\gamma^{a}$, the Clifford odd ``basis vectors'' anti-commute,
manifesting in a tensor product, $*_{T}$, with the basis in ordinary space (together
with the corresponding Hermitian conjugated partners) properties of the anti-commutation
relations postulated by Dirac for the second quantized fermion fields~\footnote{
So far, we multiply nilpotents and projectors, or products of nilpotents and projectors
forming ``basis vectors'', among themselves. With the tensor product, $*_{T}$, we
include the basis in ordinary space.}.
The creation and annihilation operators, which
include the internal space of fermions and bosons described by ``basis vectors'',
the anti-commutativity or commutativity of which determine properties of the
``basis vectors'', fulfil the postulates of the second quantized fermion and boson
fields. Basis of ordinary space commute as presented in Eq.(31).   
App.~(\ref{basis3+1})   discuses the creation and annihilation operators.  
\\
%
The Clifford even ``basis vectors'' correspondingly fulfil, in a tensor product, 
$*_{T}$,
with the basis in ordinary space, the commutation relations for the second quantized
boson fields.\\
{\bf iii.} The Clifford odd ``basis vectors'' have all the eigenvalues of the Cartan
subalgebra members equal to either $\pm \frac{1}{2}$ or to $\pm \frac{i}{2}$.\\
The Clifford even ``basis vectors'' have all the eigenvalues of the Cartan subalgebra
members ${\bf {\cal S}}^{ab}=S^{ab} + \tilde{S}^{ab}$ equal to either 
$\pm 1$ and zero or to $\pm i$ and zero.

In odd-dimensional spaces the ``basis vectors'' can not be products of only nilpotents 
and projections. As we shall see in Subsect.~\ref{dodd}, half of ``basis vectors'' can be
chosen as products of nilpotents and projectors, the rest can be obtained from the first 
half by the application of $S^{0 d}$ on the first half. 

We shall demonstrate, shortly overviewing~\cite{n2023MDPI}, that the second half of 
the ``basis vectors'' have unusual properties:
The Clifford odd ``basis vectors have properties of the Clifford even ``basis vectors'',
the Clifford  even ``basis vectors'' have properties of the Clifford odd ``basis vectors''.

\subsubsection{Clifford odd and even ``basis vectors'' in even $d$}
\label{deven}

Let us define Clifford odd and even ``basis vectors'' as products of nilpotents and projectors
in even-dimensional spaces.

\vspace{3mm}

{\bf a.} $\;\;$ {\it Clifford odd ``basis vectors''  }

\vspace{2mm}

This part overviews several papers with the same topic~(\cite{nh2021RPPNP,n2023MDPI} and
references therein).  

The Clifford odd  ``basis vectors''  must be products of an odd number of nilpotents, and 
the rest, up to $\frac{d}{2}$,  of projectors,  each nilpotent and each projector must be the 
``eigenstate'' of one of the members of the Cartan subalgebra, Eq.~(\ref{cartangrasscliff}), 
correspondingly are the ``basis vectors'' eigenstates of all the members of the Lorentz 
algebra: $S^{ab}$'s determine $2^{\frac{d}{2}-1}$ 
members of one family, $\tilde{S}^{ab}$'s transform  each member of one family to 
the same member of the rest of $2^{\frac{d}{2}-1}$ families. 

Let us call the Clifford odd ``basis vectors''  $\hat{b}^{m \dagger}_{f}$,  if it is the 
$m^{th}$ membership  of the family $f$. The Hermitian conjugated partner of 
$\hat{b}^{m \dagger}_{f}$ is called $\hat{b}^{m}_{f} \,(=
(\hat{b}^{m \dagger}_{f})^{\dagger}$.

Let us start in  $d=2(2n+1)$ with the ``basis vector'' $\hat{b}^{1 \dagger}_{1}$ 
which is the product of only nilpotents, all the rest members belonging to the $f=1$ 
family follow by the application of $S^{01}$, $S^{03}$, $ \dots, S^{0d}, S^{15}$,
$\dots, S^{1d}, S^{5 d}\dots, S^{d-2\, d}$. They are presented on the left-hand side.
Their Hermitian conjugated partners are presented on the right-hand side. 
The algebraic product mark $*_{A}$ among nilpotents and projectors is skipped.
\begin{small}
\begin{eqnarray}
\label{allcartaneigenvec}
&& \qquad  \qquad \qquad \qquad \qquad \qquad    d=2(2n+1)\, ,\nonumber\\
&& \hat{b}^{1 \dagger}_{1}=\stackrel{03}{(+i)}\stackrel{12}{(+)} \stackrel{56}{(+)}
\cdots \stackrel{d-1 \, d}{(+)}\,,\qquad  \qquad \qquad \quad \quad
\hat{b}^{1}_{1}=\stackrel{03}{(-i)}\stackrel{12}{(-)}\cdots \stackrel{d-1 \, d}{(-)}\,,
\nonumber\\
&&\hat{b}^{2 \dagger}_{1} = \stackrel{03}{[-i]} \stackrel{12}{[-]} 
\stackrel{56}{(+)} \cdots \stackrel{d-1 \, d}{(+)}\,,\qquad \qquad \qquad \qquad\;\;
\hat{b}^{2 }_{1} = \stackrel{03}{[-i]} \stackrel{12}{[-]} 
\stackrel{56}{(-)} \cdots \stackrel{d-1 \, d}{(-)}\,,\nonumber\\
&& \cdots \qquad  \qquad \qquad \qquad \qquad  \qquad \qquad \qquad \qquad\;
\cdots \nonumber\\
&&\hat{b}^{2^{\frac{d}{2}-1} \dagger}_{1} = \stackrel{03}{[-i]} \stackrel{12}{[-]} 
\stackrel{56}{(+)} \dots \stackrel{d-3\,d-2}{[-]}\;\stackrel{d-1\,d}{[-]}\,, \qquad
\hat{b}^{2^{\frac{d}{2}-1} \dagger}_{1} = \stackrel{03}{[-i]} \stackrel{12}{[-]} 
\stackrel{56}{(-)} \stackrel{78}{[-]} \dots \stackrel{d-3\,d-2}{[-]}\;\stackrel{d-1\,d}{[-]}\,,
\nonumber\\
&& \cdots\,, \qquad \qquad  \qquad \qquad \qquad \qquad \qquad \qquad  \cdots\,.
\end{eqnarray}
\end{small}

In $d=4n$ the choice of the starting ``basis vector''  with maximal number of nilpotents
must have one projector
\begin{small}
\begin{eqnarray}
\label{allcartaneigenvec4n}
&& \qquad \qquad \qquad \qquad\qquad \qquad  d=4n\, ,\nonumber\\
&& \hat{b}^{1 \dagger}_{1}=\stackrel{03}{(+i)}\stackrel{12}{(+)}
\cdots \stackrel{d-1 \, d}{[+]}\,,\qquad \qquad \qquad \qquad\quad 
\hat{b}^{1}_{1}=\stackrel{03}{(-i)}\stackrel{12}{(-)}
\cdots \stackrel{d-1 \, d}{[+]}\,
\nonumber\\
&&\hat{b}^{2 \dagger}_{1} = \stackrel{03}{[-i]} \stackrel{12}{[-]} 
\stackrel{56}{(+)} \cdots \stackrel{d-1 \, d}{[+]}\,,\qquad \qquad \qquad \qquad
\hat{b}^{2 }_{1} = \stackrel{03}{[-i]} \stackrel{12}{[-]} 
\stackrel{56}{(-)} \cdots \stackrel{d-1 \, d}{[+]}\,,
\nonumber\\
&& \cdots \,,  \qquad \qquad \qquad \qquad\qquad  \qquad \qquad  \qquad  \quad
  \cdots\,,         \nonumber\\
&&\hat{b}^{2^{\frac{d}{2}-1} \dagger}_{1} = \stackrel{03}{[-i]} \stackrel{12}{[-]} 
\stackrel{56}{(-)} \dots \stackrel{d-3\,d-2}{[-]}\;\stackrel{d-1\,d}{[+]}\,, \qquad
\hat{b}^{2^{\frac{d}{2}-1} }_{1} = \stackrel{03}{[-i]} \stackrel{12}{[-]} 
\stackrel{56}{(-)} \dots \stackrel{d-3\,d-2}{[-]}\;\stackrel{d-1\,d}{[+]}\,, 
\nonumber\\
&& \cdots\,,\qquad \qquad \qquad \qquad\qquad  \qquad \qquad  \quad  \quad
  \cdots\,.
\end{eqnarray}
\end{small}
The Hermitian conjugated partners of the Clifford odd ``basis vectors''  
$\hat{b}^{m \dagger}_{1}$, presented in Eq.~(
\ref{allcartaneigenvec4n}) on the right-hand side, follow
if all nilpotents $\stackrel{ab}{(k)}$ of  $\hat{b}^{m \dagger}_{1}$ are transformed 
into $\eta^{aa} \stackrel{ab}{(-k)}$.

For either $d=2(2n+1)$ or for $d=4n$ all the $2^{\frac{d}{2}-1}$ families follow by 
applying $\tilde{S}^{ab}$'s on all the members of the starting family. (Or one can find 
the starting $ \hat{b}^{1 \dagger}_{f}$ for all families $f$ and then generate all the members  
$\hat{b}^{m}_{f}$ from  $\hat{b}^{1\dagger}_{f}$ by the application of $S^{ab}$
on the starting member.)

It is not difficult to see that all the ``basis vectors''   within any family, as well as  the 
``basis vectors'' among families, are orthogonal; that is, their algebraic product is zero.
The same is true within their Hermitian conjugated partners. Both can be proved by
the  algebraic multiplication using  Eqs.~(\ref{usefulrel}, \ref{graficcliff0}). 

\begin{eqnarray}
\hat{b}^{m \dagger}_f *_{A} \hat{b}^{m `\dagger }_{f `}&=& 0\,, 
\quad \hat{b}^{m}_f *_{A} \hat{b}^{m `}_{f `}= 0\,, \quad \forall m,m',f,f `\,. 
\label{orthogonalodd}
\end{eqnarray}

When we choose the vacuum state equal to
\begin{eqnarray}
\label{vaccliffodd}
|\psi_{oc}>= \sum_{f=1}^{2^{\frac{d}{2}-1}}\,\hat{b}^{m}_{f}{}_{*_A}
\hat{b}^{m \dagger}_{f} \,|\,1\,>\,,
\end{eqnarray}
for one of members $m$, which can be anyone of the odd irreducible 
representations $f$
it follows that the Clifford odd ``basis vectors''  obey the relations
%
\begin{eqnarray}
\label{almostDirac}
\hat{b}^{m}_{f} {}_{*_{A}}|\psi_{oc}>&=& 0.\, |\psi_{oc}>\,,\nonumber\\
\hat{b}^{m \dagger}_{f}{}_{*_{A}}|\psi_{oc}>&=&  |\psi^m_{f}>\,,\nonumber\\
\{\hat{b}^{m}_{f}, \hat{b}^{m'}_{f `}\}_{*_{A}+}|\psi_{oc}>&=&
 0.\,|\psi_{oc}>\,, \nonumber\\
\{\hat{b}^{m \dagger}_{f}, \hat{b}^{m' \dagger}_{f  `}\}_{*_{A}+}|\psi_{oc}>
&=& 0. \,|\psi_{oc}>\,,\nonumber\\
\{\hat{b}^{m}_{f}, \hat{b}^{m' \dagger}_{f `}\}_{*_{A}+}|\psi_{oc}>
&=& \delta^{m m'} \,\delta_{f f `}|\psi_{oc}>\,,
\end{eqnarray}
while  the normalization 
$<\psi_{oc}| \hat{b}^{m \dagger}_{f}\, *_{A}\,\hat{b}^{m \dagger}_{f}
*_{A}|\psi_{oc}> = 1$ 
is used and the anti-commutation
relation mean $\{\hat{b}^{m \dagger}_{f}, \hat{b}^{m' \dagger}_{f  `}\}_{*_{A}+}=$
$\hat{b}^{m \dagger}_{f} \,*_A\, \hat{b}^{m' \dagger}_{f  `}+
\hat{b}^{m' \dagger}_{f `} \,*_A \,\hat{b}^{m \dagger}_{f }$.

If we write the creation and annihilation operators for fermions as the tensor, 
$*_{T}$, products of ``basis vectors'' and the basis in ordinary space, the creation 
and annihilation operators fulfil Dirac's anti-commutation postulates since the 
``basis vectors'' transfer their anti-commutativity to creation and annihilation 
operators; the ordinary basis namely commute as presented in Eqs.~(\ref{creatorp}, 
\ref{eigenvalue10}). 
Describing the internal space of fermions with the Clifford odd ``basis vectors'',
makes creation operators fulfilling the Dirac postulates for the second quantized 
fermion fields: No postulates are needed.
The creation and annihilation operators for fermions and 
bosons are discussed in App.~\ref{basis3+1}, in the part with the title 
``Creation and annihilation operators''. 

It turns out, therefore, that not only the Clifford odd ``basis vectors'' offer 
the description of the internal space of fermions, they explain the second 
quantization postulates for fermions as well. 

Table~\ref{Table Clifffourplet.},  presented in Subsect.~\ref{cliffordoddevenbasis5+1},
illustrates the properties of the Clifford odd ``basis vectors''  on the case of $d=(5+1)$.

\vspace{3mm}

{\bf b.} $\;\;$ {\it Clifford even ``basis vectors'' } 

\vspace{2mm}
This part proves that the Clifford even ``basis vectors'' are in
even-dimensional
spaces offering the description of the internal spaces of boson fields --- the gauge
fields of the corresponding Clifford odd ``basis vectors": It is a new recognition,
offering a new understanding of the second quantized fermion and {\bf boson}
fields~\cite{n2022epjc}.

The Clifford even ``basis vectors'' must be products of an even number of 
nilpotents and the rest, up to $\frac{d}{2}$, of projectors; each nilpotent and 
each projector is chosen to be the ``eigenstate'' of one of the members of the 
Cartan subalgebra of the Lorentz algebra, ${\bf {\cal S}}^{ab}= S^{ab} + 
\tilde{S}^{ab}$, Eq.~(\ref{cartangrasscliff}). Correspondingly the ``basis 
vectors'' are the eigenstates of all the members of the Cartan subalgebra of 
the Lorentz algebra.

The Clifford even ``basis vectors'' appear in two groups, each group has  
$2^{\frac{d}{2}-1}\times $ $2^{\frac{d}{2}-1}$ members. The members 
of one group can not be reached from the members of another group by 
either $S^{ab}$'s or $\tilde{S}^{ab}$'s or both.

$S^{ab}$ and $\tilde{S}^{ab}$ generate from the starting ``basis vector'' of
each group all the $2^{\frac{d}{2}-1} \times$ $2^{\frac{d}{2}-1}$ members.
Each group contains the Hermitian conjugated partner of any member;
$2^{\frac{d}{2}-1}$ members of each group are products of only (self adjoint)
projectors.

Let us call the Clifford even ``basis vectors''  ${}^{i}\hat{\cal A}^{m \dagger}_{f}$,
where $i=(I,II)$ denotes the two groups of Clifford even ``basis vectors'', while
$m$ and $f$ determine membership of ``basis vectors'' in any of the two groups, $I$
or $II$.
%
\begin{eqnarray}
\label{allcartaneigenvecevenI} 
d&=&2(2n+1)\nonumber\\
{}^I\hat{{\cal A}}^{1 \dagger}_{1}=\stackrel{03}{(+i)}\stackrel{12}{(+)}\cdots 
\stackrel{d-1 \, d}{[+]}\,,\qquad &&
{}^{II}\hat{{\cal A}}^{1 \dagger}_{1}=\stackrel{03}{(-i)}\stackrel{12}{(+)}\cdots 
\stackrel{d-1 \, d}{[+]}\,,\nonumber\\
{}^I\hat{{\cal A}}^{2 \dagger}_{1}=\stackrel{03}{[-i]}\stackrel{12}{[-]} 
\stackrel{56}{(+)} \cdots \stackrel{d-1 \, d}{[+]}\,, \qquad  && 
{}^{II}\hat{{\cal A}}^{2 \dagger}_{1}=\stackrel{03}{[+i]}\stackrel{12}{[-]} 
\stackrel{56}{(+)} \cdots \stackrel{d-1 \, d}{[+]}\,,
\nonumber\\ 
{}^I\hat{{\cal A}}^{3 \dagger}_{1}=\stackrel{03}{(+i)} \stackrel{12}{(+)} 
\stackrel{56}{(+)} \cdots \stackrel{d-3\,d-2}{[-]}\;\stackrel{d-1\,d}{(-)}\,, \qquad &&
{}^{II}\hat{{\cal A}}^{3 \dagger}_{1}=\stackrel{03}{(-i)} \stackrel{12}{(+)} 
\stackrel{56}{(+)} \cdots \stackrel{d-3\,d-2}{[-]}\;\stackrel{d-1\,d}{(-)}\,,  \nonumber\\
\dots \qquad && \dots \nonumber\\
d&=&4n\nonumber\\
{}^I\hat{{\cal A}}^{1 \dagger}_{1}=\stackrel{03}{(+i)}\stackrel{12}{(+)}\cdots 
\stackrel{d-1 \, d}{(+)}\,,\qquad &&
{}^{II}\hat{{\cal A}}^{1 \dagger}_{1}=\stackrel{03}{(-i)}\stackrel{12}{(+)}\cdots 
\stackrel{d-1 \, d}{(+)}\,,
\nonumber\\
{}^I\hat{{\cal A}}^{2 \dagger}_{1}= \stackrel{03}{[-i]}\stackrel{12}{[-i]} 
\stackrel{56}{(+)} \cdots \stackrel{d-1 \, d}{(+)}\,, \qquad &&
{}^{II}\hat{{\cal A}}^{2 \dagger}_{1}= \stackrel{03}{[+i]}\stackrel{12}{[-i]} 
\stackrel{56}{(+)} \cdots \stackrel{d-1 \, d}{(+)}\,,  \nonumber\\ 
{}^I\hat{{\cal A}}^{3 \dagger}_{1}=\stackrel{03}{(+i)} \stackrel{12}{(+)} 
\stackrel{56}{(+)} \cdots \stackrel{d-3\,d-2}{[-]}\;\stackrel{d-1\,d}{[-]}\,, \qquad &&
{}^{II}\hat{{\cal A}}^{3 \dagger}_{1}=\stackrel{03}{(-i)} \stackrel{12}{(+)} 
\stackrel{56}{(+)} \cdots \stackrel{d-3\,d-2}{[-]}\;\stackrel{d-1\,d}{[-]}\,\nonumber\\
\dots \qquad && \dots 
\end{eqnarray}
%
There are $2^{\frac{d}{2}-1}\times \,2^{\frac{d}{2}-1}$ Clifford  even ``basis vectors'' of
the kind ${}^{I}{\hat{\cal A}}^{m \dagger}_{f}$ and  there are $2^{\frac{d}{2}-1}$
$\times \, 2^{\frac{d}{2}-1}$ Clifford  even ``basis vectors'' of the kind
${}^{II}{\hat{\cal A}}^{m \dagger}_{f}$.

Table~\ref{Table Clifffourplet.}, presented in Subsect.~\ref{cliffordoddevenbasis5+1}, 
illustrates properties of the Clifford odd and Clifford even ``basis vectors'' on the 
case of $d=(5+1)$. Looking at this  case it is easy to evaluate properties of 
either even or odd ``basis vectors''.  We shall discuss in this subsection the 
general case  by carefully inspecting properties of both kinds of ``basis vectors''.

The Clifford even ``basis vectors''  belonging to two different groups are orthogonal due 
to the fact that they differ in the sign of one nilpotent or one projector, or the algebraic
product of a member of one group with a member of another group gives zero according
to the first two lines of Eq.~(\ref{graficcliff0}): $\stackrel{ab}{(k)}\stackrel{ab}{[k]} =0$, 
$\stackrel{ab}{[k]}\stackrel{ab}{(-k)} =0$, 
$\stackrel{ab}{[k]}\stackrel{ab}{[-k]} =0$.
\begin{eqnarray}
\label{AIAIIorth}
{}^{I}{\hat{\cal A}}^{m \dagger}_{f} *_A {}^{II}{\hat{\cal A}}^{m \dagger}_{f} 
&=&0={}^{II}{\hat{\cal A}}^{m \dagger}._{f} *_A 
{}^{I}{\hat{\cal A}}^{m \dagger}_{f}\,.
\end{eqnarray}
The members of each of these two groups have the property 
\begin{eqnarray}
\label{ruleAAI}
{}^{i}{\hat{\cal A}}^{m \dagger}_{f} \,*_A\, {}^{i}{\hat{\cal A}}^{m' \dagger}_{f `}
\rightarrow  \left \{ \begin{array} {r}
 {}^{i}{\hat{\cal A}}^{m \dagger}_{f `}\,, i=(I,II) \\
{\rm or \,zero}\,.
\end{array} \right.
\end{eqnarray}
For a chosen ($m, f, f `$) there is only one $m'$ (out of  $2^{\frac{d}{2}-1}$)
which gives nonzero contribution. 

%

Two ``basis vectors'', ${}^{i}{\hat{\cal A}}^{m \dagger}_{f}$  and 
${}^{i}{\hat{\cal A}}^{m' \dagger}_{f '}$, the algebraic product, $*_{A}$, of which 
 gives non zero contribution, ``scatter'' into the third one 
 ${}^{i}{\hat{\cal A}}^{m \dagger}_{f `}$, for $i=(I,II)$. 

\vspace{2mm} 

\begin{small}
Let us treat a particular case in $d=2(2n+1)$-dimensional internal space, like:\\
${}^{I}{\hat{\cal A}}^{m \dagger}_{f}=\stackrel{03}{(+i)}\stackrel{12}{(+)} 
\stackrel{56}{(+)}\dots \stackrel{d-3\, d-2}{(+)}\stackrel{d-1\, d}{[+]} *_{A}$
$\stackrel{03}{[-i]}\stackrel{12}{[-]} \stackrel{56}{(-)}\dots 
\stackrel{d-3\, d-2}{(-)}\stackrel{d-1\, d}{[+]}$ $\rightarrow \;\stackrel{03}{(+i)}
\stackrel{12}{(+)} \stackrel{56}{[+]}\dots \stackrel{d-3\, d-2}{[+]}
\stackrel{d-1\, d}{[+]}$, what follows if the first two lines of Eq.~(\ref{graficcliff0}) 
are taken into account. 
The eigenvalues of the Cartan subalgebra members of  $\stackrel{03}{(+i)}\stackrel{12}{(+)} 
\stackrel{56}{(+)}\dots \stackrel{d-3\, d-2}{(+)}\stackrel{d-1\, d}{[+]}$ are 
($i,1,1,1,\dots,1,0$), of  $\stackrel{03}{[-i]}\stackrel{12}{[-]} \stackrel{56}{(-)}\dots 
\stackrel{d-3\, d-2}{(-)}\stackrel{d-1\, d}{[+]}$ are ($0, 0, -1,-1,\dots,  -1, 0$),  and
of $\stackrel{03}{(+i)}\stackrel{12}{(+)} 
\stackrel{56}{[+]}\dots \stackrel{d-3\, d-2}{[+]}\stackrel{d-1\, d}{[+]}$ are
($i,1,0,0,\dots,0,0$). The sum of the Cartan subalgebra eigenvalues of the two scattered  
Clifford even ``basis vectors'' leads to the eigenvalues ($i,1,0,0,\dots,0,0$) of the third 
Clifford even ``basis vector''.
\end{small}

\vspace{2mm}
 
It remains to evaluate the algebraic application, $*_{A}$, of the Clifford even ``basis vectors'' 
${}^{I,II}{\hat{\cal A}}^{m \dagger}_{f }$ on the Clifford odd ``basis vectors'' 
$ \hat{b}^{m' \dagger}_{f `} $. One finds, taking into account Eq.~(\ref{graficcliff0}),
   for ${}^{I}{\hat{\cal A}}^{m \dagger}_{f }$
\begin{eqnarray}
\label{calIAb1234gen}
{}^{I}{\hat{\cal A}}^{m \dagger}_{f } \,*_A \, \hat{b}^{m' \dagger }_{f `}
\rightarrow \left \{ \begin{array} {r} \hat{b }^{m \dagger}_{f `}\,, \\
{\rm or \,zero}\,,
\end{array} \right.
\end{eqnarray}

For each ${}^{I}{\hat{\cal A}}^{m \dagger}_{f}$  there are among 
$2^{\frac{d}{2}-1}\times \;2^{\frac{d}{2}-1}$ members of the Clifford odd 
``basis vectors'' (describing the internal space of fermion fields) 
$2^{\frac{d}{2}-1}$ members, $\hat{b}^{m' \dagger}_{f `}$, fulfilling the
relation of Eq.~(\ref{calIAb1234gen}). All the rest ($2^{\frac{d}{2}-1}\times 
\, (2^{\frac{d}{2}-1}-1)$)  Clifford odd ``basis vectors'' give zero contributions.
 Or equivalently, there are 
$ 2^{\frac{d}{2}-1}$ pairs of quantum numbers $(f,m')$ for which 
$\hat{b }^{m \dagger}_{f `}\ne 0$. \\

\vspace{2mm}

Taking into account Eq.~(\ref{graficcliff0}) one finds
\begin{eqnarray}
\label{calbIA1234gen}
 \hat{b}^{m \dagger }_{f } *_{A} {}^{I}{\hat{\cal A}}^{m'  \dagger}_{f `} = 0\,, \quad
 \forall (m, m`, f, f `)\,.
\end{eqnarray}
\vspace{2mm}
\begin{small}
Let us treat a particular case in $d=2(2n+1)$-dimensional space:\\
${}^{I}{\hat{\cal A}}^{m \dagger}_{f} (\equiv \stackrel{03}{(+i)}\stackrel{12}{(+)} 
\stackrel{56}{(+)}\dots \stackrel{d-3\, d-2}{(+)}\stackrel{d-1\, d}{[+]} ) *_{A}$
 $\hat{b}^{m' \dagger }_{f `} (\equiv \stackrel{03}{(-i)}\stackrel{12}{(-)} 
\stackrel{56}{(-)}\dots \stackrel{d-3\, d-2}{(-)}\stackrel{d-1\, d}{(+)}) \rightarrow
\hat{b}^{m \dagger }_{f `} (\equiv \stackrel{03}{[+i]}\stackrel{12}{[+]} 
\stackrel{56}{[+]}\dots \stackrel{d-3\, d-2} {[+]}\stackrel{d-1\, d}{(+)}$. 
%
The ${\cal {\bf{S}}}^{ab}$ (meaning ${\cal {\bf{S}}}^{03}, {\cal {\bf{S}}}^{12},
{\cal {\bf{S}}}^{56}, \dots {\cal {\bf{S}}}^{d-1\,d}$) say for  the above case
that the boson field with the quantum numbers $(i, 1, 1, \dots, 1, 0)$ when ``scattering''
on the fermion field with the Cartan subalgebra quantum numbers ($S^{03}, S^{1,2}, 
S^{56}\dots S^{d-3\, d-2}, S^{d-1\,d}$) $=(-\frac{i}{2}, -\frac{1}{2},  -\frac{1}{2},
\dots, -\frac{1}{2}, \frac{1}{2})$, and the family quantum numbers $(-\frac{i}{2}, 
-\frac{1}{2},  -\frac{1}{2}, \dots,$ $-\frac{1}{2}, \frac{1}{2})$ transfers to the fermion 
field its quantum numbers $(i, 1, 1, \dots, 1, 0)$, transforming fermion family members 
quantum numbers to   $ (\frac{i}{2}, \frac{1}{2}, \frac{1}{2},
\dots, \frac{1}{2}, \frac{1}{2})$, leaving family quantum numbers unchanged. 
\end{small}
\vspace{2mm}

Eqs.~(\ref{calIAb1234gen}, \ref{calbIA1234gen}) demonstrates that 
${}^{I}{\hat{\cal A}}^{m \dagger}_{f}$, 
applying on $\hat{b}^{m' \dagger }_{f `} $, transforms the Clifford odd ``basis vector''
into another Clifford  odd   ``basis vector'' of the same family, transferring to the
   Clifford odd ``basis vector''  integer spins, or gives zero.
   
 For  ``scattering'' the Clifford even ``basis vectors'' 
${}^{II}{\hat{\cal A}}^{m \dagger}_{f }$ on the Clifford odd ``basis vectors'' 
$ \hat{b}^{m' \dagger}_{f `} $ it follows 
%
\begin{eqnarray}
\label{calIIAb1234gen}
{}^{II}{\hat{\cal A}}^{m \dagger}_{f } \,*_A \, \hat{b}^{m' \dagger }_{f `}= 0\,,\;\;
\forall (m,m',f,f `)\,,
\end{eqnarray}
while we get 
\begin{eqnarray}
\label{calbIIA1234gen}
\hat{b}^{m \dagger }_{f } *_{A} {}^{II}{\hat{\cal A}}^{m' \dagger}_{f `} \,
\rightarrow \left \{ \begin{array} {r} \hat{b }^{m \dagger}_{f ``}\,, \\
{\rm or \,zero}\,,
\end{array} \right.
\end{eqnarray}
For each $\hat{b}^{m \dagger}_{f}$ 
 there are among $2^{\frac{d}{2}-1}$
$\times \;2^{\frac{d}{2}-1}$ members of the Clifford even ``basis vectors'' 
(describing the internal space of boson  fields) \,,  
${}^{II}{\hat{\cal A}}^{m' \dagger}_{f `}$,  
$2^{\frac{d}{2}-1}$ members (with appropriate $f `$ and $m'$) 
fulfilling the relation of Eq.~(\ref{calbIIA1234gen}) while  $f ``$ runs over
$(1 - 2^{\frac{d}{2}-1})$.

All the rest ($2^{\frac{d}{2}-1}\times \;(2^{\frac{d}{2}-1}-1)$)  Clifford even``basis 
vectors'' give zero contributions.

 Or equivalently, there are 
$ 2^{\frac{d}{2}-1}$ pairs of quantum numbers $(f  ',m')$ for which 
$\hat{b }^{m \dagger}_{f }$ and ${}^{II}{\hat{\cal A}}^{m' \dagger}_{f `}$ give 
non zero contribution. 
\space{2mm}
\begin{small}
Let us treat a particular case in $d=2(2n+1)$-dimensional space:\\
 $\hat{b}^{m \dagger }_{f } (\equiv \stackrel{03}{(-i)}\stackrel{12}{(-)} 
\stackrel{56}{(-)}\dots \stackrel{d-3\, d-2}{(-)}\stackrel{d-1\, d}{(+)})*_{A}$
${}^{II}{\hat{\cal A}}^{m `\dagger}_{f `} (\equiv \stackrel{03}{(+i)}\stackrel{12}{(+)} 
\stackrel{56}{(+)}\dots \stackrel{d-3\, d-2}{(+)}\stackrel{d-1\, d}{[-]} ) \rightarrow$
 $\hat{b}^{m \dagger }_{f  `'} (\equiv \stackrel{03}{[-i]}\stackrel{12}{[-]} $
$\stackrel{56}{[-]}\dots \stackrel{d-3\, d-2}{[-]}\stackrel{d-1\, d}{(+)}) $
%
 When the fermion field with the Cartan subalgebra family members quantum numbers 
 ($S^{03}, S^{12}, S^{56} \dots $ $S^{d-3\, d-2}, S^{d-1\,d}$) 
 $=(-\frac{i}{2}, -\frac{1}{2}, 
  -\frac{1}{2}, \dots, -\frac{1}{2}, \frac{1}{2})$  and family quantum numbers 
  ($\tilde{S}^{03}, \tilde{S}^{12}, \tilde{S}^{56}\dots \tilde{S}^{d-3\, d-2},
  \tilde{S}^{d-1\,d}$) $ (-\frac{i}{2}, -\frac{1}{2},  -\frac{1}{2}, \dots, -\frac{1}{2}, 
  \frac{1}{2})$ 
 ``absorbs'' a boson field  with the Cartan subalgebra  quantum numbers 
 ${\cal {\bf{S}}}^{ab}$ (meaning ${\cal {\bf{S}}}^{03}, {\cal {\bf{S}}}^{12},
{\cal {\bf{S}}}^{56}, \dots {\cal {\bf{S}}}^{d-1\,d}$)  equal to 
 $(i, 1, 1, \dots, 1, 0)$,  the fermion field changes the family quantum numbers 
  ($\tilde{S}^{03}, \tilde{S}^{12}, \tilde{S}^{56}\dots \tilde{S}^{d-3\, d-2},
  \tilde{S}^{d-1\,d}$)  to $ (\frac{i}{2}, \frac{1}{2},  \frac{1}{2}, \dots, \frac{1}{2}, 
  \frac{1}{2})$, keeping family members quantum numbers unchanged. \\ 
 \end{small}
\vspace{2mm}

Eqs.~(\ref{calIIAb1234gen}, \ref{calbIIA1234gen}) demonstrate that 
${}^{II}{\hat{\cal A}}^{m' \dagger}_{f'}$, ``absorbed'' by 
$\hat{b}^{m \dagger }_{f } $, transforms the Clifford odd ``basis vector''
into the Clifford  odd   ``basis vector'' of the same family member and of 
another family, or gives zero.
   
The Clifford even ``basis vectors'' offer the description
of the internal space of the gauge fields of the corresponding fermion fields.

While the Clifford odd ``basis vectors'', $\hat{b}^{m \dagger}_{f}$, offer the
description of the internal space of the second quantized anti-commuting fermion
fields, appearing in families, the Clifford even ``basis vectors'',
${}^{I,II}{\hat{\cal A}}^{m \dagger}_{f }$, offer the description of the internal
space of the second quantized commuting boson fields, having no families and
appearing in two groups. One of the two groups,
${}^{I}{\hat{\cal A}}^{m \dagger}_{f }$, transferring their integer quantum
numbers to the Clifford odd ``basis vectors'', $\hat{b}^{m \dagger}_{f}$,
changes the family members quantum numbers leaving the family quantum
numbers unchanged. The second group, transferring their integer quantum
numbers to the Clifford odd ``basis vector'', changes the family quantum
numbers leaving the family members quantum numbers unchanged.

{\it Both groups of Clifford even ``basis vectors'' manifest as the gauge fields
of the corresponding fermion fields: One concerning the family members
quantum numbers, the other concerning the family quantum numbers.}

We shall discus properties of the Clifford even and odd ``basis vectors'' for $d=(5+1)$-
dimensional internal spaces in Subsect.~\ref{cliffordoddevenbasis5+1} in more details.

\subsubsection{Clifford odd and even ``basis vectors'' in $d$ odd}
\label{dodd}
 %
 %

Let us shortly overview properties of the fermion and boson ``basis vectors'' in odd
dimensional spaces, as presented in Ref.~\cite{n2023MDPI}, Subsect.~2.2.

In even dimensional spaces the Clifford odd ``basis vectors'' fulfil the postulates for 
the second quantized fermion fields, Eq.~(\ref{almostDirac}), and the Clifford even 
''basis vectors'' have the properties of the internal spaces of their corresponding gauge 
fields, Eqs.~(\ref{ruleAAI}, \ref{calIAb1234gen}, \ref{calbIIA1234gen}). In odd 
dimensional spaces, the Clifford odd and even ''basis vectors'' have unusual properties 
resembling properties of the internal spaces of the Faddeev--Popov ghosts, as we 
described in~\cite{n2023MDPI}.

In $d=(2n+1)$-dimensional cases, $n=1,2,\dots$,  half of the ``basis vectors'', 
$2^{\frac{2n}{2}-1}$ $\times \,2^{\frac{2n}{2}-1}$, can be taken from the 
$2n$-dimensional part of space, presented in Eqs.~(\ref{allcartaneigenvec}, 
\ref{allcartaneigenvec4n}, \ref{allcartaneigenvecevenI}, \ref{ruleAAI}).  

The rest  of  the ``basis vectors'' in odd dimensional spaces, $2^{\frac{2n}{2}-1}$ 
$\times \, 2^{\frac{2n}{2}-1}$, follow if  $S^{0 \,2n+1}$  is applied on these half of the 
``basis vectors''. Since $S^{0 \,2n+1}$ are Clifford even operators,  they do not change 
the oddness or evenness of the ``basis vectors''.

For the Clifford odd ``basis vectors'',  the $2^{\frac{d-1}{2}-1}$ members appearing in 
$2^{\frac{d-1}{2}-1}$ families and representing the part which is the same as in even,
$d=2n$, dimensional space are present on the left-hand side of 
Eq.~(\ref{allcartaneigenvecbdgen}), the part obtained by  applying $S^{0 \,2n+1}$ on 
the one of the left-hand side is presented on the right hand side.  Below  the ``basis 
vectors'' and their Hermitian conjugated partners are presented.
\begin{small}
\begin{eqnarray}
\label{allcartaneigenvecbdgen}
 d=&&2(2n+1)+1\, \nonumber\\
 \hat{b}^{1 \dagger}_{1}=\stackrel{03}{(+i)}\stackrel{12}{(+)} \stackrel{56}{(+)}
\cdots \stackrel{d-2 \, d-1}{(+)} \,,\quad 
&& \hat{b}^{1 \dagger}_{2^{\frac{d-1}{2}-1}+1}=\stackrel{03}{[-i]}
\stackrel{12}{(+)} \stackrel{56}{(+)} \cdots \stackrel{d-2 \, d-1}{(+)} \gamma^{d}\,,
\nonumber\\
%
\cdots \quad 
&& \cdots\nonumber\\
\hat{b}^{2^{\frac{d-1}{2}-1} \dagger}_{1} = \stackrel{03}{[-i]} \stackrel{12}{[-]} 
\stackrel{56}{(+)} \dots \stackrel{d-2\,d-1}{[-]}\;, \quad
&&\hat{b}^{2^{\frac{d-1}{2}-1} \dagger}_{2^{\frac{d-1}{2}-1}+1} = \stackrel{03}{(+i)} \stackrel{12}{[-]} 
\stackrel{56}{(+)} \dots \stackrel{d-2\,d-1}{[-]}\; \gamma^{d}\,, \nonumber\\
\cdots \quad
&& \cdots \,,\nonumber\\
&& \cdots \,,\nonumber\\
 \hat{b}^{1}_{1}=\stackrel{03}{(-i)}\stackrel{12}{(-)} \stackrel{56}{(-)}
\cdots \stackrel{d-2 \, d-1}{(-)} \,,\quad 
&& \hat{b}^{1}_{2^{\frac{d-1}{2}-1}+1}=\stackrel{03}{[+i]}\stackrel{12}{(-)} \stackrel{56}{(-)}
\cdots \stackrel{d-2 \, d-1}{(-)} \gamma^{d}\,,\nonumber\\
\cdots \quad
&& \cdots \,.
\end{eqnarray}
\end{small}

The application of $S^{0d}$ or $\tilde{S}^{0d}$ on the left-hand side of the 
``basis vectors'' (and the Hermitian conjugated partners of both) generate the whole 
set of $2\times 2^{d-2}$ members of the Clifford odd ``basis vectors'' and their 
Hermitian conjugated partners in $d=(2n+1)$- dimensional space
 appearing on the left-hand side and the right-hand sides of 
 Eq.~(\ref{allcartaneigenvecbdgen}).\\
 

It is not difficult to see that $ \hat{b}^{m \dagger}_{2^{\frac{d-1}{2}-1}+k}$ and
$ \hat{b}^{m'}_{2^{\frac{d-1}{2}-1}+k'}$ on the right-hand side of
Eq.~(\ref{allcartaneigenvecbdgen}) obtain properties of the two groups (they are
orthogonal to each other; the algebraic products, $*_{A}$, of a member from one
group, and any member of another group give zero)
with the Hermitian conjugated partners within the same group;
they have properties of the Clifford even ``basis vectors'' from the point of view of
the Hermiticity property: The operators $\gamma^a$ are up to a constant the self-adjoint
operators, while $S^{0 d}$ transform one nilpotent into a projector. 

$S^{ab}$ do not change the Clifford oddness of $ \hat{b}^{m \dagger}_{f}$, and
$ \hat{b}^{m}_{f}$;  $ \hat{b}^{m \dagger}_{f}$ remain to be Clifford odd objects,
however, with the properties of boson fields.\\

Let us find the  Clifford even ``basis vectors'' in odd dimensional space $d=2(2n+1) +1$.
\begin{small}
\begin{eqnarray}
\label{allcartaneigenvecAdgen}
 d=&&2(2n+1)+1\, \nonumber\\
 {}^{I}{\bf {\cal A}}^{1 \dagger}_{1} =\stackrel{03}{(+i)}\stackrel{12}{(+)} \stackrel{56}{(+)} \cdots \stackrel{d-2 \, d-1}{[+]} \,,\quad 
&& {}^{I}{\bf {\cal A}}^{1 \dagger}_{2^{\frac{d-1}{2}-1}+1}=\stackrel{03}{[-i]}\stackrel{12}{(+)} \stackrel{56}{(+)}
\cdots \stackrel{d-2 \, d-1}{[+]} \gamma^{d}\,,\nonumber\\
%
\cdots \quad 
&& \cdots\nonumber\\
 {}^{I}{\bf {\cal A}}^{2^{\frac{d-1}{2}-1} \dagger}_{1} = \stackrel{03}{[-i]} 
 \stackrel{12}{[-]} \stackrel{56}{[-]} \dots \stackrel{d-2\,d-1}{[+]}\;, \quad
&& {}^{I}{\bf {\cal A}}^{2^{\frac{d-1}{2}-1} \dagger}_{2^{{d-1}{2}-1}+1} = 
\stackrel{03}{(+i)} \stackrel{12}{[-]} 
\stackrel{56}{[-]} \dots \stackrel{d-2\,d-1}{[+]}\; \gamma^{d}\,, \nonumber\\
\cdots \quad
&& \cdots \,,\nonumber\\
\cdots \quad 
&& \cdots\nonumber\\
 {}^{II}{\bf {\cal A}}^{1 \dagger}_{1} =\stackrel{03}{(-i)}\stackrel{12}{(+)} \stackrel{56}{(+)} 
 \cdots \stackrel{d-2 \, d-1}{[+]} \,,\quad 
&& {}^{II}{\bf {\cal A}}^{1 \dagger}_{2^{\frac{d-1}{2}-1}+1}=\stackrel{03}{[+i]}\stackrel{12}{(+)} 
\stackrel{56}{(+)}
\cdots \stackrel{d-2 \, d-1}{[+]} \gamma^{d}\,,\nonumber\\
\cdots \quad
&& \cdots \,.
\end{eqnarray}
\end{small}
The right hand side of Eq.~(\ref{allcartaneigenvecbdgen}), although anti-commuting, is resembling
the properties of the Clifford even ``basis vectors'' on the left hand side of 
Eq.~(\ref{allcartaneigenvecAdgen}), while the right-hand side of 
Eq.~(\ref{allcartaneigenvecAdgen}), although commuting, resembles the properties of the Clifford 
odd  ``basis vectors'', from the left hand side of Eq.~(\ref{allcartaneigenvecbdgen}): $\gamma^a$ 
are up to a constant the self adjoint operators, while $S^{0 d}$ transform one nilpotent into a 
projector (or one projector into a nilpotent). However, $S^{ab}$ do not change Clifford evenness 
of  ${}^{I}{\bf {\cal A}}^{m \dagger}_{f}, i=(I,II)$. 

\vspace{2mm}
\begin{small}
 For illustration let me copy the special case for $d=(4+1)$ from Subsect.3.2.2. of 
 Ref.~\cite{n2023MDPI}.
 \end{small}
\begin{tiny}
\begin{eqnarray}
\label{allcartaneigenvecbd4+1}
 d=&&4+1\, \nonumber\\
 && {\rm Clifford\;  odd} \nonumber\\
 \hat{b}^{1 \dagger}_{1}=\stackrel{03}{(+i)}\stackrel{12}{[+]}\,,\;\, 
 \hat{b}^{1 \dagger}_{2}=\stackrel{03}{[+i]}\stackrel{12}{(+)}\,,
 &&\hat{b}^{1\dagger}_{3} =\stackrel{03}{[-i]}\stackrel{12}{[+]
 } \gamma^{5}\,,
 \; \, \hat{b}^{1 \dagger}_{4} =\stackrel{03}{(-i)}\stackrel{12}{(+)}\gamma^5 \,,\nonumber\\
\hat{b}^{2 \dagger}_{1}=\stackrel{03}{[-i]}\stackrel{12}{(-)}\,,\;\, 
 \hat{b}^{2 \dagger}_{2}=\stackrel{03}{(-i)}\stackrel{12}{[-]}\,,
 &&\hat{b}^{2\dagger}_{3}=\stackrel{03}{(+i)}\stackrel{12}{(-)} \gamma^{5}\,, \; \, 
\hat{b}^{2\dagger}_{4}= \stackrel{03}{[+i]}\stackrel{12}{[-]}\gamma^5 \,,\nonumber\\
\hat{b}^{1}_{1} = \stackrel{03}{(-i)}\stackrel{12}{[+]} \,, \;\,
\hat{b}^{1}_{2} = \stackrel{03}{[+i]} \stackrel{12}{(-)} \,,
&&\hat{b}^{1}_{3} = \stackrel{03}{[+i]}\stackrel{12}{[+]} \gamma^{5}\,,\;\,
\hat{b}^{1}_{4}= \stackrel{03}{(-i)}\stackrel{12}{(-)} \gamma^5 \,,\nonumber\\
\hat{b}^{2}_{1} = \stackrel{03}{[-i]}\stackrel{12}{(+)} \,, \;\,
\hat{b}^{2}_{2} = \stackrel{03}{(+i)} \stackrel{12}{[-]} \,,
&&\hat{b}^{2}_{3} = \stackrel{03}{(+i)}\stackrel{12}{(+)} \gamma^{5}\,,\;\,
\hat{b}^{2}_{4}= \stackrel{03}{[-i]}\stackrel{12}{[-]} \gamma^5\,,\nonumber\\
&& \nonumber\\
&& {\rm Clifford\;  even} \nonumber\\
{}^{I}{\bf {\cal A}}^{1\dagger}_{1}= \stackrel{03}{[+i]}\stackrel{12}{[+]} \,,\;\,
{}^{I}{\bf {\cal A}}^{1\dagger}_{2}= \stackrel{03}{(+i)}\stackrel{12}{(+)} \,,
&&{}^{I}{\bf {\cal A}}^{1}_{3}= \stackrel{03}{(-i)}\stackrel{12}{[+]}\gamma^5\,,\;\,
{}^{I}{\bf {\cal A}}^{1}_{4}= \stackrel{03}{[-i]}\stackrel{12}{(+)}\gamma^5\,, 
\nonumber\\
{}^{I}{\bf {\cal A}}^{2\dagger}_{1}= \stackrel{03}{(-i)}\stackrel{12}{(-i)} \,,\;\,
{}^{I}{\bf {\cal A}}^{2\dagger}_{2}= \stackrel{03}{[-i]}\stackrel{12}{[-]} \,,
&&{}^{I}{\bf {\cal A}}^{2}_{3}= \stackrel{03}{[+i]}\stackrel{12}{(-)}
\gamma^5\,,\;\,
{}^{I}{\bf {\cal A}}^{2}_{4}= \stackrel{03}{(+i)}\stackrel{12}{[-]}\, \gamma^5\,, 
\nonumber\\
&&\nonumber\\
{}^{II}{\bf {\cal A}}^{1\dagger}_{1}= \stackrel{03}{[-i]}\stackrel{12}{[+]} \,,\;\,
{}^{II}{\bf {\cal A}}^{1\dagger}_{2}= \stackrel{03}{(-i)}\stackrel{12}{(+)} \,,
&&{}^{II}{\bf {\cal A}}^{1\dagger}_{3}= \stackrel{03}{(+i)}\stackrel{12}{[+]}
\gamma^5\,,\;\,
{}^{II}{\bf {\cal A}}^{1\dagger}_{4}= \stackrel{03}{[+i]}\stackrel{12}{(+)}\gamma^5\,, 
\nonumber\\
{}^{II}{\bf {\cal A}}^{2\dagger}_{1}= \stackrel{03}{(+i)}\stackrel{12}{(-)} \,,\;\,
{}^{II}{\bf {\cal A}}^{2\dagger}_{2}= \stackrel{03}{[+i]}\stackrel{12}{[-]} \,,
&&{}^{II}{\bf {\cal A}}^{2\dagger}_{3}= \stackrel{03}{[-i]}\stackrel{12}{(-)}
\gamma^5\,,\;\,
{}^{II}{\bf {\cal A}}^{2\dagger}_{4}= \stackrel{03}{(-i)}\stackrel{12}{[-]}\,
\gamma^5 \,.
\end{eqnarray}
\end{tiny}
\begin{small}
It can clearly be seen that the left-hand side of the Clifford odd ``basis vectors'' and the 
right-hand side of the Clifford even ``basis vectors'', although the  former are the Clifford 
odd objects and the latter are Clifford even objects, have similar properties~\cite{n2023MDPI}. 
\end{small}


\vspace{1mm}
\subsection{Example demonstrating properties of  Clifford odd and even ``basis vectors'' 
for $d=(5+1)$}
\label{cliffordoddevenbasis5+1}

\vspace{2mm}

Subsect.~\ref{cliffordoddevenbasis5+1} demonstrates the properties of the Clifford odd and even ``basis vectors'' in the special case when $d=(5+1)$ to clear up the
relations of the Clifford odd and even ``basis vectors'' to fermion and boson fields, respectively.

Table~\ref{Table Clifffourplet.} presents the $64 \,(=2^{d=6})$ ``eigenvectors" of the Cartan
subalgebra members of the Lorentz algebra, $S^{ab}$ and ${\bf \cal{S}}^{ab}$, Eq.~(\ref{cartangrasscliff}).\\

The Clifford odd ``basis vectors'' --- they appear in $4 \,(=2^{\frac{d=6}{2}-1})$
families, each family has $4$ members --- are products of an odd number of nilpotents,
either of three or one. They appear  
in  the group  named $odd \,I \,\hat{b}^{m\dagger}_f$. 
Their Hermitian conjugated partners appear in the second group  named 
$odd \,II \,\,\hat{b}^m_f$. Within each of these two groups the members are mutually 
orthogonal (which can be checked by using Eq.~(\ref{graficcliff0}));  
$\hat{b}^{m\dagger} _f *_{A} \hat{b}^{m'\dagger} _{f `} =0$ for all $(m,m', f, f `)$.
 Equivalently,  $\hat{b}^{m} _f *_{A} \hat{b}^{m'} _{f `} =0$ for all $(m,m', f, f `)$. 
The ``basis vectors'' and their Hermitian  conjugated partners are normalized as
\begin{eqnarray}
< \psi_{oc}| \hat{b}^{m} _f *_{A} \hat{b}^{m' \dagger} _{f `} |\psi_{oc} >=
\delta^{m m'} \delta_{f f `}\,, 
\label{Cliffnormalizationodd}
\end{eqnarray}
since the vacuum state $|\psi_{oc} >= \frac{1}{\sqrt{2^{\frac{d=6}{2}-1}}}$
$(\stackrel{03}{[-i]} \stackrel{12}{[-]}\stackrel{56}{[-]}+ \stackrel{03}{[-i]} 
\stackrel{12}{[+]}\stackrel{56}{[+]} + \stackrel{03}{[+i]} \stackrel{12}{[-]}
\stackrel{56}{[+]}+ \stackrel{03}{[+i]} \stackrel{12}{[+]}\stackrel{56}{[-]})$
is normalized to one: $< \psi_{oc}|\psi_{oc} >=1$.

The more extended overview of the properties of the Clifford odd ``basis vectors'' and 
their Hermitian conjugated partners for the case $d=(5+1)$  can be found in 
Ref.~\cite{nh2021RPPNP}.

\vspace{2mm}

The Clifford even ``basis vectors'' are products of an even number of nilpotents --- of 
either two  or none in this case. They are presented  in Table~\ref{Table Clifffourplet.} 
in two groups, each with $16\, (=2^{\frac{d=6}{2}-1}\times 2^{\frac{d=6}{2}-1})$  
members, as $even \, {I}\,{\bf {\cal A}}^{m \dagger}_{f}$ and 
$even \, II \, {\bf {\cal A}}^{m \dagger}_{f}$. One can easily check, using 
Eq.~(\ref{graficcliff0}), that the algebraic product 
${}^{I} {\bf {\cal A}}^{m \dagger}_{f} *_{A}$
${}^{II} {\bf {\cal A}}^{m' \dagger}_{f `} =0 = 
{}^{II} {\bf {\cal A}}^{m \dagger}_{f} *_{A}$
${}^{I} {\bf {\cal A}}^{m' \dagger}_{f `}, \forall\, (m,m',f. f `)$, 
Eq.~(\ref{AIAIIorth}).
An overview of the Clifford even ``basis vectors'' and their Hermitian conjugated 
partners for the case $d=(5+1)$ can be found in Ref.~\cite{n2022epjc}.

\begin{table*}
\begin{small}
\caption{\label{Table Clifffourplet.}  $2^d=64$ ``eigenvectors" of the Cartan subalgebra
of the Clifford odd and even algebras --- the superposition of odd and
even products of $\gamma^{a}$'s --- in $d=(5+1)$-dimensional space are presented,
divided into four groups. The first group, $odd \,I$, is chosen to represent ``basis vectors",
named ${\hat b}^{m \dagger}_f$, appearing in $2^{\frac{d}{2}-1}=4$ 
``families" ($f=1,2,3,4$), each ''family'' with $2^{\frac{d}{2}-1}=4$
``family'' members ($m=1,2,3,4$).
The second group, $odd\,II$, contains Hermitian conjugated partners of the first
group for each family separately, ${\hat b}^{m}_f=$
$({\hat b}^{m \dagger}_f)^{\dagger}$. Either $odd \,I$ or $odd \,II$ are products
of an odd number of nilpotents (one or three) and projectors (two or none).
The ``family" quantum numbers of ${\hat b}^{m \dagger}_f$, that is the eigenvalues of
$(\tilde{S}^{03}, \tilde{S}^{12},\tilde{S}^{56})$, are for the first {\it odd I }
group appearing above each ``family", the quantum
numbers of the family members $(S^{03}, S^{12}, S^{56})$ are 
written in the last three columns. 
For the Hermitian conjugated partners of {\it odd I}, presented in the group {\it odd II},
the quantum numbers $(S^{03}, S^{12}, S^{56})$ are presented above each group of the
Hermitian conjugated partners, the last three columns 
tell eigenvalues of $(\tilde{S}^{03}, \tilde{S}^{12},\tilde{S}^{56})$.
The two groups with the even number of $\gamma^a$'s, {\it even \,I} and {\it even \,II},
each group has their Hermitian conjugated partners within its group,
have the quantum numbers $f$, that is the eigenvalues of
$(\tilde{S}^{03}, \tilde{S}^{12},\tilde{S}^{56})$, written above column of
four members, the quantum numbers of the members, $(S^{03}, S^{12}, S^{56})$, are
written in the last three columns. To find the quantum numbers of $({\cal {\bf S}}^{03},
{\cal {\bf S}}^{12}, {\cal {\bf S}}^{56})$ one has to take into account that
${\cal {\bf S}}^{ab}$ $= S^{ab} + \tilde{S}^{ab} $.
 \vspace{2mm}}
 \end{small}
\begin{tiny}
\begin{center}
  \begin{tabular}{|c|c|c|c|c|c|r|r|r|}
\hline
$ $&$$&$ $&$ $&$ $&&$$&$$&$$\\
$''basis\, vectors'' $&$m$&$ f=1$&$ f=2 $&$ f=3 $&
$ f=4 $&$$&$$&$$\\ 
$(\tilde{S}^{03}, \tilde{S}^{12}, \tilde{S}^{56})$&$\rightarrow$&$(\frac{i}{2},- \frac{1}{2},-\frac{1}{2})$&$(-\frac{i}{2},-\frac{1}{2},\frac{1}{2})$&
$(-\frac{i}{2},\frac{1}{2},-\frac{1}{2})$&$(\frac{i}{2},\frac{1}{2},\frac{1}{2})$&$S^{03}$
 &$S^{12}$&$S^{56}$\\ 
\hline
$ $&$$&$ $&$ $&$ $&&$$&$$&$$\\ 
$odd \,I\; {\hat b}^{m \dagger}_f$&$1$& 
$\stackrel{03}{(+i)}\stackrel{12}{[+]}\stackrel{56}{[+]}$&
                        $\stackrel{03}{[+i]}\stackrel{12}{[+]}\stackrel{56}{(+)}$ & 
                        $\stackrel{03}{[+i]}\stackrel{12}{(+)}\stackrel{56}{[+]}$ &  
                        $\stackrel{03}{(+i)}\stackrel{12}{(+)}\stackrel{56}{(+)}$ &
                        $\frac{i}{2}$&$\frac{1}{2}$&$\frac{1}{2}$\\ 
$$&$2$&    $[-i](-)[+] $ & $(-i)(-)(+) $ & $(-i)[-][+] $ & $[-i][-](+) $ &$-\frac{i}{2}$&
$-\frac{1}{2}$&$\frac{1}{2}$\\ 
$$&$3$&    $[-i] [+](-)$ & $(-i)[+][-] $ & $(-i)(+)(-) $ & $[-i](+)[-] $&$-\frac{i}{2}$&
$\frac{1}{2}$&$-\frac{1}{2}$\\ 
$$&$4$&    $(+i)(-)(-)$ & $[+i](-)[-] $ & $[+i][-](-) $ & $(+i)[-][-]$&$\frac{i}{2}$&
$-\frac{1}{2}$&$-\frac{1}{2}$\\ 
\hline
$ $&$$&$ $&$ $&$ $&&$$&$$&$$\\ 
$(S^{03}, S^{12}, S^{56})$&$\rightarrow$&$(-\frac{i}{2}, \frac{1}{2},\frac{1}{2})$&
$(\frac{i}{2},\frac{1}{2},-\frac{1}{2})$&
$(\frac{i}{2},- \frac{1}{2},\frac{1}{2})$&$(-\frac{i}{2},-\frac{1}{2},-\frac{1}{2})$&
$\tilde{S}^{03}$
&$\tilde{S}^{12}$&$\tilde{S}^{56}$\\ 
&&
$\stackrel{03}{\;\,}\;\;\,\stackrel{12}{\;\,}\;\;\,\stackrel{56}{\;\,}$&
$\stackrel{03}{\;\,}\;\;\,\stackrel{12}{\;\,}\;\;\,\stackrel{56}{\;\,}$&
$\stackrel{03}{\;\,}\;\;\,\stackrel{12}{\;\,}\;\;\,\stackrel{56}{\;\,}$&
$\stackrel{03}{\;\,}\;\;\,\stackrel{12}{\;\,}\;\;\,\stackrel{56}{\;\,}$&
&&\\
\hline
$ $&$$&$ $&$ $&$ $&&$$&$$&$$\\ 
$odd\,II\; {\hat b}^{m}_f$&$1$ &$(-i)[+][+]$ & $[+i][+](-)$ & $[+i](-)[+]$ & $(-i)(-)(-)$&
$-\frac{i}{2}$&$-\frac{1}{2}$&$-\frac{1}{2}$\\ 
$$&$2$&$[-i](+)[+]$ & $(+i)(+)(-)$ & $(+i)[-][+]$ & $[-i][-](-)$&
$\frac{i}{2}$&$\frac{1}{2}$&$-\frac{1}{2}$\\ 
$$&$3$&$[-i][+](+)$ & $(+i)[+][-]$ & $(+i)(-)(+)$ & $[-i](-)[-]$&
$\frac{i}{2}$&$-\frac{1}{2}$&$\frac{1}{2}$\\ 
$$&$4$&$(-i)(+)(+)$ & $[+i](+)[-]$ & $[+i][-](+)$ & $(-i)[-][-]$&
$-\frac{i}{2}$&$\frac{1}{2}$&$\frac{1}{2}$\\ 
\hline
&&&&&&&&\\ 
\hline
$ $&$$&$ $&$ $&$ $&&$$&$$&$$\\ 
$(\tilde{S}^{03}, \tilde{S}^{12}, \tilde{S}^{56})$&$\rightarrow$&
$(-\frac{i}{2},\frac{1}{2},\frac{1}{2})$&$(\frac{i}{2},-\frac{1}{2},\frac{1}{2})$&
$(-\frac{i}{2},-\frac{1}{2},-\frac{1}{2})$&$(\frac{i}{2},\frac{1}{2},-\frac{1}{2})$&
$S^{03}$&$S^{12}$&$S^{56}$\\ 
&& 
$\stackrel{03}{\;\,}\;\;\,\stackrel{12}{\;\,}\;\;\,\stackrel{56}{\;\,}$&
$\stackrel{03}{\;\,}\;\;\,\stackrel{12}{\;\,}\;\;\,\stackrel{56}{\;\,}$&

$\stackrel{03}{\;\,}\;\;\,\stackrel{12}{\;\,}\;\;\,\stackrel{56}{\;\,}$&
$\stackrel{03}{\;\,}\;\;\,\stackrel{12}{\;\,}\;\;\,\stackrel{56}{\;\,}$&
&&\\ 
\hline
$ $&$$&$ $&$ $&$ $&&$$&$$&$$\\ 
$even\,I \; {}^{I}{\cal A}^{m}_f$&$1$&$[+i](+)(+) $ & $(+i)[+](+) $ & $[+i][+][+] $ & $(+i)(+)[+] $ &$\frac{i}{2}$&
$\frac{1}{2}$&$\frac{1}{2}$\\ 
$$&$2$&$(-i)[-](+) $ & $[-i](-)(+) $ & $(-i)(-)[+] $ & $[-i][-][+] $ &$-\frac{i}{2}$&
$-\frac{1}{2}$&$\frac{1}{2}$\\ 
$$&$3$&$(-i)(+)[-] $ & $[-i][+][-] $ & $(-i)[+](-) $ & $[-i](+)(-) $&$-\frac{i}{2}$&
$\frac{1}{2}$&$-\frac{1}{2}$\\ 
$$&$4$&$[+i][-][-] $ & $(+i)(-)[-] $ & $[+i](-)(-) $ & $(+i)[-](-) $&$\frac{i}{2}$&
$-\frac{1}{2}$&$-\frac{1}{2}$\\ 
\hline
$ $&$$&$ $&$ $&$ $&&$$&$$&$$\\ 
$(\tilde{S}^{03}, \tilde{S}^{12}, \tilde{S}^{56})$&$\rightarrow$&
$(\frac{i}{2},\frac{1}{2},\frac{1}{2})$&$(-\frac{i}{2},-\frac{1}{2},\frac{1}{2})$&
$(\frac{i}{2},-\frac{1}{2},-\frac{1}{2})$&$(-\frac{i}{2},\frac{1}{2},-\frac{1}{2})$&
$S^{03}$&$S^{12}$&$S^{56}$\\ 
&& 
$\stackrel{03}{\;\,}\;\;\,\stackrel{12}{\;\,}\;\;\,\stackrel{56}{\;\,}$&
$\stackrel{03}{\;\,}\;\;\,\stackrel{12}{\;\,}\;\;\,\stackrel{56}{\;\,}$&
$\stackrel{03}{\;\,}\;\;\,\stackrel{12}{\;\,}\;\;\,\stackrel{56}{\;\,}$&
$\stackrel{03}{\;\,}\;\;\,\stackrel{12}{\;\,}\;\;\,\stackrel{56}{\;\,}$&
&&\\ 
\hline
$ $&$$&$ $&$ $&$ $&&$$&$$&$$\\ 
$even\,II \; {}^{II}{\cal A}^{m}_f$&$1$& $[-i](+)(+) $ & $(-i)[+](+) $ & $[-i][+][+] $ & 
$(-i)(+)[+] $ &$-\frac{i}{2}$&
$\frac{1}{2}$&$\frac{1}{2}$\\ 
$$&$2$&    $(+i)[-](+) $ & $[+i](-)(+) $ & $(+i)(-)[+] $ & $[+i][-][+] $ &$\frac{i}{2}$&
$-\frac{1}{2}$&$\frac{1}{2}$ \\ 
$$&$3$&    $(+i)(+)[-] $ & $[+i][+][-] $ & $(+i)[+](-) $ & $[+i](+)(-) $&$\frac{i}{2}$&
$\frac{1}{2}$&$-\frac{1}{2}$\\ 
$$&$4$&    $[-i][-][-] $ & $(-i)(-)[-] $ & $[-i](-)(-) $ & $(-i)[-](-) $&$-\frac{i}{2}$&
$-\frac{1}{2}$&$-\frac{1}{2}$\\ 
\hline
 \end{tabular}
\end{center}
\end{tiny}
\end{table*}

While the Clifford odd ``basis vectors'' are (chosen to be) left handed, 
$\Gamma^{(5+1)}= -1$,  their Hermitian conjugated partners have opposite 
handedness, Eq.~\ref{Gamma} in App.~\ref{A}~\footnote{
Let us check the hadedness of the chosen representation: 
$\Gamma^{5+1} \hat{b}^{1 \dagger}_1 (\equiv \stackrel{03}{(+i)}\stackrel{12}{[+]}\stackrel{56}{[+]})=
\sqrt{(-1)^5} i^{3} (\frac{2}{i})^3 S^{03}S^{12}S^{56} (\stackrel{03}{(+i)}\stackrel{12}{[+]}\stackrel{56}{[+]})=
\frac{i^{4} 2^{3}}{i^{3}} \frac{i}{2} \frac{1}{2}\frac{1}{2} (\stackrel{03}{(+i)}\stackrel{12}{[+]}\stackrel{56}{[+]})
= -1 (\stackrel{03}{(+i)}\stackrel{12}{[+]}\stackrel{56}{[+]})$).}.

While the Clifford odd ``basis vectors'' have half integer eigenvalues of the Cartan subalgebra
members, Eq.~(\ref{cartangrasscliff}), that is of $S^{03}, S^{12}, S^{56}$ in this particular 
case of $d=(5+1)$,  the Clifford even ``basis vectors'' have integer spins, obtained by 
${\bf {\cal S}}^{03}= S^{03}+ \tilde{S}^{03}$, ${\bf {\cal S}}^{12}= 
S^{12} +\tilde{S}^{12}$, 
${\bf {\cal S}}^{56}= S^{56}+ \tilde{S}^{56}$.

Let us check what does the algebraic application, $*_A$, of 
${}^{I}{\hat{\cal A}}^{m=1 \dagger}_{f=4}$, for example, presented in 
Table~\ref{Table Clifffourplet.} in the first line of the fourth column of $even\,I$, do on 
the Clifford odd  ``basis vector'' $\hat{b}^{m=2 \dagger}_{f=2}$, presented in
 $odd \,I$  as the second member of the second column. 
 (This can easily be evaluated by taking into 
account Eq.~(\ref{graficcliff0}) for any $m$.)

\begin{small}
\begin{eqnarray}
&&{}^{I}{\hat{\cal A}}^{1 \dagger}_{4} (\equiv \stackrel{03}{(+i)}
\stackrel{12}{(+)} \stackrel{56}{[+]})  *_{A} \hat{b}^{2 \dagger}_{2} 
(\equiv \stackrel{03}{(-i)} \stackrel{12}{(-)} \stackrel{56}{(+)}) \rightarrow
\hat{b}^{1 \dagger}_{2} (\equiv \stackrel{03}{[+i]}
\stackrel{12}{[+]} \stackrel{56}{(+)}) \,,
%
%
\label{calAb1}
\end{eqnarray}
\end{small}
%
The sign $\rightarrow$ means that the relation is valid up to the constant.
The Hermitian conjugated partner of ${}^{I}{\hat{\cal A}}^{1 \dagger}_{4}$ is 
${}^{I}{\hat{\cal A}}^{2 \dagger}_{3}$. 

Let us check the Cartan subalgebra quantum numbers of this ``scattering'':
  ${}^{I}{\hat{\cal A}}^{1 \dagger}_{4}$  has  $({\bf \cal{S}}^{03},  
  {\bf \cal{S}}^{12}, {\bf \cal{S}}^{56}) = (i,1,0)$, 
 $\hat{b}^{2 \dagger}_{2}$  has $(S^{03},  S^{12}, S^{56})
 = (-\frac{i}{2}, -\frac{1}{2},  \frac{1}{2})$
and  $(\tilde{S}^{03},  \tilde{S}^{12}, \tilde{S}^{56}) = (-\frac{i}{2}, -\frac{1}{2},  
\frac{1}{2})$, and $\hat{b}^{1 \dagger}_{2} $ has $(S^{03},  S^{12}, 
S^{56})= (\frac{i}{2}, \frac{1}{2},  \frac{1}{2})$
and $(\tilde{S}^{03}, \tilde{S}^{12}, \tilde{S}^{56})= (-\frac{i}{2}, -\frac{1}{2},  
\frac{1}{2})$. This means that Clifford even ``basis vector''  changes the family 
members quantum numbers of the Clifford odd ``basis vector'', leaving the family 
quantum numbers unchanged. 

One can find that the  algebraic  application, $*_A$,  of
 ${}^{I}{\hat{\cal A}}^{1 \dagger}_{3} (\equiv \stackrel{03}{[+i]}
\stackrel{12}{[+]} \stackrel{56}{[+]}) $ on $\hat{b}^{1 \dagger}_{1}$ leads to
the same  family member of the same family $f=1$, namely to
$\hat{b}^{1 \dagger}_{1} $.

Calculating the eigenvalues of the Cartan subalgebra members,
Eq.~(\ref{cartangrasscliff}), before and after the algebraic multiplication, $*_A$, 
assures us that ${}^{I}{\hat{\cal A}}^{m \dagger}_{3}$ carry the integer
eigenvalues of the Cartan subalgebra members, namely of
${\cal {\bf S}}^{ab}$ $= S^{ab} + \tilde{S}^{ab} $, since they transfer to
the Clifford odd ``basis vector'' integer eigenvalues of the Cartan subalgebra
members, changing the Clifford odd ``basis vector'' into another Clifford odd
``basis vector'' of the same family.

We,  therefore, confirm that the algebraic application 
 of ${}^{I}{\hat{\cal A}}^{m \dagger}_{3} $, $m=1,2,3,4$,
on $\hat{b}^{1 \dagger}_{1}$ transforms $\hat{b}^{1 \dagger}_{1}$ into
$\hat{b}^{m \dagger}_{1}$, $m=(1,2,3,4)$.
\begin{small} 
Similarly we find that the algebraic application of ${}^{I}{\hat {\cal A}}^{m}_4,$ 
$m=(1,2,3,4)$ on $\hat{b}^{2 \dagger}_{1}$ transforms $\hat{b}^{2 \dagger}_{1}$ 
into $\hat{b}^{m \dagger}_{1}, m=(1,2,3,4)$.
The algebraic application of ${}^{I}{\hat {\cal A}}^{m}_2,$ $m=(1,2,3,4)$ on   
$\hat{b}^{3 \dagger}_{1}$ transforms $\hat{b}^{3 \dagger}_{1}$ into
$\hat{b}^{m \dagger}_{1}, m=(1,2,3,4)$.
And the algebraic application of ${}^{I}{\hat {\cal A}}^{m}_1,$ $m=(1,2,3,4)$ on   
$\hat{b}^{4 \dagger}_{1}$ transforms $\hat{b}^{4 \dagger}_{1}$ into
$\hat{b}^{m \dagger}_{1}, m=(1,2,3,4)$.


One easily checks Eq.~(\ref{calbIA1234gen}) if taking into account Eq.~(\ref{graficcliff0}); 
like: $\hat{b}^{1 \dagger}_{1} *_{A} {}^{I}{\hat {\cal A}}^{m}_{4} =0$, $(m=(1,2,3,4))$, 
since either $(\stackrel{03}{(+i)})^2=0$ or $\stackrel{12}{[+]} *_{A} 
\stackrel{12}{[-]}=0$ or$\stackrel{56}{[+]} \stackrel{56}{(-)} =0$. 

Similarly, one can check Eq.~(\ref{calIIAb1234gen}) by evaluating, for example,
$ {}^{II}{\hat {\cal A}}^{m}_4 *_{A} \hat{b}^{1 \dagger}_{1}$, since
either $\stackrel{12}{(+)} *_{A} \stackrel{12}{[+]}=0$ or
$\stackrel{12}{[-]} \stackrel{56}{[+]} =0$. 
\end{small}
\vspace{2mm}

Let us check the validity of Eq.~(\ref{calbIIA1234gen}) on the case: 
$\hat{b}^{4 \dagger}_{1} *_{A} {}^{II}{\hat {\cal A}}^{m}_4 =$
$\hat{b}^{4 \dagger}_{3}$ for $m=1$, and zero for $m=(2,3,4)$, while
$\hat{b}^{4 \dagger}_{1} *_{A} {}^{II}{\hat {\cal A}}^{1}_f =$
($\hat{b}^{4 \dagger}_{4},  \hat{b}^{4 \dagger}_{2}, \hat{b}^{4 \dagger}_{1}, 
\hat{b}^{4 \dagger}_{3}$)  for $ f=(1,2,3,4)$. All ${}^{II}{\hat {\cal A}}^{m}_f$ 
giving non zero contributions,  
keep the family member quantum numbers of the Clifford odd ``basis vectors''
unchanged, changing the family quantum number. All the rest give zero 
contribution.

\vspace{2mm}
 
 The statements of Eq.~(\ref{calIAb1234gen}, \ref{calbIA1234gen}, \ref{calIIAb1234gen},
  \ref{calbIIA1234gen}), are, therefore, demonstrated on the case
 of $d=(5+1)$.
 
\vspace{2mm}

The Cartan subalgebra has in $d=(5+1)$-dimensional space $3$ members. 
To  illustrate  that the  Clifford even ``basis vectors'' have the properties of the gauge 
fields of the corresponding  Clifford odd ``basis vectors'' let us study properties of the
 $SU(3)$ $\times U(1)$ subgroups of the Clifford odd and Clifford even ``basis vectors''. 
 We need the relations between $S^{ab}$ and ($ \tau^{3},  \tau^{8},  \tau^{`}$)
 \begin{eqnarray}
 \label{so64 5+1}
 \tau^{3}: = &&\frac{1}{2} \,(%
 -S^{1\,2} - iS^{0\,3})\, , \qquad 
\tau^{8}= \frac{1}{2\sqrt{3}} (-i S^{0\,3} + S^{1\,2} -  2 S^{5\;6})\,,\nonumber\\
 \tau' = &&-\frac{1}{3}(-i S^{0\,3} + S^{1\,2} + S^{5\,6})\,.
%
 \end{eqnarray}

 The corresponding relations for ($\tilde{\tau}^{3}, \tilde{\tau}^{8}$,
 $\tilde{\tau}'$) can be read from Eq.~(\ref{so64 5+1}), if replacing $S^{ab}$  
 by $\tilde{S}^{ab}$.

 The corresponding relations for superposition of the Cartan subalgebra elements 
($\tau', \tau^3, \tau^8$)  for ${\bf {\cal S}}^{ab}=S^{ab}+ \tilde{S}^{ab}$
follow if  in Eq.~(
\ref{so64 5+1})  $S^{ab}$ is replaced by ${\bf {\cal S}}^{ab}$.

\vspace{2mm}

In Tables~(\ref{oddcliff basis5+1.}, \ref{Cliff basis5+1even I.}) the Clifford odd and
even ``basis vectors'' ($\hat{b}^{m \dagger}_{f}$ and ${}^{I}{\hat {\cal A}}^{m}_f$,
respectively) are presented as products of nilpotents (odd number of nilpotents for
$\hat{b}^{m \dagger}_{f}$ and even number of nilpotents for
${}^{I}{\hat {\cal A}}^{m}_f$) and projectors: Like in Table~\ref{Table Clifffourplet.}.
Besides the eigenvalues of the Cartan subalgebra members of Eq.~(\ref{cartangrasscliff}) 
also ($ \tau^{3}, \tau^{8}, \tau^{`}$) are added on both tables. In 
Table~\ref{oddcliff basis5+1.} also ($ \tilde{\tau}^{3},
\tilde{\tau}^{8}, \tilde{\tau}^{`}$) are written. In Fig.~(\ref{FigSU3U1odd})
only one  family is presented; all four families have the same ($ \tau^{3}, \tau^{8}, 
\tau^{`}$), they only distinguish in ($ \tilde{\tau}^{3},
\tilde{\tau}^{8}, \tilde{\tau}^{`}$) .

\vspace{2mm}

The corresponding table for  the Clifford even ``basis vectors'' 
${}^{II}{\hat {\cal A}}^{m}_f$ are not presented.  ${}^{II}{\hat {\cal A}}^{m}_f$
carry, namely, the same quantum numbers ($\tau^{3}, \tau^{8}, \tau^{`}$) as
${}^{I}{\hat {\cal A}}^{m}_f$. There are only products of nilpotents and projectors
which distinguish among ${}^{I}{\hat {\cal A}}^{m}_f$ and 
${}^{II}{\hat {\cal A}}^{m}_f$, causing differences in properties with respect to
 the Clifford odd ``basis vectors''; 
 ${}^{II}{\hat {\cal A}}^{m'}_{f `}$  transform $\hat{b}^{m \dagger}_{f}$ with a 
 family member $m$ of particular family $f$ into $\hat{b}^{m \dagger}_{f''}$  of
the same family member $m$ of another family $f''$.  
${}^{I}{\hat {\cal A}}^{m}_f$  transform a family member of particular family 
$\hat{b}^{m' \dagger}_{f `}$ into another family member  $m$ of the same family 
$\hat{b}^{m \dagger}_{f `}$.
 (Let us remind the reader that the $SO(5,1)$ group and the $SU(3), U(1)$ subgroups 
 have the same number of commuting operators, but  different number of  generators; 
 $SO(5,1)$ has $15$ generators, $SU(3)$ and $U(1)$ have  together $9$ generators.)


%
\begin{table}
\begin{tiny}
\begin{small}
\caption{\label{oddcliff basis5+1.} The 
''basis vectors'' $\hat{b}^{m\dagger}_{f}$ are presented for $d= (5+1)$-dimensional
case. Each $\hat{b}^{m \dagger}_{f}$ is a product of projectors
and of an odd number of nilpotents and is the "eigenvector" of all the Cartan
subalgebra members, ($S^{03}$, $S^{12}$, $S^{56}$) and ($\tilde{S}^{03}$,
$\tilde{S}^{12}$, $\tilde{S}^{56}$), Eq.~(\ref{cartangrasscliff}),
$m$ counts the members of each family, while $f$
determines the family quantum numbers (the eigenvalues of ($\tilde{S}^{03}$,
$\tilde{S}^{12}$, $\tilde{S}^{56}$)).
This table also presents in the columns ($8^{th}, 9^{th}, 10^{th}$)
the eigenvalues of the three commuting
operators ($\tau^3, \tau^8$ and $\tau'$) of the subgroups $SU(3)\times U(1)$,
Eq.~(\ref{so64 5+1}),
as well as (in the last three columns) the corresponding ($\tilde{\tau}^3, \tilde{\tau}^8, \tilde{\tau}'$).
$\Gamma^{(3+1)}$ $= i\gamma^0 \gamma^1 \gamma^2 \gamma^3$
is written in the $7^{th}$ column. $\Gamma^{(5+1)}= -1$ ($=-\gamma^0
\gamma^1 \gamma^2 \gamma^3\gamma^5\gamma^6$).
Operators $\hat{b}^{m \dagger}_{f}$ and $\hat{b}^{m}_{f}$
fulfil the anti-commutation relations of Eq.~(\ref{almostDirac}).
\vspace{2mm}}
\end{small}
\label{cliff basis5+1.}
\begin{center}
 \begin{tabular}{|r|lr|r|r|r|r|r|r|r|r|r|r|r|r|r|} 
 \hline
$\, f $&$m $&$\hat{b}^{ m \dagger}_f$
&$S^{03}$&$ S^{1 2}$&$S^{5 6}$&$\Gamma^{3+1}$ &
$\tau^3$&$\tau^8$&$\tau'$& $\tilde{S}^{03}$&$\tilde{S}^{1 2}$& $\tilde{S}^{5 6}$&
$\tilde{\tau}^3$&$\tilde{\tau}^8$&$\tilde{\tau}^{`}$ \\
\hline
$I$&$1$&
$\stackrel{03}{(+i)}\,\stackrel{12}{[+]}| \stackrel{56}{[+]}$&
$\frac{i}{2}$&$\frac{1}{2}$&$\frac{1}{2}$&$1$&
$0$&$0$&$-\frac{1}{2}$&$\frac{i}{2}$&$-\frac{1}{2}$&$-\frac{1}{2}$&$\frac{1}{2}$&
$\frac{1}{2\sqrt{3}}$&$\frac{1}{6}$\\
$$ &$2$&
$\stackrel{03}{[-i]}\,\stackrel{12}{(-)}|\stackrel{56}{[+]}$&
$-\frac{i}{2}$&$-\frac{1}{2}$&$\frac{1}{2}$&$1$
&$0$&$-\frac{1}{\sqrt{3}}$&$\frac{1}{6}$&$\frac{i}{2}$&$-\frac{1}{2}$&$-\frac{1}{2}$&
$\frac{1}{2}$&$\frac{1}{2\sqrt{3}}$&$\frac{1}{6}$\\
$$ &$3$&
$\stackrel{03}{[-i]}\,\stackrel{12}{[+]}|\stackrel{56}{(-)}$&
$-\frac{i}{2}$&$ \frac{1}{2}$&$-\frac{1}{2}$&$-1$
&$-\frac{1}{2}$&$\frac{1}{2\sqrt{3}}$&$\frac{1}{6}$&$\frac{i}{2}$&$-\frac{1}{2}$&
$-\frac{1}{2}$&$\frac{1}{2}$&$\frac{1}{2\sqrt{3}}$&$\frac{1}{6}$\\
$$ &$4$
&$\stackrel{03}{(+i)}\, \stackrel{12}{(-)}|\stackrel{56}{(-)}$&
$\frac{i}{2}$&$- \frac{1}{2}$&$-\frac{1}{2}$&$-1$
&$\frac{1}{2}$&$\frac{1}{2\sqrt{3}}$&$\frac{1}{6}$&$\frac{i}{2}$&$-\frac{1}{2}$&
$-\frac{1}{2}$&$\frac{1}{2}$&$\frac{1}{2\sqrt{3}}$&$\frac{1}{6}$\\ 
\hline 
$II$&$1$
&$\stackrel{03}{[+i]}\,\stackrel{12}{[+]}| \stackrel{56}{(+)}$&
$\frac{i}{2}$&
$\frac{1}{2}$&$\frac{1}{2}$&$1$&
$0$&$0$&$-\frac{1}{2}$&$-\frac{i}{2}$&$-\frac{1}{2}$&$\frac{1}{2}$&
$0$&$-\frac{1}{\sqrt{3}}$&$\frac{1}{6}$
\\
$$ &$2$
&$\stackrel{03}{(-i)}\,\stackrel{12}{(-)}|\stackrel{56}{(+)}$&
$-\frac{i}{2}$&$-\frac{1}{2}$&$\frac{1}{2}$&$1$
&$0$&$-\frac{1}{\sqrt{3}}$&$\frac{1}{6}$&$-\frac{i}{2}$&$-\frac{1}{2}$&$\frac{1}{2}$&
$0$&$-\frac{1}{\sqrt{3}}$&$\frac{1}{6}$\\
$$ &$3$
&$\stackrel{03}{(-i)}\,\stackrel{12}{[+]}|\stackrel{56}{[-]}$&
$-\frac{i}{2}$&$ \frac{1}{2}$&$-\frac{1}{2}$&$-1$
&$-\frac{1}{2}$&$\frac{1}{2\sqrt{3}}$&$\frac{1}{6}$&$-\frac{i}{2}$&$-\frac{1}{2}$&
$\frac{1}{2}$&$0$&$-\frac{1}{\sqrt{3}}$&$\frac{1}{6}$\\
$$ &$4$
&$\stackrel{03}{[+i]}\, \stackrel{12}{(-)}|\stackrel{56}{[-]}$&
$\frac{i}{2}$&$- \frac{1}{2}$&$-\frac{1}{2}$&$-1$
&$\frac{1}{2}$&$\frac{1}{2\sqrt{3}}$&$\frac{1}{6}$&$-\frac{i}{2}$&$-\frac{1}{2}$&
$\frac{1}{2}$&
$0$&$-\frac{1}{\sqrt{3}}$&$\frac{1}{6}$\\ 
\hline
$III$&$1$
&$\stackrel{03}{[+i]}\,\stackrel{12}{(+)}| \stackrel{56}{[+]}$&
$\frac{i}{2}$&
$\frac{1}{2}$&$\frac{1}{2}$&$1$&
 $0$&$0$&$-\frac{1}{2}$&$-\frac{i}{2}$&$\frac{1}{2}$&$-\frac{1}{2}$&
$-\frac{1}{2}$&$\frac{1}{2\sqrt{3}}$&$\frac{1}{6}$\\
$$ &$2$
&$\stackrel{03}{(-i)}\,\stackrel{12}{[-]}|\stackrel{56}{[+]}$&
$-\frac{i}{2}$&$-\frac{1}{2}$&$\frac{1}{2}$&$1$
&$0$&$-\frac{1}{\sqrt{3}}$&$\frac{1}{6}$&$-\frac{i}{2}$&$\frac{1}{2}$&$-\frac{1}{2}$&
$-\frac{1}{2}$&$\frac{1}{2\sqrt{3}}$&$\frac{1}{6}$\\
$$ &$3$
&$\stackrel{03}{(-i)}\,\stackrel{12}{(+)}|\stackrel{56}{(-)}$&
$-\frac{i}{2}$&$ \frac{1}{2}$&$-\frac{1}{2}$&$-1$
&$-\frac{1}{2}$&$\frac{1}{2\sqrt{3}}$&$\frac{1}{6}$&$-\frac{i}{2}$&$\frac{1}{2}$&$-\frac{1}{2}$&$-\frac{1}{2}$&$\frac{1}{2\sqrt{3}}$&$\frac{1}{6}$\\
$$ &$4$
&$\stackrel{03}{[+i]} \stackrel{12}{[-]}|\stackrel{56}{(-)}$&
$\frac{i}{2}$&$- \frac{1}{2}$&$-\frac{1}{2}$&$-1$
&$\frac{1}{2}$&$\frac{1}{2\sqrt{3}}$&$\frac{1}{6}$&$-\frac{i}{2}$&$\frac{1}{2}$&
$-\frac{1}{2}$&$-\frac{1}{2}$&$\frac{1}{2\sqrt{3}}$&$\frac{1}{6}$\\
\hline
$IV$&$1$
&$\stackrel{03}{(+i)}\,\stackrel{12}{(+)}| \stackrel{56}{(+)}$&
$\frac{i}{2}$&$\frac{1}{2}$&$\frac{1}{2}$&$1$
&$0$&$0$&$-\frac{1}{2}$&$\frac{i}{2}$&$\frac{1}{2}$&
$\frac{1}{2}$&$0$&$0$&$-\frac{1}{2}$\\
$$ &$2$
&$\stackrel{03}{[-i]}\,\stackrel{12}{[-]}|\stackrel{56}{(+)}$&
$-\frac{i}{2}$&$-\frac{1}{2}$&$\frac{1}{2}$&$1$
&$0$&$-\frac{1}{\sqrt{3}}$&$\frac{1}{6}$&$\frac{i}{2}$&$\frac{1}{2}$&$\frac{1}{2}$&
$0$&$0$&$-\frac{1}{2}$\\
$$ &$3$
&$\stackrel{03}{[-i]}\,\stackrel{12}{(+)}|\stackrel{56}{[-]}$&
$-\frac{i}{2}$&$ \frac{1}{2}$&$-\frac{1}{2}$&$-1$
&$-\frac{1}{2}$&$\frac{1}{2\sqrt{3}}$&$\frac{1}{6}$&$\frac{i}{2}$&$\frac{1}{2}$&$\frac{1}{2}$&$0$&$0$&$-\frac{1}{2}$\\
$$ &$4$
&$\stackrel{03}{(+i)}\,\stackrel{12}{[-]}|\stackrel{56}{[-]}$&
$\frac{i}{2}$&$- \frac{1}{2}$&$-\frac{1}{2}$&$-1$
&$\frac{1}{2}$&$\frac{1}{2\sqrt{3}}$&$\frac{1}{6}$&$\frac{i}{2}$&$\frac{1}{2}$&$\frac{1}{2}$&$0$&$0$&$-\frac{1}{2}$\\
\hline
 \end{tabular}
 \end{center}
\end{tiny}
\end{table}

\begin{figure}
  \centering
   \includegraphics[width=0.45\textwidth]{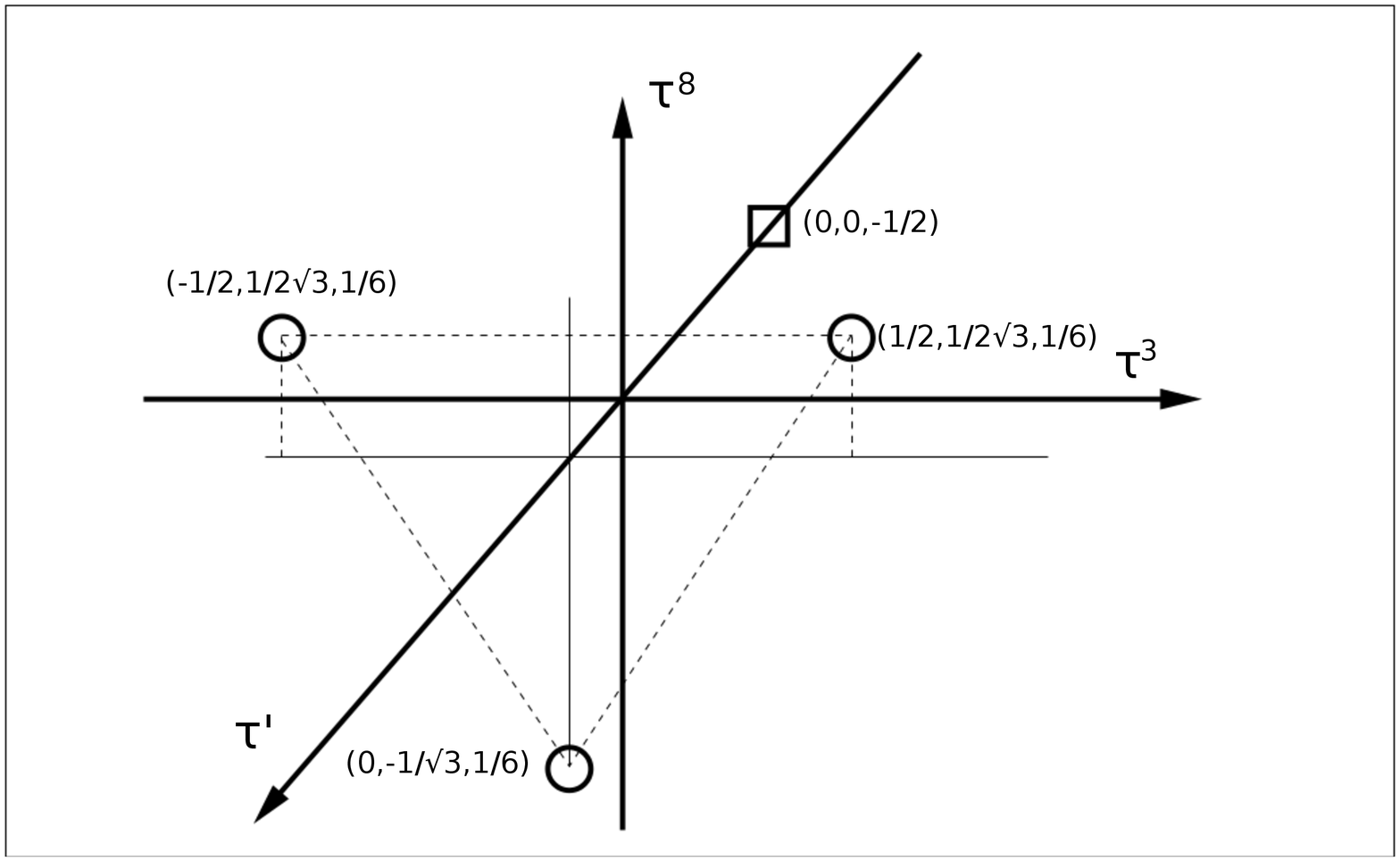}
  \caption{\label{FigSU3U1odd} 
The representations of the subgroups $SU(3)$ and $U(1)$ of the group $SO(5,1)$, the properties of which appear in 
Tables~(\ref{Table Clifffourplet.}, \ref{oddcliff basis5+1.}) for the Clifford odd ``basis vectors'', are
presented. ($\tau^3, \tau^8$, $\tau'$) can be calculated if using
Eq.~(\ref{so64 5+1}).
On the abscissa axis, on the ordinate axis and on the third axis, the eigenvalues of the
superposition of the three Cartan subalgebra members, ($\tau^3$%
, $\tau^8$
, $\tau'$), 
are presented. One notices one triplet,
denoted by ${\bf \bigcirc}$ with the values $\tau'=\frac{1}{6}$, ($\tau^3=-\frac{1}{2},
\tau^8=\frac{1}{2\sqrt{3}}, \tau'=\frac{1}{6})$, ($\tau^3=\frac{1}{2},
\tau^8=\frac{1}{2\sqrt{3}}, \tau'=\frac{1}{6}$), ($\tau^3=0,
\tau^8=-\frac{1}{\sqrt{3}}, \tau'=\frac{1}{6}$), respectively, and one singlet denoted
by the square. 
($\tau^3=0, \tau^8=0, \tau'=-\frac{1}{2}$).
The triplet and the singlet appear in four families, with the family quantum numbers
presented in the last three columns of Table~\ref{oddcliff basis5+1.}.}.
\end{figure}

In the case that the group $SO(5,1)$ --- manifesting as $SU(3) \times U(1)$ and
representing the colour group with quantum numbers ($\tau^3$, $\tau^8$) and the
``fermion'' group with the quantum number $\tau^{,}$ --- is embedded into
$SO(13,1)$ the triplet would represent quarks (and antiquarks), and the singlet leptons
(and antileptons).

The corresponding gauge fields, presented in Table~\ref{Cliff basis5+1even I.} and Fig.~\ref{FigSU3U1even}, if
belonging to the sextet, would transform the triplet of quarks among themselves, changing
the colour and leaving the ``fermion'' quantum number equal to $\frac{1}{6}$. \\


Table~\ref{Cliff basis5+1even I.} presents the Clifford even ``basis vectors'' 
${}^{I}{\hat{\cal A}}^{m \dagger}_{f}$ for $d=(5+1)$ with the properties: 

i. They are products of an even number of nilpotents,  $\stackrel{ab}{(k)}$, 
with the rest up to $\frac{d}{2}$ of projectors,~$\stackrel{ab}{[k]}$. 

ii. Nilpotents and projectors are eigenvectors of the Cartan 
subalgebra members ${\bf {\cal S}}^{ab}$ $= S^{ab} + \tilde{S}^{ab} $, Eq.~(\ref{cartangrasscliff}), carrying the integer eigenvalues 
of the Cartan subalgebra members.

iii. They have their Hermitian conjugated partners within the same group  of
${}^{I}{\hat{\cal A}}^{m \dagger}_{f}$ (with $2^{\frac{d}{2}-1}$ 
$\times$ $2^{\frac{d}{2}-1}$ members).

iv. They have properties of the boson gauge fields. 
When  the Clifford even ``basis vectors'', ${}^{I}{\hat{\cal A}}^{m \dagger}_{f}$,  apply  
on the Clifford odd ``basis vectors'' (offering the description of 
the fermion fields) they transform the Clifford odd ``basis vectors'' into
another Clifford odd ``basis vectors'' of the same family, transferring to the Clifford odd 
``basis vectors'' the integer spins with respect to the  $SO(d-1,1)$ group, 
while with respect to subgroups of the  $SO(d-1,1)$ group they transfer appropriate
superposition of the eigenvalues (manifesting the properties of the adjoint representations
of the corresponding subgroups.)

\begin{small}
If, for example, ${}^{I}{\hat{\cal A}}^{1 \dagger}_{3}$ applies on a singlet 
$\hat{b}^{1 \dagger}_{1}$ keeps the internal space of $\hat{b}^{1 \dagger}_{1}$ 
unchanged (it can change only momentum), while if 
${}^{I}{\hat{\cal A}}^{2 \dagger}_{3}$ applies on $\hat{b}^{1 \dagger}_{1}$ 
transforms it to a member of a triplet, to $\hat{b}^{2 \dagger}_{1}$.
\end{small}

%
\begin{small}
\begin{figure}
  \centering
   \includegraphics[width=0.45\textwidth]{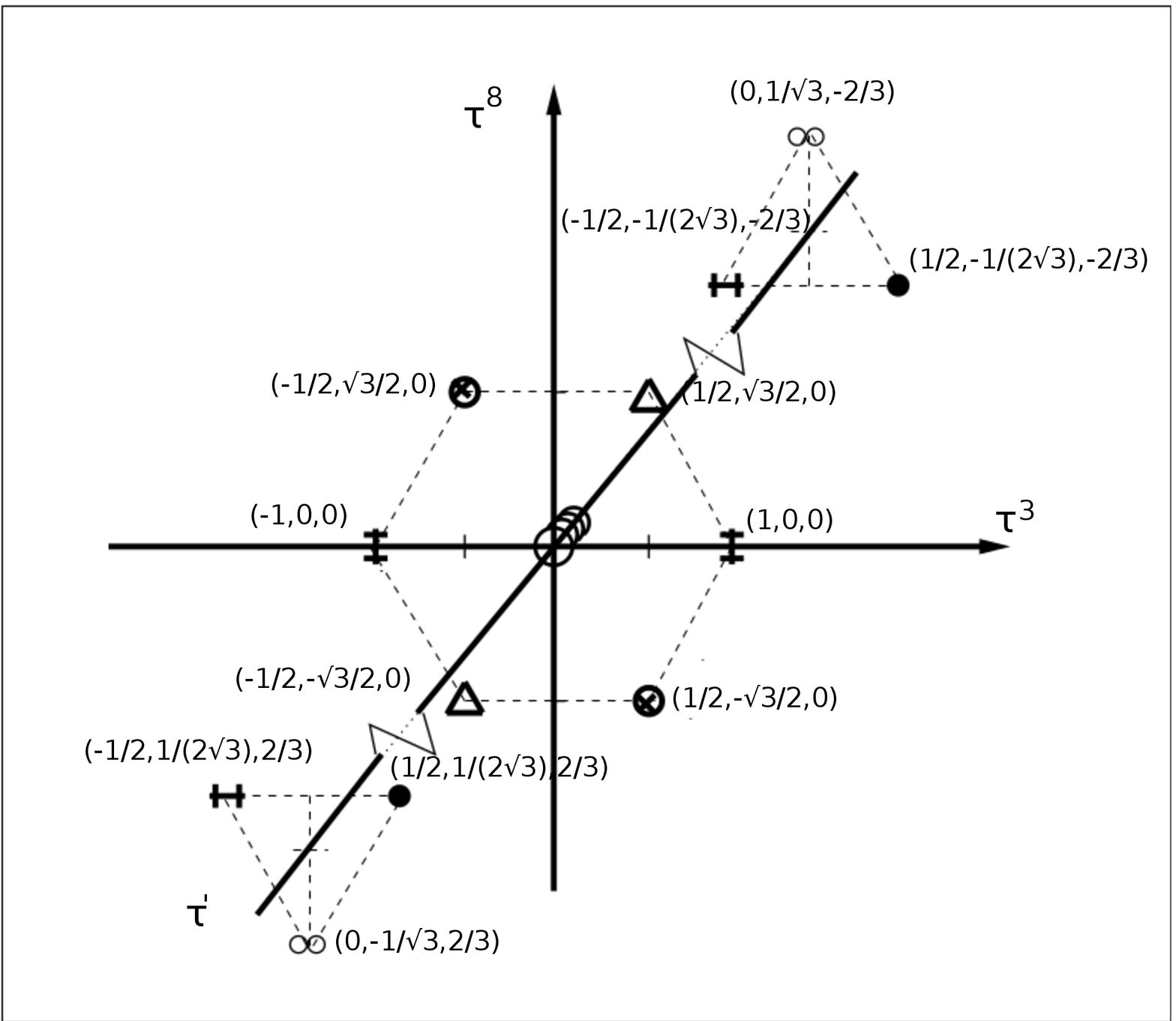}
  \caption{\label{FigSU3U1even} 
The Clifford even ''basis vectors'' ${}^{I}{\hat{\cal A}}^{m\dagger}_{f}$ in the case that
$d=(5+1)$ are presented concerning the eigenvalues of the commuting operators
of the subgroups $SU(3)$ and $U(1)$ of the group $SO(5,1)$, Eq.~(\ref{so64 5+1}): (${\cal \bf{\tau}}^3$, 
${\cal \bf{\tau}}^8$, 
${\cal \bf{\tau}}'$). 
Their properties appear in Table~\ref{Cliff basis5+1even I.}.
The abscissa axis carries the eigenvalues of ${\cal \bf{\tau}}^3$, the ordinate axis carries the
eigenvalues of ${\cal \bf{\tau}}^8$ and the third axis carries the eigenvalues of ${\cal \bf{\tau}}'$.
One notices four singlets with (${\cal \bf{\tau}}^3=0, {\cal \bf{\tau}}^8=0, {\cal \bf{\tau}}'=0$),
denoted by $\bigcirc$, representing four self adjoint Clifford even ''basis vectors''
${}^{I}{\hat{\cal A}}^{m \dagger}_{f}$, with ($f=1, m=4$), ($f=2, m=3$),
($f=3, m=1$), ($f=4, m=2$)\,,
one sextet of three pairs, Hermitian conjugated to each other, with $ {\cal \bf{\tau}}'=0$, denoted
by $\bigtriangleup$ (${}^{I}{\hat{\cal A}}^{2\dagger}_{1}$ with
($ {\cal \bf{\tau}}'=0, {\cal \bf{\tau}}^3=-\frac{1}{2}, {\cal \bf{\tau}}^8=-\frac{3}{2\sqrt{3}}$)
and ${}^{I}{\hat{\cal A}}^{4\dagger}_{4}$ with ($ {\cal \bf{\tau}}'=0, {\cal \bf{\tau}}^3=\frac{1}{2},
{\cal \bf{\tau}}^8=\frac{3}{2\sqrt{3}}$)),
by $\ddagger$ (${}^{I}{\hat{\cal A}}^{3\dagger}_{1}$ with ($ {\cal \bf{\tau}}'=0, {\cal \bf{\tau}}^3=-1,
{\cal \bf{\tau}}^8=0$) and ${}^{I}{\hat{\cal A}}^{4\dagger}_{2}$ with $ {\cal \bf{\tau}}'=0, {\cal \bf{\tau}}^3=1, {\cal \bf{\tau}}^8=0$)), and by $\otimes$
(${}^{I}{\hat{\cal A}}^{2\dagger}_{2}$ with (${\cal \bf{\tau}}'=0, {\cal \bf{\tau}}^3=\frac{1}{2},
{\cal \bf{\tau}}^8=-\frac{3}{2\sqrt{3}}$)
and ${}^{I}{\hat{\cal A}}^{3\dagger}_{4}$ with (${\cal \bf{\tau}}'=0, {\cal \bf{\tau}}^3=- \frac{1}{2},
{\cal \bf{\tau}}^8=\frac{3}{2\sqrt{3}}$)),
and one triplet, denoted by $\star \star$ (${}^{I}{\hat{\cal A}}^{4\dagger}_{3}$ with
(${\cal \bf{\tau}}'=\frac{2}{3}, {\cal \bf{\tau}}^3=\frac{1}{2}, {\cal \bf{\tau}}^8=\frac{1}{2\sqrt{3}}$)),
by $\bullet$ (${}^{I}{\hat{\cal A}}^{3\dagger}_{3}$ with (${\cal \bf{\tau}}'=\frac{2}{3},
{\cal \bf{\tau}}^3= -\frac{1}{2}, {\cal \bf{\tau}}^8=\frac{1}{2\sqrt{3}}$)), and by $\odot \odot$
(${}^{I}{\hat{\cal A}}^{2\dagger}_{3}$ with
(${\cal \bf{\tau}}'=\frac{2}{3}, {\cal \bf{\tau}}^3=0, {\cal \bf{\tau}}^8=-\frac{1}{\sqrt{3}}$)),
as well as one antitriplet, Hermitian conjugated to triplet, denoted by $\star \star$
(${}^{I}{\hat{\cal A}}^{1 \dagger}_{1}$ with (${\cal \bf{\tau}}'=-\frac{2}{3},
{\cal \bf{\tau}}^3=-\frac{1}{2}, {\cal \bf{\tau}}^8=-\frac{1}{2\sqrt{3}}$)),
by $\bullet$ (${}^{I}{\hat{\cal A}}^{1\dagger}_{2}$ with (${\cal \bf{\tau}}'=-\frac{2}{3},
{\cal \bf{\tau}}^3= \frac{1}{2}, {\cal \bf{\tau}}^8=- \frac{1}{2\sqrt{3}}$)), and by $\odot \odot$
(${}^{I}{\hat{\cal A}}^{4 \dagger}_{1}$ with (${\cal \bf{\tau}}'=-\frac{2}{3}, {\cal \bf{\tau}}^3=0,
{\cal \bf{\tau}}^8=\frac{1}{\sqrt{3}}$)).}
\end{figure}
\end{small}

We can see that ${}^{I}{\hat{\cal A}}^{m \dagger}_{3}$  with $(m=2,3,4)$,
if applied on the $SU(3)$ singlet $\hat{b}^{1 \dagger}_{4}$ with (${\cal \bf{\tau}}'= 
-\frac{1}{2}, {\cal \bf{\tau}}^3=0, {\cal \bf{\tau}}^8=0$), transforms it to  $\hat{b}^{m(=2,3,4) \dagger}_{4}$, 
respectively, which are members of the $SU(3 )$ triplet. All these Clifford even 
``basis vectors'' have ${\cal \bf{\tau}}'$ equal to $\frac{2}{3}$, changing correspondingly 
${\cal \bf{\tau}}'= -\frac{1}{2}$ into 
${\cal \bf{\tau}}'=\frac{1}{6}$ and  bringing the needed values of ${\cal \bf{\tau}}^3$ and ${\cal \bf{\tau}}^8$. 

\vspace{2mm}

In Table~\ref{Cliff basis5+1even I.} we find $(6+4)$ Clifford even ``basis vectors'' 
${}^{I}{\hat{\cal A}}^{m \dagger}_{f}$ with ${\cal \bf{\tau}}'=0$. 
Six of them are Hermitian 
conjugated to each other --- the Hermitian conjugated partners are denoted by the 
same geometric figure on the third column. Four of them are self-adjoint and 
correspondingly with (${\cal \bf{\tau}} '=0, {\cal \bf{\tau}}^3=0, {\cal \bf{\tau}}^8=0$), 
denoted in the third column of Table~\ref{Cliff basis5+1even I.} by $\bigcirc$.
The rest $6$  Clifford even ``basis vectors'' belong to one triplet with ${\cal \bf{\tau}} '=
\frac{2}{3}$ and $({\cal \bf{\tau}}^3, {\cal \bf{\tau}}^8)$ equal to [$(0,- \frac{1}{\sqrt{3}})$, 
$(-\frac{1}{2},  \frac{1}{2\sqrt{3}})$, $(\frac{1}{2}, \frac{1}{2\sqrt{3}})$] and one antitriplet 
with ${\cal \bf{\tau}}'=-\frac{2}{3}$ and ($({\cal \bf{\tau}}^3, {\cal \bf{\tau}}^8)$ equal to
 [$(-\frac{1}{2}, -\frac{1}{2\sqrt{3}})$, $(\frac{1}{2}, -\frac{1}{2\sqrt{3}})$, 
$(0, \frac{1}{\sqrt{3}})$]. 

Each triplet has Hermitian conjugated partners in anti-triplet and opposite. In 
Table~\ref{Cliff basis5+1even I.} the Hermitian conjugated partners of the triplet and 
antitriplet are denoted by the same signum:  (${}^{I}{\hat{\cal A}}^{1 \dagger}_{1}$, 
${}^{I}{\hat{\cal A}}^{4 \dagger}_{3}$) by
$\star \star$, (${}^{I}{\hat{\cal A}}^{1 \dagger}_{2}$, 
 ${}^{I}{\hat{\cal A}}^{3 \dagger}_{3}$)  by $\bullet$, and 
 (${}^{I}{\hat{\cal A}}^{2 \dagger}_{3}$, ${}^{I}{\hat{\cal A}}^{1 \dagger}_{4}$) 
by $\odot \odot$.

The octet, two triplets and four singlets are presented in Fig.~\ref{FigSU3U1even}.


\vspace{2mm}

%
\begin{table}
\begin{tiny}
\caption{The Clifford even ``basis vectors'' ${}^{I}{\hat{\cal A}}^{m \dagger}_{f}$,
each of them is the product of projectors and an even number of nilpotents, and each is
the eigenvector of all the Cartan subalgebra members, ${\cal S}^{03}$,
${\cal S}^{12}$, ${\cal S}^{56}$, Eq.~(\ref{cartangrasscliff}), are presented for
$d= (5+1)$-dimensional case. Indexes $m$ and $f$ determine
$2^{\frac{d}{2}-1}\times 2^{\frac{d}{2}-1}$ different members
${}^{I}{\hat{\cal A}}^{m \dagger}_{f}$.
In the third column the ``basis vectors'' ${}^{I}{\hat{\cal A}}^{m \dagger}_{f}$
which are Hermitian conjugated partners to each other (and can therefore annihilate
each other) are pointed out with the same symbol. For example, with $\star \star$
are equipped the first member with $m=1$ and $f=1$ and the last member of $f=3$
with $m=4$.
The sign $\bigcirc$ denotes the Clifford even ``basis vectors'' which are self-adjoint
$({}^{I}{\hat{\cal A}}^{m \dagger}_{f})^{\dagger}$
$={}^{I}{\hat{\cal A}}^{m' \dagger}_{f `}$. It is obvious that ${}^{\dagger}$
has no meaning, since ${}^{I}{\hat{\cal A}}^{m \dagger}_{f}$ are self adjoint or
are Hermitian conjugated partner to another ${}^{I}{\hat{\cal A}}^{m' \dagger}_{f `}$.
This table also represents the eigenvalues of the three commuting operators
${\cal {\bf \tau}}^3, {\cal {\bf \tau}}^8$ and ${\cal {\bf \tau}}' $ of the subgroups
$SU(3)\times U(1)$.
\vspace{3mm}
}
\label{Cliff basis5+1even I.} 
 %
 \begin{center}
 \begin{tabular}{|r|r|r|r|r|r|r|r|r|r|}
 \hline
$\, f $&$m $&$*$&${}^{I}\hat{\cal A}^{m \dagger}_f$
&${\cal S}^{03}$&$ {\cal S}^{1 2}$&${\cal S} ^{5 6}$&
${\cal {\bf \tau}}^3$&${\cal {\bf \tau}}^8$&${\cal {\bf \tau}}'$
\\
\hline
%
$I$&$1$&$\star \star$&$
\stackrel{03}{[+i]}\,\stackrel{12}{(+)} \stackrel{56}{(+)}$&
$0$&$1$&$1$
&$-\frac{1}{2}$&$-\frac{1}{2\sqrt{3}}$&$-\frac{2}{3}$
\\
$$ &$2$&$\bigtriangleup$&$
\stackrel{03}{(-i)}\,\stackrel{12}{[-]}\,\stackrel{56}{(+)}$&
$- i$&$0$&$1$
&$-\frac{1}{2}$&$-\frac{3}{2\sqrt{3}}$&$0$
\\
$$ &$3$&$\ddagger$&$
\stackrel{03}{(-i)}\,\stackrel{12}{(+)}\,\stackrel{56}{[-]}$&
$-i$&$ 1$&$0$
&$-1$&$0$&$0$
\\
$$ &$4$&$\bigcirc$&$
\stackrel{03}{[+i]}\,\stackrel{12}{[-]}\,\stackrel{56}{[-]}$&
$0$&$0$&$0$
&$0$&$0$&$0$
\\
\hline 
$II$&$1$&$\bullet$&$
\stackrel{03}{(+i)}\,\stackrel{12}{[+]}\, \stackrel{56}{(+)}$&
$i$&$0$&$1$
&$\frac{1}{2}$&$-\frac{1}{2\sqrt{3}}$&$-\frac{2}{3}$\\
$$ &$2$&$\otimes$&$
\stackrel{03}{[-i]}\,\stackrel{12}{(-)}\,\stackrel{56}{(+)}$&
$0$&$-1$&$1$
&$\frac{1}{2}$&$-\frac{3}{2\sqrt{3}}$&$0$
\\
$$ &$3$&$\bigcirc$&$
\stackrel{03}{[-i]}\,\stackrel{12}{[+]}\,\stackrel{56}{[-]}$&
$0$&$ 0$&$0$
&$0$&$0$&$0$
\\
$$ &$4$&$\ddagger$&$
\stackrel{03}{(+i)}\, \stackrel{12}{(-)}\,\stackrel{56}{[-]}$&
$i$&$-1$&$0$
&$1$&$0$&$0$
\\ 
%
%
 \hline
$III$&$1$&$\bigcirc$&$
\stackrel{03}{[+i]}\,\stackrel{12}{[+]}\, \stackrel{56}{[+]}$&
$0$&$0$&$0$
&$0$&$0$&$0$\\
$$ &$2$&$\odot \odot$&$
\stackrel{03}{(-i)}\,\stackrel{12}{(-)}\,\stackrel{56}{[+]}$&
$-i$&$-1$&$0$
&$0$&$-\frac{1}{\sqrt{3}}$&$\frac{2}{3}$\\
$$ &$3$&$\bullet$&$
\stackrel{03}{(-i)}\,\stackrel{12}{[+]}\,\stackrel{56}{(-)}$&
$-i$&$ 0$&$-1$
&$-\frac{1}{2}$&$\frac{1}{2\sqrt{3}}$&$\frac{2}{3}$
\\
$$ &$4$&$\star \star$&$
\stackrel{03}{[+i]} \stackrel{12}{(-)}\,\stackrel{56}{(-)}$&
$0$&$- 1$&$- 1$
&$\frac{1}{2}$&$\frac{1}{2\sqrt{3}}$&$\frac{2}{3}$
\\
\hline
$IV$&$1$&$\odot \odot $&$
\stackrel{03}{(+i)}\,\stackrel{12}{(+)}\, \stackrel{56}{[+]}$&
$i$&$1$&$0$
&$0$&$\frac{1}{\sqrt{3}}$&$-\frac{2}{3}$
\\
$$ &$2$&$\bigcirc$&$
\stackrel{03}{[-i]}\,\stackrel{12}{[-]}\,\stackrel{56}{[+]}$&
$0$&$0$&$0$
&$0$&$0$&$0$
\\
$$ &$3$&$\otimes$&$
\stackrel{03}{[-i]}\,\stackrel{12}{(+)}\,\stackrel{56}{(-)}$&
$0$&$ 1$&$-1$
&$-\frac{1}{2}$&$\frac{3}{2\sqrt{3}}$&$0$
\\
$$ &$4$&$\bigtriangleup$&$
\stackrel{03}{(+i)}\, \stackrel{12}{[-]}\,\stackrel{56}{(-)}$&
$i$&$0$&$-1$
&$\frac{1}{2}$&$\frac{3}{2\sqrt{3}}$&$0$\\ 
\hline 
 \end{tabular}
 \end{center}
\end{tiny}
\end{table}

Fig.~\ref{FigSU3U1even} represents the $2^{\frac{d}{2}-1}\times 2^{\frac{d}{2}-1}$
members ${}^{I}{\hat{\cal A}}^{m}_{f}$ of the Clifford even ``basis vectors'' for the
case that $d=(5+1)$. The properties of ${}^{I}{\hat{\cal A}}^{m}_{f}$ are presented
also in Table~\ref{Cliff basis5+1even I.}.
Manifesting the structure of subgroups $SU(3) \times U(1)$ of the group $SO(5,1)$ they
are represented as eigenvectors of the superposition of the Cartan subalgebra members
(${\bf {\cal S}}^{03}, {\bf {\cal S}}^{12}, {\bf {\cal S}}^{56}$), that is with
$\tau^3=\frac{1}{2} (- {\bf {\cal S}}^{12} -i {\bf {\cal S}}^{03})$,
$\tau^8=\frac{1}{2\sqrt{3}} ( {\bf {\cal S}}^{12} -i {\bf {\cal S}}^{03}-
2 {\bf {\cal S}}^{56})$, and $\tau'=- \frac{1}{3}
({\bf {\cal S}}^{12} -i {\bf {\cal S}}^{03} + {\bf {\cal S}}^{56})$.
There are four self adjoint Clifford even ``basis vectors'' with ($\tau^3=0, \tau^8=0, \tau'=0$),
one sextet of three pairs Hermitian conjugated to each other, one triplet and one
antitriplet with the members of the triplet Hermitian conjugated to the corresponding
members of the antitriplet and opposite. These $16$ members of the Clifford even
``basis vectors'' ${}^{I}{\hat{\cal A}}^{m}_{f}$ are the gauge fields ``partners'' of the
Clifford odd ``basis vectors'' $\hat{b}^{m \dagger }_{f}$, presented in Fig.~\ref{FigSU3U1odd}
for one of four families, anyone. The reader can check
that the algebraic application of ${}^{I}{\hat{\cal A}}^{m}_{f}$, belonging to
the triplet transforms applying on the Clifford odd singlet, denoted in Fig.~\ref{FigSU3U1odd}
by a square, this singlet to one of the members of the triplet, denoted in Fig.~\ref{FigSU3U1odd}
by the circle ${\bf \bigcirc}$.

\vspace{2mm}

Looking at the boson fields ${}^{I}{\hat{\cal A}}^{m \dagger}_{f}$ from the point of view
of subgroups $SU(3)\times U(1)$ of the group $SO(5+1)$ we recognize in the part
of fields forming the octet the colour gauge fields of quarks and leptons and antiquarks and
antileptons. The Clifford even ``basis vectors'' ${}^{I}{\hat{\cal A}}^{m \dagger}_{f}$
transform when applying on the Clifford odd ``basis vectors'' $\hat{b}^{m' \dagger}_{f `}$
to another (or the same) member, keeping the family member
unchanged.


We can check that although ${}^{II}{\hat{\cal A}}^{m \dagger}_{f}$ have different
structure of an even number of nilpotents, and the rest of the projectors than
${}^{I}{\hat{\cal A}}^{m \dagger}_{f}$, having correspondingly different properties
with respect to the Clifford odd ``basis vectors'': ${}^{I}{\hat{\cal A}}^{m \dagger}_{f}$
transform $\hat{b}^{m' \dagger}_{f `}$ among the family members, keeping the
family quantum numbers unchanged, ${}^{II}{\hat{\cal A}}^{m \dagger}_{f}$
transform $\hat{b}^{m \dagger}_f$ into the same member of another family,
keeping the family member's quantum number unchanged, 
both, ${}^{I}{\hat{\cal A}}^{m \dagger}_{f}$ and ${}^{II}{\hat{\cal A}}^{m \dagger}_{f}$
do have the equivalent figure and equivalent ${\cal {\bf S}}^{ab}$ and
correspondingly also $({\cal {\bf \tau}}^3, {\cal {\bf \tau}}^8,
{\cal {\bf \tau}}')$ content. 

Let us anyhow demonstrate properties of ``scattering'' of $\hat{b}^{m \dagger}_f$
on ${}^{II}{\hat{\cal A}}^{m' \dagger}_{f `}$, paying attention on $SU(3)$ and $U(1)$
substructure of $SO(5,1)$.

Let us look at	the ``scattering'' of the kind of Eq.~(\ref{calAb1})
\begin{small}
\begin{eqnarray}
&& \hat{b}^{2 \dagger}_{2} (\equiv \stackrel{03}{(-i)} \stackrel{12}{(-)} \stackrel{56}{(+)})
*_{A} {}^{II}{\hat{\cal A}}^{3 \dagger}_{1} (\equiv \stackrel{03}{(+i)}
\stackrel{12}{(+)} \stackrel{56}{[-]}) \rightarrow
\hat{b}^{2 \dagger}_{4} (\equiv \stackrel{03}{[-i]}
\stackrel{12}{[-]} \stackrel{56}{(+)}) \,,
\label{calAb2}
\end{eqnarray}
\end{small}
$ \hat{b}^{2 \dagger}_{2} (\equiv \stackrel{03}{(-i)} \stackrel{12}{(-)} \stackrel{56}{(+)}) $
has $(\tau^3=0, \tau^8=-\frac{1}{\sqrt{3}}, \tau' = \frac{1}{6}$) and
$(\tilde{\tau}^3=0, \tilde{\tau}^8=-\frac{1}{\sqrt{3}}, \tilde{\tau}' = \frac{1}{6}$).
$ \hat{b}^{2 \dagger}_{4} (\equiv \stackrel{03}{[-i]} \stackrel{12}{[-]} \stackrel{56}{(+)}) $
has $(\tau^3=0, \tau^8=-\frac{1}{\sqrt{3}}, \tau' = \frac{1}{6}$) and
$(\tilde{\tau}^3=0, \tilde{\tau}^8=0, \tilde{\tau}' = -\frac{1}{2}$).
${}^{II}{\hat{\cal A}}^{1 \dagger}_{4} (\equiv \stackrel{03}{(+i)}
\stackrel{12}{(+)} \stackrel{56}{[-]}$) has
$({\cal {\bf \tau}}^3=0, {\cal {\bf \tau}}^8=\frac{1}{\sqrt{3}}, {\cal {\bf \tau}}' =- \frac{2}{3}).$\\

If $ \hat{b}^{2 \dagger}_{2} $ absorbs $ {}^{II}{\hat{\cal A}}^{3 \dagger}_{4} (\equiv \stackrel{03}{[+i]}
\stackrel{12}{(+)} \stackrel{56}{(-)}) $ with $( {\cal {\bf \tau}}^3=-\frac{1}{2}, {\cal {\bf \tau}}^8=
\frac{3}{2\sqrt{3}}, {\cal {\bf \tau}}' = 0$) becomes $ \hat{b}^{2 \dagger}_{3}
(\equiv \stackrel{03}{(-i)} \stackrel{12}{[-]} \stackrel{56}{[+]}) $
with quantum numbers $(\tau^3=0, \tau^8=-\frac{1}{\sqrt{3}}, \tau' = \frac{1}{6}$) and
$(\tilde{\tau}^3=-\frac{1}{2}, \tilde{\tau}^8=\frac{1}{2\sqrt{3}}, \tilde{\tau}' = \frac{1}{6}$).

$ {}^{II}{\hat{\cal A}}^{3 \dagger}_{4}$ transfers its quantum numbers to
$ \hat{b}^{2 \dagger}_{2} $,
changing family and leaving the family member $m$ unchanged.
 
\vspace{1mm}
\subsection{Second quantized fermion and boson fields with internal spaces
described by Clifford ``basis vectors'' in even dimensional spaces}
\label{secondquantizedfermionsbosonsdeven}

\vspace{2mm}

We learned in the previous Subsects.~(\ref{basisvectors}, \ref{cliffordoddevenbasis5+1})
that in even dimensional spaces ($d=2(2n+1)$ or $d=4n$) the Clifford odd and the
Clifford even ``basis vectors'', which are the superposition of the Clifford odd and
the Clifford even products of $\gamma^a$'s, respectively, offer the description of
the internal spaces of fermion and boson fields.

The Clifford odd algebra offers $2^{\frac{d}{2}-1}$ ``basis vectors''
$\hat{b}^{m \dagger}_{f}$, appearing in $2^{\frac{d}{2}-1}$ families
(with the family quantum numbers determined by $\tilde{S}^{ab}= \frac{i}{2} \{ \tilde{\gamma}^a, \tilde{\gamma}^b\}_{-}$), which, together with their
$2^{\frac{d}{2}-1}\times$ $2^{\frac{d}{2}-1}$ Hermitian conjugated partners
$\hat{b}^{m}_{f}$ fulfil the postulates for the second quantized fermion fields, Eq.~(\ref{almostDirac}) in this paper, Eq.(26) in Ref.~\cite{nh2021RPPNP},
explaining the second quantization postulate of Dirac.

%
The Clifford even algebra offers $2^{\frac{d}{2}-1}\times$ $2^{\frac{d}{2}-1}$
``basis vectors'' of ${}^{I}{\hat{\cal A}}^{m \dagger}_{f}$, and the same number
of ${}^{II}{\hat{\cal A}}^{m \dagger}_{f}$, with the properties of the second
quantized boson fields manifesting as the gauge fields of fermion fields described
by the Clifford odd ``basis vectors'' $\hat{b}^{m \dagger}_{f}$.
The commutation relations of ${}^{i}{\hat{\cal A}}^{m \dagger}_{f}, i=(I,II),$ 
are commented in the last paragraph of App.~\ref{basis3+1} on a simple case of 
$d=(3+1)$.
The subgroup structure of 
$SU(3)$ can be recognized on Fig. 2, leading to the commutation relations of 
the observed colour boson gauge fields. However, further studies are needed to 
recognize what new this way of describing internal spaces of fermion and boson 
fields with the Clifford algebra is offering.

The Clifford odd and the Clifford even ``basis vectors'' are chosen to be products of
nilpotents, $\stackrel{ab}{(k)}$ (with the odd number of nilpotents if describing
fermions and the even number of nilpotents if describing bosons), and projectors,
$\stackrel{ab}{[k]}$. Nilpotents and projectors are (chosen to be) eigenvectors
of the Cartan subalgebra members of the Lorentz algebra in the internal space of
$S^{ab}$ for the Clifford odd ``basis vectors'' and of ${\bf {\cal S}}^{ab} (=
S^{ab}+ \tilde{S}^{ab}$) for the Clifford even ``basis vectors''.

\vspace{2mm}

To define the creation operators, for fermions or bosons,
besides the ``basis vectors'' defining the internal space of fermions and bosons,
the basis in ordinary space in momentum or coordinate representation is needed.
Here Ref.~\cite{nh2021RPPNP}, Subsect.~3.3 and App. J is overviewed. \\

Let us introduce the momentum part of the single-particle states. (The extended version
is presented in Ref.~\cite{nh2021RPPNP} in Subsect.~3.3 and App. J.)
\begin{eqnarray}
\label{creatorp}
|\vec{p}>&=& \hat{b}^{\dagger}_{\vec{p}} \,|\,0_{p}\,>\,,\quad
<\vec{p}\,| = <\,0_{p}\,|\,\hat{b}_{\vec{p}}\,, \nonumber\\
<\vec{p}\,|\,\vec{p}'>&=&\delta(\vec{p}-\vec{p}')=
<\,0_{p}\,|\hat{b}_{\vec{p}}\; \hat{b}^{\dagger}_{\vec{p}'} |\,0_{p}\,>\,,
\nonumber\\
&&{\rm pointing  \;out\;} \nonumber\\
<\,0_{p}\,| \hat{b}_{\vec{p'}}\, \hat{b}^{\dagger}_{\vec{p}}\,|\,0_{p}\, > &=&\delta(\vec{p'}-\vec{p})\,,
\end{eqnarray}
 with the normalization $<\,0_{p}\, |\,0_{p}\,>=1$.
While the quantized operators $\hat{\vec{p}}$ and $\hat{\vec{x}}$ commute
$\{\hat{p}^i\,, \hat{p}^j \}_{-}=0$ and $\{\hat{x}^k\,, \hat{x}^l \}_{-}=0$,
it follows for $\{\hat{p}^i\,, \hat{x}^j \}_{-}=i \eta^{ij}$. One correspondingly
finds
\begin{small}
\begin{eqnarray}
\label{eigenvalue10}
<\vec{p}\,| \,\vec{x}>&=&<0_{\vec{p}}\,|\,\hat{b}_{\vec{p}}\;
\hat{b}^{\dagger}_{\vec{x}}
|0_{\vec{x}}\,>=(<0_{\vec{x}}\,|\,\hat{b}_{\vec{x}}\;
\hat{b}^{\dagger}_{\vec{p}} \,\,
|0_{\vec{p}}\,>)^{\dagger}\, \nonumber\\
<0_{\vec{p}}\,|\{\hat{b}^{\dagger}_{\vec{p}}\,, \,
\hat{b}^{\dagger}_{\vec{p}\,'}\}_{-}|0_{\vec{p}}\,>&=&0\,,\qquad
<0_{\vec{p}}\,|\{\hat{b}_{\vec{p}}, \,\hat{b}_{\vec{p}\,'}\}_{-}|0_{\vec{p}}\,>=0\,,\qquad
<0_{\vec{p}}\,|\{\hat{b}_{\vec{p}}, \,\hat{b}^{\dagger}_{\vec{p}\,'}\}_{-}|0_{\vec{p}}\,>=0\,,
\nonumber\\
<0_{\vec{x}}\,|\{\hat{b}^{\dagger}_{\vec{x}}, \,\hat{b}^{\dagger}_{\vec{x}\,'}\}_{-}|0_{\vec{x}}\,>&=&0\,,
\qquad
<0_{\vec{x}}\,|\{\hat{b}_{\vec{x}}, \,\hat{b}_{\vec{x}\,'}\}_{-}|0_{\vec{x}}\,>=0\,,\qquad
<0_{\vec{x}}\,|\{\hat{b}_{\vec{x}}, \,\hat{b}^{\dagger}_{\vec{x}\,'}\}_{-}|0_{\vec{x}}\,>=0\,,
\nonumber\\
<0_{\vec{p}}\,|\{\hat{b}_{\vec{p}}, \,\hat{b}^{\dagger}_{\vec{x}}\}_{-}|0_{\vec{x}}\,>&=&
e^{i \vec{p} \cdot \vec{x}} \frac{1}{\sqrt{(2 \pi)^{d-1}}}\,,\quad
<0_{\vec{x}}\,|\{\hat{b}_{\vec{x}}, \,\hat{b}^{\dagger}_{\vec{p}}\}_{-}|0_{\vec{p}}\,>=
e^{-i \vec{p} \cdot \vec{x}} \frac{1}{\sqrt{(2 \pi)^{d-1}}}\,.
\end{eqnarray}.
\end{small}
The internal space of either fermion or boson fields has the finite number of ``basis
vectors'', $2^{\frac{d}{2}-1}\times 2^{\frac{d}{2}-1}$ for fermions (and the same
number of their Hermitian conjugated partners),  and twice 
$2^{\frac{d}{2}-1}\times 2^{\frac{d}{2}-1}$ for bosons, the momentum basis is
continuously infinite.\\

The creation operators for either fermions or bosons must be tensor products,
$*_{T}$, of both contributions, the ``basis vectors'' describing the internal space of
fermions or bosons and the basis in ordinary momentum or coordinate space.

The creation operators for a free massless fermion of the energy
$p^0 =|\vec{p}|$, belonging to a family $f$ and to a superposition of
family members $m$ applying on the vacuum state
$|\psi_{oc}>\,*_{T}\, |0_{\vec{p}}>$ 
can be written as~(\cite{nh2021RPPNP}, Subsect.3.3.2, and the references therein)
\begin{eqnarray}
\label{wholespacefermions}
{\bf \hat{b}}^{s \dagger}_{f} (\vec{p}) \,&=& \,
\sum_{m} c^{sm}{}_f (\vec{p}) \,\hat{b}^{\dagger}_{\vec{p}}\,*_{T}\,
\hat{b}^{m \dagger}_{f} \, \,,
\end{eqnarray}
where the vacuum state for fermions $|\psi_{oc}>\,*_{T}\, |0_{\vec{p}}> $
includes both spaces, the internal part, Eq.(\ref{vaccliffodd}), and the momentum
part, Eq.~(\ref{creatorp}) (in a tensor product for a starting single particle state
with zero momentum, from which one obtains the other single fermion states of the
same ''basis vector'' by the operator $\hat{b}^{\dagger}_{\vec{p}}$ which pushes
the momentum by an amount $\vec{p}$~\footnote{
The creation operators and their Hermitian conjugated annihilation
operators in the coordinate representation can be
read in~\cite{nh2021RPPNP} and the references therein:
$\hat{\bf b}^{s \dagger}_{f }(\vec{x},x^0)=
\sum_{m} \,\hat{b}^{ m \dagger}_{f} \, *_{T}\, \int_{- \infty}^{+ \infty} \,
\frac{d^{d-1}p}{(\sqrt{2 \pi})^{d-1}} \, c^{s m }{}_{f}\;
(\vec{p}) \; \hat{b}^{\dagger}_{\vec{p}}\;
e^{-i (p^0 x^0- \varepsilon \vec{p}\cdot \vec{x})}
$
~(\cite{nh2021RPPNP}, subsect. 3.3.2., Eqs.~(55,57,64) and the references therein).}).
\\
The creation operators and annihilation operators for fermion fields
fulfil the anti-commutation relations for the second quantized
fermion fields~\footnote{
Let us evaluate:
 $<0_{\vec{p}}\,|
\{ \hat{\bf b}^{s' }_{f `}(\vec{p'})\,,\,
\hat{\bf b}^{s \dagger}_{f }(\vec{p}) \}_{+} \,|\psi_{oc}> |0_{\vec{p}}>=
\delta^{s s'} \delta_{f f'}\,\delta(\vec{p}' - \vec{p})\,\cdot |\psi_{oc}> =$
$ <0_{\vec{p}}\,|\hat{\bf b}^{s' }_{f `}\,\hat{\bf b}^{s \dagger}_{f }\,\hat{b}_{\vec{p}'}
\hat{b}^{\dagger}_{\vec{p}} +\, \hat{b}^{\dagger}_{\vec{p}} \hat{b}_{\vec{p}'}\,
\hat{\bf b}^{s \dagger}_{f }\, \hat{\bf b}^{s' }_{f `}\,|\psi_{oc}>|0_{\vec{p}}> =$
$<0_{\vec{p}}\,|\hat{\bf b}^{s' }_{f `}\,\hat{\bf b}^{s \dagger}_{f }\,\hat{b}_{\vec{p}'}
\hat{b}^{\dagger}_{\vec{p}} \,|\psi_{oc}> |0_{\vec{p}}>$,
since, according to Eq.~(\ref{almostDirac}), $ \hat{\bf b}^{s' }_{f ` }\,|\psi_{oc}>=0.$

Let us demonstrate for free fields  
$|\vec{p}> = e^{-i \vec{p}\cdot \vec{x}} |0_{p}>= \hat{b}^{\dagger}_{\vec{p}} \,|0_{p}>$\,\,,
$\,<\vec{p}| = <0_{p}| e^{i \vec{p}\cdot \vec{x}}= <0_{p}|\, \hat{b}_{\vec{p}} \,$\\
$<\vec{p'}|\vec{p}>=<0_{p}|\,\hat{b}_{\vec{p'}}\,\hat{b}^{\dagger}_{\vec{p}} \,|0_{p}>= 
\delta{(\vec{p'}-\vec{p})}\,, $ 
$<\vec{-p'}|\vec{-p}>=
<0_{p}|\,\hat{b}^{\dagger}_{\vec{p'}}\,\hat{b}_{\vec{p}} \,|0_{p}>= 
\delta{(\vec{-p'}-(\vec{-p}))}=\delta{(\vec{p}-\vec{p'})} \,, $
consequently
$<0_{p}|\{\,\hat{b}_{\vec{p}}\,,\hat{b}^{\dagger}_{\vec{p'}}\}_{-}\,|0_{p}>=0$\,.
}
~\footnote{
 Two fermion states (formed from two creation operators applying on the vacuum 
 state) with the orthogonal basis part in ordinary space (with two different 
momenta in ordinary space in the case of free fields)  "do not meet"; 
 correspondingly, each can carry the same ``basis vector''. They must differ 
in  the internal basis if they have the identical ordinary part of the basis.
(Otherwise, the tensor product, $*_{T_{H}}$, of such two fermion states is 
zero.) 
Illustration: Let us treat an atom with many electrons. Each electron has a spin 
of either $1/2$ or $-1/2.$ Their orthogonal basis in ordinary space allows them 
to have the internal spin $\pm1/2$ (leading to total angular momentum either 
$\pm 1/2$ or larger due to the angular momentum in ordinary space). As 
mentioned in the introduction section in {\bf a.iii.} the Hilbert space of the 
second quantized fermion states is represented by the tensor products, 
$*_{T_{H}}$, of all possible members of creation operators from zero to infinity 
applying on the simple vacuum state. For any of these members the scalar product 
is obtained by multiplying from the left hand side by their Hermitian conjugated 
partner.}.\\

\begin{small}
\begin{eqnarray}
<0_{\vec{p}}\,|
\{ \hat{\bf b}^{s' }_{f `}(\vec{p'})\,,\,
\hat{\bf b}^{s \dagger}_{f }(\vec{p}) \}_{+} \,|\psi_{oc}> |0_{\vec{p}}>&=&
\delta^{s s'} \delta_{f f'}\,\delta(\vec{p}' - \vec{p})\,\cdot |\psi_{oc}> 
\,,\nonumber\\
\{ \hat{\bf b}^{s' }_{f `}(\vec{p'})\,,\,
\hat{\bf b}^{s}_{f }(\vec{p}) \}_{+} \,|\psi_{oc}> |0_{\vec{p}}>&=&0\, \cdot \,
|\psi_{oc}> |0_{\vec{p}}>
\,,\nonumber\\
\{ \hat{\bf b}^{s' \dagger}_{f '}(\vec{p'})\,,\,
\hat{\bf b}^{s \dagger}_{f }(\vec{p}) \}_{+}\, |\psi_{oc}> |0_{\vec{p}}>&=&0\, \cdot
\,|\psi_{oc}> |0_{\vec{p}}>
\,,\nonumber\\
\hat{\bf b}^{s \dagger}_{f }(\vec{p}) \,|\psi_{oc}> |0_{\vec{p}}>&=&
|\psi^{s}_{f}(\vec{p})>\,,\nonumber\\
\hat{\bf b}^{s}_{f }(\vec{p}) \, |\psi_{oc}> |0_{\vec{p}}>&=&0\, \cdot\,
\,|\psi_{oc}> |0_{\vec{p}}>\,, \nonumber\\
|p^0| &=&|\vec{p}|\,.
\label{Weylpp'comrel}
\end{eqnarray}
\end{small}
The creation operators $ \hat{\bf b}^{s\dagger}_{f }(\vec{p}) $ and their
Hermitian conjugated partners annihilation operators
$\hat{\bf b}^{s}_{f }(\vec{p}) $, creating and annihilating the single fermion
states, respectively, fulfil when applying the vacuum state,
$|\psi_{oc}> *_{T} |0_{\vec{p}}>$, the anti-commutation relations for the second quantized
fermions, postulated by Dirac (Ref.~\cite{nh2021RPPNP}, Subsect.~3.3.1,
Sect.~5).~\footnote{
The anti-commutation relations of Eq.~(\ref{Weylpp'comrel}) are valid also if we
replace the vacuum state, $|\psi_{oc}>|0_{\vec{p}}>$, by the Hilbert space of the
Clifford fermions generated by the tensor products multiplication, $*_{T_{H}}$, of
any number of the Clifford odd fermion states of all possible internal quantum
numbers and all possible momenta (that is, of any number of
$ \hat {\bf b}^{s\, \dagger}_{f} (\vec{p})$ of any
$(s,f, \vec{p})$), Ref.~(\cite{nh2021RPPNP}, Sect. 5.).}\\

To write the creation operators for boson fields, we must take into account that
boson gauge fields have the space index $\alpha$, describing the $\alpha$
component of the boson field in the ordinary space~\footnote{
In the {\it spin-charge-family} theory the Higgs's scalars origin
in the boson gauge fields with the vector index $(7,8)$, Ref.~(\cite{nh2021RPPNP},
Sect.~7.4.1, and the references therein).}.
We, therefore, add the space index $\alpha$ as follows.
\begin{eqnarray}
\label{wholespacebosons}
{\bf {}^{i}{\hat{\cal A}}^{m \dagger}_{f \alpha}} (\vec{p}) \,&=&
\hat{b}^{\dagger}_{\vec{p}}\,*_{T}\,
{}^{i}{\cal C}^{ m}{}_{f \alpha}\, {}^{i}{\hat{\cal A}}^{m \dagger}_{f} \, \,, i=(I,II)\,.
\end{eqnarray}
We treat free massless bosons of momentum $\vec{p}$ and energy $p^0=|\vec{p}|$
and of particular ``basis vectors'' ${}^{i}{\hat{\cal A}}^{m \dagger}_{f}$'s which are
eigenvectors of all the Cartan subalgebra members~\footnote{
In the general case, the energy eigenstates of bosons are in a superposition of
${\bf {}^{i}{\hat{\cal A}}^{m \dagger}_{f}}$, for either $i=I$ or $i=II$. One example,
which uses the
superposition of the Cartan subalgebra eigenstates manifesting the $SU(3)\times U(1)$
subgroups of the group $SO(5,1)$, is presented in Fig.~\ref{FigSU3U1even}.},
${}^{i}{\cal C}^{ m}{}_{f \alpha}$ carry the space index $\alpha$ of the boson
field. Creation operators operate on the vacuum state
$|\psi_{oc_{ev}}>\,*_{T}\, |0_{\vec{p}}> $ with the internal space part
just a constant, $|\psi_{oc_{ev}}>=$ $|\,1>$, and for
a starting single boson state with zero momentum from which one obtains
the other single boson states with the same ''basis vector'' by the operators
$\hat{b}^{\dagger}_{\vec{p}}$ which push the momentum by an amount
$\vec{p}$, making also ${}^{i}{\cal C}^{ m}{}_{f \alpha}$ depending on $\vec{p}$.


For the creation operators for boson fields in a coordinate
representation one finds using Eqs.~(\ref{creatorp}, \ref{eigenvalue10})
\begin{eqnarray}
{\bf {}^{i}{\hat{\cal A}}^{m \dagger}_{f \alpha}}
(\vec{x}, x^0)& =& \int_{- \infty}^{+ \infty} \,
\frac{d^{d-1}p}{(\sqrt{2 \pi})^{d-1}} \,
{}^{i}{\hat{\cal A}}^{m \dagger}_{f \alpha} (\vec{p})\,
e^{-i (p^0 x^0- \varepsilon \vec{p}\cdot \vec{x})}|_{p^0=|\vec{p}|}\,,i=(I,II)\,.
\label{Weylbosonx}
\end{eqnarray}
\vspace{2mm}

To understand what new the Clifford algebra description of the internal space
of fermion and boson fields, Eqs.~(\ref{wholespacebosons}, \ref{Weylbosonx},
\ref{wholespacefermions}), bring to our understanding of the second quantized
fermion and boson fields and what new can we learn from this offer,
we need to relate $\sum_{ab} c^{ab} \omega_{ab \alpha}$ and
$ \sum_{m f} {}^{I}{\hat{\cal A}}^{m \dagger}_{f} \,{}^{I}{\cal C}^{m}{}_{f\alpha}$,
recognizing that ${}^{I}{\hat{\cal A}}^{m \dagger}_{f} \,{}^{I}{\cal C}^{m}{}_{f\alpha}$
are eigenstates of the Cartan subalgebra members, while $\omega_{ab \alpha}$
are not. And, equivalently, we need to relate $\sum_{ab} \tilde{c}^{ab} \tilde{\omega}_{ab \alpha}$ and
$ \sum_{m f} {}^{II}{\hat{\cal A}}^{m \dagger}_{f}\, {}^{II}{\cal C}^{m}{}_{f\alpha}$.

The gravity fields, the vielbeins and the two kinds of spin connection fields,
$f^{a}{}_{\alpha}$, $\omega_{ab \alpha}$, $\tilde{\omega}_{ab \alpha}$,
respectively, are in the {\it spin-charge-family} theory
(unifying spins, charges and families of fermions and offering not only the 
explanation for all the assumptions of the {\it standard model} but also for 
the increasing number of phenomena observed so far) the only boson fields in
$d=(13+1)$, observed in $d=(3+1)$ besides as gravity also as all the other
boson fields with the Higgs's scalars included~\cite{nd2017}.

We, therefore, need to relate: 
\begin{eqnarray}
\label{relationomegaAmf0}
\{\frac{1}{2} \sum_{ab} S^{ab}\, \omega_{ab \alpha} \}
\sum_{m } \beta^{m f}\, \hat{\bf b}^{m \dagger}_{f }(\vec{p}) &{\rm related\, \,to}&
\{ \sum_{m' f '} {}^{I}{\hat{\cal A}}^{m' \dagger}_{f '} \,
{\cal C}^{m' f '}_{\alpha} \}
\sum_{m } \beta^{m f} \, \hat{\bf b}^{m \dagger}_{f }(\vec{p}) \,, \nonumber\\
&&\forall f \,{\rm and}\,\forall \, \beta^{m f}\,, \nonumber\\
{\bf \cal S}^{cd} \,\sum_{ab} (c^{ab}{}_{mf}\, \omega_{ab \alpha}) &{\rm related\, \,to}&
{\bf \cal S}^{cd}\, ({}^{I}{\hat{\cal A}}^{m \dagger}_{f}\, {\cal C}^{m f}_{\alpha})\,, \nonumber\\
&& \forall \,(m,f), \nonumber\\
&&\forall \,\,{\rm Cartan\,\,subalgebra\, \, \, member} \,{\bf \cal S}^{cd} \,.
\end{eqnarray}
Let be repeated that ${}^{I}{\hat{\cal A}}^{m \dagger}_{f } $ are chosen to be
the eigenvectors of the Cartan subalgebra members, Eq.~(\ref{cartangrasscliff}).
Correspondingly we can relate a particular ${}^{I}{\hat{\cal A}}^{m \dagger}_{f }
\,{}^{I}{\cal C}^{m}{}_{f \alpha}$ with such a superposition of $\omega_{ab \alpha}$'s,
which is the eigenvector with the same values of the Cartan subalgebra members as
there is a particular ${}^{I}{\hat{\cal A}}^{m \dagger}_{f } {\cal C}^{m f }_{\alpha}$.
We can do this in two ways:\\
{\bf i.} $\;\;$ Using the first relation in Eq.~(\ref{relationomegaAmf0}). On the left
hand side of this relation ${S}^{ab}$'s apply on $ \hat{b}^{m \dagger}_{f} $ part of
$ \hat{\bf b}^{m \dagger}_{f }(\vec{p}) $.
On the right hand side ${}^{I}{\hat{\cal A}}^{m \dagger}_{f }$ apply as well on the
same ``basis vector'' $ \hat{b}^{m \dagger}_{f} $. \\
{\bf ii.} $\;\;$ Using the second relation, in which ${\bf \cal S}^{cd}$ apply on
the left hand side on $\omega_{ab \alpha}$'s,
\begin{eqnarray}
\label{sonomega}
\, {\bf \cal S}^{cd} \,\sum_{ab}\, c^{ab}{}_{mf}\, \omega_{ab \alpha}
&=& \sum_{ab}\, c^{ab}{}_{mf}\, i \,(\omega_{cb \alpha} \eta^{ad}-
\omega_{db \alpha} \eta^{ac}+ \omega_{ac \alpha} \eta^{bd}-
\omega_{ad \alpha} \eta^{bc}),
\end{eqnarray}
on each $ \omega_{ab \alpha}$ separately; $c^{ab}{}_{mf}$ are constants to be
determined from the second relation, where on the right-hand side of this relation
${\bf \cal S}^{cd} (= S^{cd}+ \tilde{S}^{cd})$ apply on the ``basis vector''
${}^{I}{\hat{\cal A}}^{m \dagger}_{f }$ of the corresponding gauge field~\footnote{
The reader can find the relation of Eq.~(\ref{relationomegaAmf0}) demonstrated for the
case $d={3+1}$ in Ref.~\cite{n2022epjc} at the end of Sect.~3.}. \\

We must treat equivalently also ${}^{II}{\hat{\cal A}}^{m \dagger}_{f }
\,{}^{II}{\cal C}^{m}{}_{f \alpha}$ and $\tilde{\omega}_{ab \alpha}$.\\

Let us conclude this section by pointing out that either the Clifford odd ``basis vectors'',
$\hat{b}^{m \dagger}_{f}$, or the Clifford even ``basis vectors'',
${}^{i}{\hat{\cal A}}^{m \dagger}_{f}, i=(I,II) $, have each in any even $d$,
$2^{\frac{d}{2}-1}$ $\times \,2^{\frac{d}{2}-1}$ members, while $\omega_{ab \alpha}$
as well as $\tilde{\omega}_{ab \alpha}$ have each for a particular 
$\alpha$ $\frac{d}{2}(d-1)$members. It is needed to find out what new this difference 
brings into the unifying theories of the Kaluza-Klein-like kind to what the 
{\it spin-charge-family} belongs.
 %

\section{Conclusions}
\label{conclusions}

\vspace{2mm}

In the {\it spin-charge-family} theory~\cite{norma93,%
n2014matterantimatter,nd2017,JMP2013,nh2017,%
nh2018,nh2021RPPNP}
the Clifford odd algebra describes the internal space of fermion fields. The
Clifford odd ``basis vectors'' --- the superposition of odd products of $\gamma^a$'s
--- in a tensor
product with the basis in ordinary space form the creation and annihilation operators,
in which the anti-commutativity of the ``basis vectors'' is transferred to the creation
and annihilation operators for fermions, explaining the second
quantization postulates for fermion fields.

The Clifford odd ``basis vectors'' have all the properties of fermions: Half integer
spins concerning the Cartan subalgebra members of the Lorentz algebra in
the internal space of fermions in even dimensional spaces ($d=2(2n+1)$ or
$d=4n$), as discussed in Subsects.~(\ref{basisvectors},
\ref{secondquantizedfermionsbosonsdeven}) (and in App~\ref{basis3+1} in
a pedagogical way).
With respect to the subgroups of the $SO(d-1, 1)$ group the Clifford odd ``basis
vectors'' appear in the fundamental representations, as illustrated in
Subsects.~\ref{cliffordoddevenbasis5+1}.

In this article, it is demonstrated that Clifford even algebra is offering the description
of the internal space of boson fields. The Clifford even ``basis vectors'' --- the
superposition of even products of $\gamma^a$'s --- in a tensor product
with the basis in ordinary space form the creation and annihilation operators which
manifest the commuting properties of the second quantized boson fields, offering
the explanation for the second quantization postulates for boson
fields~\cite{n2021SQ,n2022epjc}.
The Clifford even ``basis vectors'' have all the properties of boson fields: Integer spins
for the Cartan subalgebra members of the Lorentz algebra in the internal space
of bosons, as discussed in Subsects.~\ref{basisvectors}.

With respect to the subgroups of the $SO(d-1, 1)$ group the Clifford even ``basis
vectors'' manifest the adjoint representations, as illustrated in
Subsect.~\ref{cliffordoddevenbasis5+1}. 

\vspace{2mm}

\begin{small}
There are two kinds of anti-commuting algebras~\cite{norma93}: The Grassmann
algebra, offering in $d$-dimensional space $2\,.\, 2^d$ operators ($2^d$ $\,\theta^a$'s
and $2^d$ $\frac{\partial}{\partial \theta_a}$'s, Hermitian conjugated to each other,
Eq.~(\ref{thetaderher0})), and the two Clifford subalgebras, each with $2^d$ operators
named $\gamma^a$'s and $\tilde{\gamma}^a$'s, respectively, \cite{norma93,nh02},
Eqs.~(\ref{thetaderanti0}-\ref{gammatildeantiher0}).

The operators in each of the two Clifford subalgebras appear in even-dimensional spaces
in two groups of $2^{\frac{d}{2}-1}\times $ $2^{\frac{d}{2}-1}$ of the Clifford odd
operators (the odd products of either $\gamma^a$'s in one subalgebra or of
$\tilde{\gamma}^a$'s in the other subalgebra), which are Hermitian conjugated
to each other: In each Clifford odd group of any of the two subalgebras, there appear
$2^{\frac{d}{2}-1}$ irreducible representation each with the $2^{\frac{d}{2}-1}$
members and the group of their Hermitian conjugated partners.

There are as well the Clifford even operators (the even products of either
$\gamma^a$'s in one subalgebra or of $\tilde{\gamma}^a$'s in another
subalgebra) which again appear in two groups of $2^{\frac{d}{2}-1}\times $
$2^{\frac{d}{2}-1}$ members each. In the case of the Clifford even objects, the
members of each group of $2^{\frac{d}{2}-1}\times $ $2^{\frac{d}{2}-1}$
members have the Hermitian conjugated partners within the same
group, Subsect.~\ref{basisvectors}, Table~\ref{Table Clifffourplet.}.

The Grassmann algebra operators are expressible with the operators of the two
Clifford subalgebras and opposite,~Eq.~(\ref{clifftheta1}). The two Clifford sub-algebras
are independent of each other, Eq.~(\ref{gammatildeantiher0}), forming two independent
spaces.

Either the Grassmann algebra~\cite{nh2018} or the two Clifford
subalgebras can be used to describe the internal space of anti-commuting objects,
if the superposition of odd products of operators
($\theta^a$'s or $\gamma^a$'s, or $ \tilde{\gamma}^a$'s) are used to describe the
internal space of these objects. The commuting objects must be a superposition of
even products of operators ($\theta^a$'s or $ \gamma^a$'s or $\tilde{\gamma}^a$'s).
\end{small}

\vspace{2mm}

No integer spin anti-commuting objects have been observed so far, and to describe the
internal space of the so far observed fermions only one of the two Clifford odd
subalgebras are needed.


The problem can be solved by reducing the two Clifford subalgebras to only one, the one
(chosen to be) determined by $\gamma^{a}$'s. The decision
that $ \tilde{\gamma}^a$'s apply on $ \gamma^a$ as follows:
$\{ \tilde{\gamma}^a B =(-)^B\, i \, B \gamma^a\}\, |\psi_{oc}>$,
Eq.~(\ref{tildegammareduced0}),
(with $(-)^B = -1$, if $B$ is a function of an odd products of $\gamma^a$'s,
otherwise $(-)^B = 1$) enables that $2^{\frac{d}{2}-1}$ irreducible representations
of $S^{ab}= \frac{i}{2}\, \{\gamma^a\,,\, \gamma^b\}_{-}$ (each with the
$2^{\frac{d}{2}-1}$ members) obtain the family quantum numbers determined by
$\tilde{S}^{ab}= \frac{i}{2}\, \{\tilde{\gamma}^a\,,\,\tilde{\gamma}^b\}_{-}$.

\vspace{2mm}

The decision to use in the {\it spin-charge-family} theory in $d=2(2n +1)$, $n\ge 3$
($d\ge (13+1)$ indeed),
the superposition of the odd products of the Clifford algebra elements $\gamma^{a}$'s
to describe the internal space of fermions which interact with gravity only
(with the vielbeins, the gauge fields of momenta, and the two kinds of the spin
connection fields, the gauge fields of $S^{ab}$ and $\tilde{S}^{ab}$, respectively),
Eq.~(\ref{wholeaction}), offers not
only the explanation for all the assumed properties of fermions and bosons in
the {\it standard model}, with the appearance of the families of quarks and leptons
and antiquarks and antileptons~(\cite{nh2021RPPNP} and the references therein) and
of the corresponding vector gauge fields and the Higgs's scalars included~\cite{nd2017},
but also for the appearance of the dark matter~\cite{gn2009} in the universe, for the
explanation of the matter/antimatter asymmetry in the
universe~\cite{n2014matterantimatter}, and for several other observed phenomena,
making several predictions~\cite{pikan2006,gmdn2007,gmdn2008,gn2013}.

\vspace{2mm}

The recognition that the use of the superposition of the even products of the Clifford 
algebra elements $\gamma^{a}$'s to describe the internal space of boson fields, what 
appears to manifest all the properties of the observed boson fields, as demonstrated 
in this article, makes clear that the Clifford algebra offers not only the explanation 
for the postulates of the second quantized anti-commuting fermion fields but also 
for the postulates of the second quantized boson fields. 

This recognition, however, offers the possibility to relate
\begin{eqnarray}
\label{relationomegaAmf01}
\{\frac{1}{2} \sum_{ab} S^{ab}\, \omega_{ab \alpha} \}
\sum_{m } \beta^{m f}\, \hat{\bf b}^{m \dagger}_{f }(\vec{p}) &{\rm \,\, to}&
\{ \sum_{m' f '} {}^{I}{\hat{\cal A}}^{m' \dagger}_{f '} \,
{}^{I}{\cal C}^{m'}{}_{f `\alpha} \}
\sum_{m } \beta^{m f} \, \hat{\bf b}^{m \dagger}_{f }(\vec{p}) \,, \nonumber\\
&&\forall f \,{\rm and}\,\forall \, \beta^{m f}\,, \nonumber\\
{\bf \cal S}^{cd} \,\sum_{ab} (c^{ab}{}_{mf}\, \omega_{ab \alpha}) &{\rm \,\, to}&
{\bf \cal S}^{cd}\, ({}^{I}{\hat{\cal A}}^{m \dagger}_{f}\, {}^{I}{\cal C}^{m}{}_{f \alpha})\,, 
\nonumber\\
&& \forall \,(m,f), \nonumber\\
&&\forall \,\,{\rm Cartan\,\,subalgebra\, \, \, member} \,{\bf \cal S}^{cd} \,,
\nonumber
\end{eqnarray}
and equivalently for ${}^{II}{\hat{\cal A}}^{m \dagger}_{f}\,{}^{II}{\cal C}^{m}{}_{f \alpha}$ 
and $\tilde{\omega}_{ab \alpha}$, what offers the possibility to replace the covariant derivative
$ p_{0 \alpha }$
$$p_{0\alpha} = p_{\alpha} - \frac{1}{2} S^{ab} \omega_{ab \alpha} -
\frac{1}{2} \tilde{S}^{ab} \tilde{\omega}_{ab \alpha}
\quad \quad \quad\quad\;$$
in Eq.~(\ref{wholeaction}) with

$$ p_{0\alpha} = p_{\alpha} -
\sum_{m f} {}^{I}{ \hat {\cal A}}^{m \dagger}_{f}\,
{}^{I}{\cal C}^{m}{}_{f \alpha} -
\sum_{m f} {}^{II}{\hat{\cal A}}^{m \dagger}_{f}\,
{}^{II}{\cal C}^{m}{}_{f \alpha}\,, $$
\noindent
where the relations among ${}^{I}{\hat{\cal A}}^{m \dagger}_{f}
{}^{I}{\cal C}^{m}_{f \alpha}$ and
${}^{II}{\hat{\cal A}}^{m \dagger}_{f}\,
{}^{II}{\cal C}^{m}_{f \alpha}$ with respect to $\omega_{ab \alpha}$
and $\tilde{\omega}_{ab \alpha}$, not discussed directly in this article, need additional
study and explanation.

Although the properties of the Clifford odd and even ``basis vectors'' and correspondingly
of the creation and annihilation operators for fermion and boson fields are, hopefully,
demonstrated in this article, yet the proposed way of the second quantization
of fields, the fermion and the boson ones needs further study to find out what new can
the description of the internal space of fermions and bosons bring into the understanding of
the second quantized fields.

This study showing up that the Clifford algebra can be used to
describe the internal spaces of fermion and boson fields equivalently, offering
correspondingly the explanation for the second quantization postulates for fermion and
boson fields is opening a new insight into the quantum field theory,
since studies of the interaction of fermion fields with boson fields and of boson fields with
boson fields so far looks very promising.

The study of properties of the second quantized boson fields, the internal space of
which is described by Clifford even algebra has just started and needs further
consideration.
\appendix

%
\section{``Basis vectors'' in $d=(3+1)$ }
\label{basis3+1}

This section, suggested by the referee, is to illustrate on a simple case of $d=(3+1)$ 
the properties of ``basis vectors'' when describing internal spaces of fermions and bosons
by the Clifford algebra: 
i. The way of constructing the``basis vectors'' for fermions which appear in families 
and for bosons which have no families.
ii. The manifestation of anti-commutativity of the second quantized fermion fields and
commutativity of the second quantized boson fields.
iii. The creation and annihilation operators, described by a tensor product, $*_{T}$,
of the ``basis vectors'' and their Hermitian conjugated partners with the basis in ordinary
space-time. 

%

This section is a short overview of some sections presented in the
article~\cite{n2023MDPI}, equipped by concrete examples of ``basis vectors''
for fermions and bosons in $d=(3+1)$.

\vspace{2mm}

{\bf ``Basis vectors''}

\vspace{2mm}

Let us start by arranging the ``basis vectors'' as a superposition of products of
(operators~\footnote{We repeat that we treat $\gamma^{a}$  as operators,
not as matrices. We write ``basis vectors'' as the superposition of products of $\gamma^{a}$.
If we want to look for a matrix representation of any operator, say $S^{ab}$, we arrange
the ``basis vectors'' into a series and write a matrix of transformations caused by
the operator. However, we do not need to look for the matrix representations of the
operators since we can directly calculate the application of any operators on
``basis vectors''.})
$\gamma^a$, each ``basis vector'' is the eigenvector of all the Cartan subalgebra
members, Eq.~(\ref{cartangrasscliff}). To achieve this, we arrange ``basis vectors'' to
be products of nilpotents and projectors, Eqs.~(\ref{nilproj}, \ref{signature0}), so that
every nilpotent and every projector is the eigenvector of one of the Cartan subalgebra
members.

\begin{small}
Example 1.\\
Let us notice that, for example, two nilpotents anti-commute, while one nilpotent and
one projector (or two projectors) commute due to
Eq.~(\ref{gammatildeantiher0}): \\
$\frac{1}{2} (\gamma^0 - \gamma^3) \frac{1}{2}
(\gamma^1 - i\gamma^2)=- \frac{1}{2} (\gamma^1 - i\gamma^2) \frac{1}{2}
(\gamma^0 - \gamma^3)$, while $\frac{1}{2} (\gamma^0 - \gamma^3) \frac{1}{2}
(1+i \gamma^1 \gamma^2)= \frac{1}{2} (1+i \gamma^1 \gamma^2) \frac{1}{2}
(\gamma^0 - \gamma^3)$.
\end{small}

\vspace{3mm}

In $d=(3+1) $ there are $16 \, (2^{d=4})$ ``eigenvectors" of the Cartan subalgebra
members ($S^{03}, S^{12}$) and (${\bf {\cal S}}^{03}, {\bf {\cal S}}^{12}$) of the
Lorentz algebras $S^{ab}$ and ${\bf {\cal S}}^{ab}$ , Eq.~(\ref{cartangrasscliff}).\\

Half of them are the Clifford odd ``basis vectors'' (and their Hermitian 
conjugated partners in a separate group), appearing in two irreducible 
representations, in two ``families'' ($2^{\frac{4}{2}-1}$, $f=(1,2)$), each with two
($2^{\frac{4}{2}-1}$, $m=(1,2)$) members, $\hat{b}^{ m \dagger}_{f}$, 
Eq.~(\ref{3+1oddb}). \\
There is 
$2^{\frac{4}{2}-1}\times $$2^{\frac{4}{2}-1} $
(Clifford odd) Hermitian conjugated partners $\hat{b}^{ m}_{f}=
(\hat{b}^{ m \dagger}_{f})^{\dagger}$ appearing in a separate group which is
not reachable by $S^{ab}$, Eq.~(\ref{3+1oddHb}).

There are two separate groups of $2^{\frac{4}{2}-1}\times 2^{\frac{4}{2}-1} $
Clifford even ''basis vectors'', ${}^{i}{\bf {\cal A}}^{m \dagger}_{f}, i=(I, II)$,
the $2^{\frac{4}{2}-1}$ members of each are self-adjoint, the rest have their
Hermitian conjugated partners within the same group, 
Eqs.~(\ref{3+1evenAI}, \ref{3+1evenAII}).\\
All the members of each group are reachable by $S^{ab}$ or $\tilde{S}^{ab}$
from any starting ''basis vector'' ${}^{i}{\bf {\cal A}}^{1\dagger}_{1}$.

\begin{small}
Example 2.\\
$\hat{b}^{ m=1 \dagger}_{f=1}=\stackrel{03}{(+i)}\stackrel{12}{[+]}
(=\frac{1}{2} (\gamma^0 - \gamma^3) \frac{1}{2} (1+i\gamma^1 \gamma^2))$ 
is a Clifford odd ``basis vector'', its Hermitian conjugated partner, 
Eq.~(\ref{gammatildeantiher0}), is 
$\hat{b}^{ m=1 }_{f=1}=\stackrel{03}{(-i)}\stackrel{12}{[+]}(=\frac{1}{2} 
(\gamma^0 +\gamma^3) \frac{1}{2} (1+i\gamma^1 \gamma^2)$, not reachable 
by either  $S^{ab}$ or by $\tilde{S}^{ab}$ from any of two members in any of 
two ``families'' of the group of $\hat{b}^{ m \dagger}_{f}$, presented in 
Eq.~(\ref{3+1oddb}).\\
 ${}^{I}{\bf {\cal A}}^{m=1 \dagger}_{f=1} (=\stackrel{03}{[+i]}\stackrel{12}{[+]}
= \frac{1}{2} (1+\gamma^0 \gamma^3) \frac{1}{2} (1+i\gamma^1 \gamma^2)$ 
is self-adjoint,
${}^{I}{\bf {\cal A}}^{m=2 \dagger}_{f=1} (=\stackrel{03}{(-i)}\stackrel{12}{(-)}
= \frac{1}{2} (\gamma^0 +\gamma^3) (\gamma^1 -i\gamma^2)$. Its Hermitian 
conjugated partner, belonging to the same group, is 
${}^{I}{\bf {\cal A}}^{m=1 \dagger}_{f=2}$  and is reachable from 
${}^{I}{\bf {\cal A}}^{m=1 \dagger}_{f=1}$ by the application of $\tilde{S}^{01}$,
since $\tilde{\gamma}^{0}*_{A}\stackrel{03}{[+i]}=i\stackrel{03}{(+i)}$ and
$\tilde{\gamma}^{1}*_{A}\stackrel{12}{[+]}=i\stackrel{12}{(+)}$.
\end{small}

\vspace{2mm}

{\bf Clifford odd ``basis vectors''}

\vspace{2mm}

Let us first present the Clifford odd anti-commuting ``basis vectors'', appearing in
two ``families'' ${\hat b}^{m \dagger}_{f}$, and their Hermitian conjugated partners
$({\hat b}^{m \dagger}_f)^{\dagger}$. Each member of the two groups is a
product of one nilpotent and one projector. We choose the right-handed
Clifford odd ``basis vectors''~\footnote{
We could choose the left-handed Clifford odd ``basis vectors'' by exchanging
the role of `basis vectors'' and their Hermitian conjugated partners.}. Clifford odd
``basis vectors'' appear in two families, each family has two members~\footnote{
In the case of $d=(1+1)$, we would have one family with one member only, which
must be nilpotent.}. Let us notice that members of each of two families have the 
same quantum numbers ($S^{03}\,,S^{12}$). They distinguish in ``family'' quantum 
numbers ($\tilde{S}^{03}\,,\tilde{S}^{12}$).
\begin{small}
\begin{eqnarray}
\label{3+1oddb}
\begin{array} {ccrr}
f=1&f=2&&\\
\tilde{S}^{03}=\frac{i}{2}, \tilde{S}^{12}=-\frac{1}{2}&
\;\;\tilde{S}^{03}=-\frac{i}{2}, \tilde{S}^{12}=\frac{1}{2}\;\;\; &S^{03}\, &S^{12}\\
\hat{b}^{ 1 \dagger}_{1}=\stackrel{03}{(+i)}\stackrel{12}{[+]}&
\hat{b}^{ 1 \dagger}_{2}=\stackrel{03}{[+i]}\stackrel{12}{(+)}&\frac{i}{2}&
\frac{1}{2}\\
\hat{b}^{ 2 \dagger}_{1}=\stackrel{03}{[-i]}\stackrel{12}{(-)}&
\hat{b}^{ 2 \dagger}_{2}=\stackrel{03}{(-i)}\stackrel{12}{[-]}&-\frac{i}{2}&
-\frac{1}{2}\,.
\end{array}
\end{eqnarray}
\end{small}
We find for their Hermitian conjugated partners 
\begin{small}
\begin{eqnarray}
\label{3+1oddHb}
\begin{array} {ccrr}
S^{03}=- \frac{i}{2}, S^{12}=\frac{1}{2}&
\;\;S^{03}=\frac{i}{2}, S^{12}=-\frac{1}{2}\;\;&\tilde{S}^{03} &\tilde{S}^{12}\\
\hat{b}^{ 1 }_{1}=\stackrel{03}{(-i)}\stackrel{12}{[+]}&
\hat{b}^{ 1 }_{2}=\stackrel{03}{[+i]}\stackrel{12}{(-)}&-\frac{i}{2}&
-\frac{1}{2}\\
\hat{b}^{ 2 }_{1}=\stackrel{03}{[-i]}\stackrel{12}{(+)}&
\hat{b}^{ 2 }_{2}=\stackrel{03}{(+i)}\stackrel{12}{[-]}&\frac{i}{2}&
\frac{1}{2}\,.
\end{array}
\end{eqnarray}
\end{small}
The vacuum state $|\psi_{oc}>$, Eq.~(\ref{vaccliffodd1}), on which the Clifford odd
''basis vectors'' apply is equal to:
$|\psi_{oc}>= \frac{1}{\sqrt{2}} (\stackrel{03}{[-i]}\stackrel{12}{[+]}
+\stackrel{03}{[+i]}\stackrel{12}{[-]} )$.

Let us recognize that the Clifford odd ''basis vectors'' anti-commute due to the odd
number of nilpotents, Example 1. And they are orthogonal according to Eqs.~(\ref{graficcliff0},
\ref{graficcliff1}, \ref{graficfollow1}):
$\hat{b}^{ m \dagger}_{f} *_{A} \hat{b}^{ m' \dagger}_{f '}=0$.


\begin{small}
Example 3.\\
According to the vacuum state presented above, one finds that, for example, 
$\hat{b}^{1 \dagger}_1 (=\stackrel{03}{(+i)}\stackrel{12}{[+]})|\psi_{oc}>$
is $\hat{b}^{1 \dagger}_1$ back, since $\stackrel{03}{(+i)}\stackrel{12}{[+]} *_{A}
\stackrel{03}{[-i]}\stackrel{12}{[+]}=\stackrel{03}{(+i)}\stackrel{12}{[+]}$, 
according to Eq.~(\ref{graficcliff0}), while  $\stackrel{03}{(-i)}\stackrel{12}{[+]} *_{A}
\stackrel{03}{[-i]}\stackrel{12}{[+]}=0$ (due to $(\gamma^0 + \gamma^3) 
(1-\gamma^0 \gamma^3) =0$).\\
Let us apply $S^{01}$ and $\tilde{S}^{01}$ on some of the
``basis vectors'' $\hat{b}^{m \dagger}_{f}$, say $\hat{b}^{1 \dagger}_{1}$.\\
When applying $S^{01}=\frac{i}{2}\gamma^0 \gamma^1$ on 
$\frac{1}{2} (\gamma^0 - \gamma^3) \frac{1}{2} (1+ i\gamma^1 \gamma^2)
(\equiv \stackrel{03}{(+i)} \stackrel{12}{[+]})$ we 
get $- \frac{i}{2}  \frac{1}{2} (1-\gamma^0 \gamma^3) 
\frac{1}{2} (\gamma^1-i \gamma^2) (\equiv(- \frac{i}{2} \stackrel{03}{[-i]} 
 \stackrel{12}{(-)})$.\\
 When applying $\tilde{S}^{01}=\frac{i}{2}\tilde{\gamma}^0 \tilde{\gamma}^1$ on 
$\frac{1}{2} (\gamma^0 - \gamma^3) \frac{1}{2} (1+ i\gamma^1 \gamma^2)
(\equiv \stackrel{03}{(+i)} \stackrel{12}{[+]})$ we get, according to 
Eq.~(\ref{tildegammareduced0}), or if using Eq.~(\ref{usefulrel}), 
$- \frac{i}{2}  \frac{1}{2} (1+\gamma^0 \gamma^3) 
\frac{1}{2} (\gamma^1+i \gamma^2) (\equiv(- \frac{i}{2} \stackrel{03}{[+i]} 
 \stackrel{12}{(+)})$.\\
\end{small}


 It then follows, after using Eqs.~(\ref{usefulrel}, \ref{graficcliff0}, \ref{graficcliff1}, 
 \ref{graficfollow1}) or  just the starting relation, Eq.~(\ref{gammatildeantiher0}),
 and taking into account the above concrete evaluations, 
 the relations of Eq.~(\ref{almostDirac}) for our particular case
\begin{small}
\begin{eqnarray}
\label{3+1oddDirac}
\hat{b}^{ m \dagger}_{f}*_{A} |\psi_{oc}>&=&|\psi^m_{f}>\,,\nonumber\\
\hat{b}^{ m }_{f}*_{A} |\psi_{oc}>&=& 0\cdot|\psi_{oc}>\,,\nonumber\\
\{\hat{b}^{ m \dagger}_{f}, \hat{b}^{ m'\dagger}_{f '}\}_{-}*_{A}|\psi_{oc}>&=&0\cdot|\psi_{oc}>\,,
\nonumber\\
\{\hat{b}^{ m }_{f}, \hat{b}^{ m'}_{f '}\}_{-}*_{A}|\psi_{oc}>&=&0\cdot|\psi_{oc}>\,,\nonumber\\
\{\hat{b}^{ m }_{f}, \hat{b}^{ m'\dagger}_{f '}\}_{-}*_{A}|\psi_{oc}>&=&\delta^{m m'}\delta_{f f `}|\psi_{oc}>\,.
\end{eqnarray}
\end{small}
%
The last relation of Eq.~(\ref{3+1oddDirac}) takes into account that each 
``basis vector'' carries  the ``family'' quantum number, determined by 
$\tilde{S}^{ab}$ of the Cartan subalgebra members, Eq.~(\ref{cartangrasscliff}), 
and the appropriate normalization of ``basis vectors'', Eqs.~(\ref{3+1oddb}, \ref{3+1oddHb}).

\vspace{2mm}

{\bf Clifford even ``basis vectors''}

\vspace{2mm}

Besides $2^{\frac{4}{2}-1}\times 2^{\frac{4}{2}-1} $ Clifford odd
``basis vectors'' and the same number of their Hermitian conjugated partners,
Eqs.~(\ref{3+1oddb}, \ref{3+1oddHb}), the Clifford algebra objects offer two
groups of $2^{\frac{4}{2}-1}\times 2^{\frac{4}{2}-1} $ Clifford even ''basis
vectors'', the members of the group ${}^{I}{\bf {\cal A}}^{m \dagger}_{f}$
and ${}^{II}{\bf {\cal A}}^{m \dagger}_{f}$, which have Hermitian conjugated
partners within the same group or are self-adjoint~\footnote {
Let be repeated that ${\bf {\cal S}}^{ab}=S^{ab} + \tilde{S}^{ab} $~\cite{n2022epjc}.}.
We have the group ${}^{I}{\bf {\cal A}}^{m \dagger}_{f}$, $m=(1,2), f=(1,2)$, the
members of which are Hermitian conjugated to each other or are self-adjoint,
\begin{small}
\begin{eqnarray}
\label{3+1evenAI}
\begin{array} {crrcrr}
&{\bf {\cal S}}^{03}&{\bf {\cal S}}^{12}&&{\bf {\cal S}}^{03}&{\bf {\cal S}}^{12}\\
{}^{I}{\bf {\cal A}}^{1 \dagger}_{1}= \stackrel{03}{[+i]}\stackrel{12}{[+]}&0&0&\,,\quad
{}^{I}{\bf {\cal A}}^{1 \dagger}_{2}= \stackrel{03}{(+i)}\stackrel{12}{(+)}&i&1\\
{}^{I}{\bf {\cal A}}^{2 \dagger}_{1}= \stackrel{03}{(-i)}\stackrel{12}{(-)}&-i&-1&\,,\quad
{}^{I}{\bf {\cal A}}^{2 \dagger}_{2}= \stackrel{03}{[-i]}\stackrel{12}{[-]}&0&0\,,
\end{array}
\end{eqnarray}
\end{small}
and the group ${}^{II}{\bf {\cal A}}^{m \dagger}_{f}$, $m=(1,2), f=(1,2)$, the
members of which are either Hermitian conjugated to each other or are self adjoint
\begin{small}
\begin{eqnarray}
\label{3+1evenAII}
\begin{array} {crrcrr}
&{\bf {\cal S}}^{03}&{\bf {\cal S}}^{12}&&{\bf {\cal S}}^{03}&{\bf {\cal S}}^{12}\\
{}^{II}{\bf {\cal A}}^{1 \dagger}_{1}= \stackrel{03}{[+i]}\stackrel{12}{[-]}&0&0&\,,\quad
{}^{II}{\bf {\cal A}}^{1 \dagger}_{2}= \stackrel{03}{(+i)}\stackrel{12}{(-)}&i&-1\\
{}^{II}{\bf {\cal A}}^{2 \dagger}_{1}= \stackrel{03}{(-i)}\stackrel{12}{(+)}&-i&1&\,,\quad
{}^{II}{\bf {\cal A}}^{2 \dagger}_{2}= \stackrel{03}{[-i]}\stackrel{12}{[+]}&0&0\,.
\end{array}
\end{eqnarray}
\end{small}
The Clifford even ``basis vectors'' have no families. The two groups, 
$ {}^{I}{\bf {\cal A}}^{m \dagger}_{f}$ and ${}^{II}{\bf {\cal A}}^{m \dagger}_{f}$ 
(they are not reachable from one another by ${\bf {\cal S}}^{ab}$), are orthogonal (which 
can easily be checked, since  $ \stackrel{ab}{(\pm k)} *_{A}  \stackrel{ab}{(\pm k)}=0$, 
and $ \stackrel{ab}{[\pm k]} *_{A}  \stackrel{ab}{[\mp k]}=0$).
\begin{eqnarray}
\label{3+1AIAIIorth}
{}^{I}{\bf {\cal A}}^{m \dagger}_{f} *_{A} {}^{II}{\bf {\cal A}}^{m' \dagger}_{f `} 
=0, \quad{\rm for \;any } \;(m, m', f, f `)\,.
\end{eqnarray}

\vspace{2mm}

{\bf Application of ${}^{i}{\bf {\cal A}}^{m \dagger}_{f}, i=(I,II)$ on $\hat{b}^{m \dagger}_{f}$}

\vspace{2mm}

Let us demonstrate the application of $ {}^{i}{\bf {\cal A}}^{m \dagger}_{f}, i=(I,II)$,
on the Clifford odd ``basis vectors'' $\hat{b}^{m \dagger}_{f}$, Eqs.~(\ref{calIAb1234gen},
\ref{calbIIA1234gen}), for our particular case $d=(3+1)$ and compare the result with the
result of application $S^{ab}$ and $\tilde{S}^{ab}$ on $\hat{b}^{m \dagger}_{f}$
evaluated above in Example 3. We found, for example, that
$S^{01}(=\frac{i}{2}\gamma^0 \gamma^1) *_{A} \hat{b}^{1 \dagger}_1(=$
$\frac{1}{2} (\gamma^0 - \gamma^3) \frac{1}{2} (1+ i\gamma^1 \gamma^2)
(=\stackrel{03}{(+i)} \stackrel{12}{[+]})=$
$- \frac{i}{2} \frac{1}{2} (1-\gamma^0 \gamma^3)
\frac{1}{2} (\gamma^1-i \gamma^2) (=(- \frac{i}{2} \stackrel{03}{[-i]}
\stackrel{12}{(-i)})= - \frac{i}{2} \hat{b}^{2 \dagger}_{1}$.

Applying ${}^{I}{\bf {\cal A}}^{2 \dagger}_{1} (=\stackrel{03}{(-i)}\stackrel{12}{(-)})
*_{A} \, \hat{b}^{1 \dagger}_1(=$
$\stackrel{03}{(+i)} \stackrel{12}{[+]})=-\stackrel{03}{[-i]} \stackrel{12}{(-)}$, which is
$- \hat{b}^{2 \dagger}_1$, presented in Eq.~(\ref{3+1oddb}). We obtain in both cases
the same result, up to the factor $\frac{i}{2}$ (in front of $\gamma^{0} \gamma^{1}$ in
$S^{01}$). In the second case one sees that ${}^{I}{\bf {\cal A}}^{2 \dagger}_{1}$
(carrying ${\cal S}^{03}=-i, {\cal S}^{12}=-1$) transfers these quantum numbers to
$ \hat{b}^{1 \dagger}_1 $ (carrying ${S}^{03}=\frac{i}{2}, {S}^{12}=\frac{1}{2}$)
what results in $ \hat{b}^{2 \dagger}_1 $ (carrying ${S}^{03}=\frac{-i}{2}, {S}^{12}
=\frac{-1}{2}$).

We can check what the application of the rest three
$ {}^{I}{\bf {\cal A}}^{m \dagger}_{f}$,
do when applying on $ \hat{b}^{m \dagger}_f $. The self-adjoint member carrying
${\cal S}^{03}=0, {\cal S}^{12}=0$, either gives $ \hat{b}^{m \dagger}_f $
back, or gives zero, according to Eq.~(\ref{graficcliff0}). The Clifford even ``basis
vectors'', carrying non zero ${\cal S}^{03}$ and ${\cal S}^{12}$ transfer their
internal values to $ \hat{b}^{m \dagger}_f $ or give zero. In all cases
$ {}^{I}{\bf {\cal A}}^{m \dagger}_{f}$ transform a ``family'' member to another
or the same ``family'' member of the same ``family''.

\begin{small}
Example 4.:\\
${}^{I}{\bf {\cal A}}^{1 \dagger}_{1} (=\stackrel{03}{[+i]}\stackrel{12}{[+]}) *_{A}
\,\hat{b}^{1 \dagger}_1(=$ $\stackrel{03}{(+i)} \stackrel{12}{[+]})=\hat{b}^{1 \dagger}_1(=$  
$\stackrel{03}{(+i)} \stackrel{12}{[+]})$\,,\quad
${}^{I}{\bf {\cal A}}^{1 \dagger}_{1} (=\stackrel{03}{[+i]}\stackrel{12}{[+]}) *_{A}
\,\hat{b}^{1 \dagger}_2(=$ $\stackrel{03}{[+i]} \stackrel{12}{(+)})=\hat{b}^{1 \dagger}_2(=$  
$\stackrel{03}{[+i]} \stackrel{12}{(+)})$\,,\\
${}^{I}{\bf {\cal A}}^{2 \dagger}_{1}  (=\stackrel{03}{(-i)}\stackrel{12}{(-)}) *_{A}
\,\hat{b}^{1 \dagger}_2(=$ $\stackrel{03}{[+i]} \stackrel{12}{(+)})=-\hat{b}^{2 \dagger}_2(=$  
$\stackrel{03}{(-i)} \stackrel{12}{[-]})$\,,\quad
${}^{I}{\bf {\cal A}}^{2 \dagger}_{1}  (=\stackrel{03}{(-i)}\stackrel{12}{(-)}) *_{A}
\hat{b}^{2 \dagger}_2 (=$  $\stackrel{03}{(-i)} \stackrel{12}{[-]})=0$.\\
\end{small}

\vspace{2mm}

One easily sees that the application of $  {}^{II}{\bf {\cal A}}^{m \dagger}_{f}$ on 
$ \hat{b}^{m' \dagger}_{f `}$ gives zero for all $(m,m',f, f ' )$ (due to 
$\stackrel{ab}{[\pm k]} *_{A} \stackrel{ab}{[\mp k]} =0$,  
$\stackrel{ab}{[\pm k]} *_{A} \stackrel{ab}{(\mp k)} = 0$, and similar applications).

We realised in Example 3. that  the application of $\tilde{S}^{01}=
\frac{i}{2}\tilde{\gamma}^0 \tilde{\gamma}^1$ on $\hat{b}^{1 \dagger}_1$
gives $(- \frac{i}{2} \stackrel{03}{[+i]}  \stackrel{12}{(+i)}) = 
-\frac{i}{2} \hat{b}^{1 \dagger}_{2}$.

Let us algebraically,  $*_{A} $, apply $  {}^{II}{\bf {\cal A}}^{2 \dagger}_{1}
(= \stackrel{03}{(-i)} \stackrel{12}{(+)}$), with quantum numbers $({\cal S}^{03}, 
{\cal S}^{12})=(-i,1)$, from the right hand side the Clifford odd ``basis vector'' 
$\hat{b}^{1 \dagger}_{1}$. This application causes the transition of 
$\hat{b}^{1 \dagger}_{1}$ (with quantum numbers $(\tilde{S}^{03}, \tilde{S}^{12})=
(\frac{i}{2}, -\frac{1}{2})$ (see Eq.~(\ref{signature0})) into $\hat{b}^{1 \dagger}_{2}$ 
(with quantum numbers $(\tilde{S}^{03}, \tilde{S}^{12})=(-\frac{i}{2}, \frac{1}{2})$). 
$  {}^{II}{\bf {\cal A}}^{2 \dagger}_{1}$ obviously transfers its quantum numbers to
Clifford odd ``basis vectors'', keeping $m$ unchanged, and changing the ``family'' quantum 
number:
$\hat{b}^{1 \dagger}_{1} *_{A}  {}^{II}{\bf {\cal A}}^{2 \dagger}_{1}=
\hat{b}^{1 \dagger}_{2}$.

\vspace{2mm}

We can conclude: The internal space of the Clifford even ``basis vectors'' has 
properties of the gauge fields of the Clifford odd ``basis vectors''; 
${}^{I}{\bf {\cal A}}^{m \dagger}_{f}$ transform ``family'' members of the Clifford odd 
``basis vectors'' among themselves, keeping the ``family'' quantum number unchanged, 
${}^{II}{\bf {\cal A}}^{m \dagger}_{f}$ transform a particular ``family'' member into 
the same ``family'' member of another ``family''.

\vspace{3mm}

{\bf Creation and annihilation operators}

\vspace{3mm}

To define creation and annihilation operators for fermion and boson fields, we must 
include besides the internal space, the ordinary space, presented in Eq.~(\ref{creatorp}), 
which defines the momentum or coordinate part of fermion and boson fields.

\vspace{2mm}

We define the creation operators for the single particle fermion states as a tensor
product, $*_{T}$, of the Clifford odd ``basis vectors'' and the basis in ordinary space,
Eq.~(\ref{wholespacefermions}):\\
${\bf \hat{b}}^{s \dagger}_{f} (\vec{p}) =
\sum_{m} c^{sm}{}_f (\vec{p}) \,\hat{b}^{\dagger}_{\vec{p}}\,*_{T}\,
\hat{b}^{m \dagger}_{f}$. The annihilation operators are their Hermitian conjugated
partners.

We have seen in Example 1. that Clifford odd ``basis vectors'' (having odd products
of nilpotents) anti-commute. The commuting objects $\hat{b}^{\dagger}_{\vec{p}}$
(multiplying the ``basis vectors'') do not change the Clifford oddness of 
${\bf \hat{b}}^{s \dagger}_{f} (\vec{p})$.
The two Clifford odd objects, ${\bf \hat{b}}^{s \dagger}_{f} (\vec{p})$ and
${\bf \hat{b}}^{s' \dagger}_{f `} (\vec{p'}) $, keep their anti-commutativity,
fulfilling the anti-commutation relations as presented in Eq.~(\ref{Weylpp'comrel}).
Correspondingly we do not need to postulate anti-commutation relations of Dirac.
The Clifford odd ``basis vectors'' in a tensor product with the basis in ordinary
space explain the second quantized postulates for fermion fields.

The Clifford odd ``basis vectors'' contribute for each $\vec{p}$ a finite number
of ${\bf \hat{b}}^{s \dagger}_{f} (\vec{p})$, the ordinary basis offers infinite
possibilities~\footnote{
An infinitesimally small difference between $\vec{p}$ and
$\vec{p'}$ makes two creation operators ${\bf \hat{b}}^{s \dagger}_{f} (\vec{p})$
and ${\bf \hat{b}}^{s \dagger}_{f } (\vec{p'}) $ with the same ``basis vector''
describing the internal space of fermion fields
still fulfilling the anti-commutation relations (as we learn from atomic physics;
two electrons can carry the same spin if they distinguish in the coordinate part of
the state).}.

\vspace{2mm}

Recognizing that internal spaces of fermion fields and their corresponding
boson gauge fields are describable in even dimensional spaces by the Clifford
odd and even ``basis vectors'', respectively, it becomes evidently that when
including the basis in ordinary space, we must take into account
that boson gauge fields have the space index $\alpha$, which describes the
$\alpha$ component of the boson fields in ordinary space.

We multiply, therefore, as presented in Eq.~(\ref{wholespacebosons}),
the Clifford even ``basis vectors'' with the coefficient
${}^{i}{\cal C}^{ m}{}_{f \alpha}$ carrying the space index $\alpha$ so
that the creation operators
${\bf {}^{i}{\hat{\cal A}}^{m \dagger}_{f \alpha}} (\vec{p})=
\hat{b}^{\dagger}_{\vec{p}}\,*_{T}\,
{}^{i}{\cal C}^{ m}{}_{f \alpha}\, {}^{i}{\hat{\cal A}}^{m \dagger}_{f} \,
\,, i=(I,II)$ carry the space index $\alpha$~\footnote{
Requiring the local phase symmetry for the fermion part of the action,
Eq.~(\ref{wholeaction}), would lead to the requirement of the existence of the
boson fields with the space index $\alpha$.}.
The self-adjoint ``basis vectors'', like
($ {}^{i}{\hat{\cal A}}^{1 \dagger}_{1 \alpha},
{}^{i}{\hat{\cal A}}^{2 \dagger}_{2 \alpha}, i=(I,II)$), do not change
quantum numbers of the Clifford odd ``basis vectors'', since they have
internal quantum numbers equal to zero.

In higher dimensional space,
like in $d=(5+1)$, ${}^{I}{\hat{\cal A}}^{1 \dagger}_{3}$, presented in
Table~\ref{Cliff basis5+1even I.}, could represent the internal space of a
photon field, which transfers to, for example, a fermion and anti-fermion
pair with the internal space described by ($\hat{b}^{1 \dagger}_{1}$,
$\hat{b}^{3 \dagger}_{1}$), presented in Table~\ref{oddcliff basis5+1.},
the momentum in ordinary space.

The subgroup structure of $SU(3)$ gauge fields can be recognized in
Fig.~\ref{FigSU3U1even}.

\vspace{3mm}

Properties of the gauge fields $ {}^{i}{\hat{\cal A}}^{m \dagger}_{f \alpha}$
need further studies. 
 


%




In even dimensional spaces, the Clifford odd and even ``basis vectors'', describing internal
spaces of fermion and boson fields, offer the explanation for the second quantized
postulates for fermion and boson fields~[17].


%
\section{Discussion on the open questions of the {\it standard model} and answers offered
by the {\it spin-charge-family} theory}
\label{OpenQuestionsSM}

\begin{small}

There are many suggestions in the literature for unifying charges in larger groups, 
adding additional groups for describing families ~\cite{Geor,FritzMin,PatiSal,GeorGlas,Cho}, 
or by going to higher dimensional spaces of the Kaluza-Kline like theories~\cite{KaluzaKlein,Witten,Duff,App,SapTin,Wetterich,mil,zelenaknjiga}, what also 
the {\it spin-charge-family} is.\\

Let me present some open questions of the {\it standard model} and briefly tell the 
answers offered by the {\it spin-charge family} theory. \\

\noindent
{\bf A.} Where do fermions --- quarks and leptons and antiquarks and antileptons ---
and their families originate?\\
The answer offered by the {\it spin-charge-family} theory: In $d= (13+1)$ one 
irreducible representation of $SO(13,1)$ analysed with
respect to subgroups $SO(7,1)$ (containing subgroups of $SO(3,1)\times SU(2)\times
SU(2)$) and $SO(6)$ (containing subgroups of $SU(3)\times U(1)$) offers the Clifford
odd ``basis vectors'', describing the internal spaces of quarks and leptons and
antiquarks and antileptons, Table~\ref{Table so13+1.}, as assumed by the
{\it standard model}. The Clifford odd ``basis vectors'' appear in families.\\
{\bf B.} Why are charges of quarks so different from charges of leptons, and why
have left-handed family members so different charges from the right-handed ones? \\
The answer offered by the {\it spin-charge-family} theory: The $SO(7,1)$ part
of the ``basis vectors'' is identical for quarks and leptons and identical for
antiquarks and antileptons, Table~\ref{Table so13+1.}, they distinguish only in
the $SU(3)$, the colour or anticolour part, and in the fermion or antifermion $U(1)$
quantum numbers. All families have the same content of $SO(7,1), SU(3)$
and $U(1)$ with respect to $S^{ab}$. They distinguish only in the family
quantum number, determined by $\tilde{S}^{ab}$. The difference
between left-handed and right-handed members appears due to the difference in
one quantum numbers of the two $SU(2)$ groups, as seen in
Table~\ref{Table so13+1.}.\\
{\bf C.} Why do family members --- quarks and leptons --- manifest such different
masses if they all start as massless, as (elegantly) assumed by the {\it standard model}?\\
The answer offered by the {\it spin-charge-family} theory: Masses of quarks and leptons
are in this theory determined by the spin connection fields $\omega_{st \sigma}$, the
gauge fields of $S^{ab}$~\footnote{
The three $U(1)$ singlets, the gauge fields of the ``fermion'' quantum number $\tau^{4}$,
of the hypercharge $Y$, and of the electromagnetic charge $Q$, determine the
difference in masses of quarks and leptons, presented in Table~\ref{Table so13+1.}, 
Ref.~(\cite{nh2021RPPNP}, Sect, 6.2.2, Eq.~(108))},
and by $\tilde{\omega}_{st \sigma}$, the gauge fields of $\tilde{S}^{ab}$, which
are the same for quarks and leptons~\footnote{
The two times two $\widetilde{SU}(2)$ triplets are the same for quarks and leptons, forming
two groups of four families. Ref.~(\cite{nh2021RPPNP}, Sect, 6.2.2, Eq.~(108).}.
Triplets and singlets are scalar gauge fields with the space index $\sigma= (7,8)$. They
have, with respect to the space index, the quantum numbers of the Higgs scalars,
Ref.~(\cite{nh2021RPPNP}, Table 8, Eq.~(110,111)). \\
%
%
{\bf D.} What is the origin of boson fields, of vector fields which are the gauge fields
of fermions, and the Higgs' scalars and the Yukawa couplings? Have all boson
fields, with gravity and scalar fields included a common origin? \\
The answer offered by the {\it spin-charge-family} theory: In a simple starting
action, Eq.~(\ref{wholeaction}), boson fields origin in gravity --- in vielbeins and
two kinds of spin connection fields, $\omega_{ab\alpha}$ and $
\tilde{\omega}_{ab \alpha}$, in $d=(13 +1)$ --- and manifest in $d=(3+1)$
as vector gauge fields, $\alpha= (0,1,2,3)$, or scalar gauge fields,
$\alpha \ge 5$~\cite{nd2017}, (\cite{nh2021RPPNP}, Sect. 6 and references
therein). Boson gauge fields are massless as there are fermion fields. The breaks
of the starting symmetry makes some gauge fields massive.
This article describes the internal space of boson fields by the Clifford
even basis vectors, manifesting as the boson gauge fields of the corresponding
fermion fields described by the Clifford odd ``basis vectors''. The description of the
boson fields with the Clifford even ``basis vectors'' confirms the
existence of two kinds of spin connection fields as we see in
Sects.~\ref{basisvectors} and\ref{cliffordoddevenbasis5+1}, but also open a door to 
a new understanding of gravity. According to the starting action, 
Eq.~(\ref{wholeaction}), all gauge fields start in $d\ge (13+1)$ as gravity. \\ 
%
{\bf E.} How are scalar fields connected with the origin of families? How many scalar 
fields determine properties of the so far (and others possibly be) observed fermions and
of weak bosons? \\
The answer offered by the {\it spin-charge-family} theory:
The interaction between quarks and leptons and the scalar gauge fields, which at the
electroweak brake obtain constant values, causes that quarks and leptons and the weak
bosons become massive. There are three singlets, they distinguish among quarks and 
leptons, and two triplets, they do not distinguish among quarks and leptons, which 
give masses to the lower four families~\footnote{
There are the same three singlets and two additional triplets, which determine the masses
of the upper four families- explaining the existence of the dark matter.}.\\
%
%
%
%
%
%
{\bf F.} Where does the {\it dark matter} originate?\\
The answer offered by the {\it spin-charge-family} theory: The theory predicts two 
groups of four families at low energy. The stable of the upper four groups
are candidates to form the dark matter~\cite{gn2009}.\\
%
%
{\bf G.} Where does the ``ordinary" matter-antimatter asymmetry originate?\\
The answer offered by the {\it spin-charge-family} theory: The theory predicts
scalars triplets and antitriplets with the space index
$\alpha=(9,10,11,12,13,14)$~\cite{n2014matterantimatter}.\\
{\bf H.} How can we understand the second quantized fermion and boson fields?\\
The answer offered by the {\it spin-charge-family} theory: The main contribution
of this article, Sect.~\ref{creationannihilation}, is the description of the internal
spaces of fermion and boson fields with the superposition of odd (for fermions) and
even (for bosons) products of $\gamma^{a}$. The
corresponding creation and annihilation operators, which are tensor, $*_{T}$,
products of (finite number) ``basis vectors'' and (infinite) basis in ordinary space
inherit anti-commutativity or commutativity from the corresponding ``basis
vectors'', explaining the postulates for the second quantized fermion and boson
fields. \\
%
%
{\bf I.} What is the dimension of space? $(3+1)?$, $((d-1)+1)?$, $\infty?$ \\
The answer offered by the {\it spin-charge-family} theory: We observe
$(3+1)$-dimensional space. In order that
one irreducible representation (one family) of the Clifford odd ``basis vectors'',
analysed with respect to subgroups $SO(3,1)\times$ $SO(4)$ $\times SU(3)$
$\times U(1)$ of the group $SO(13,1)$ includes all quarks and leptons and
antiquarks and antileptons, the space must have $d\ge (13+1)$. (Since the only
``elegantly'' acceptable numbers are $0$ and $\infty$, the space-time could be
$\infty$.)\\
The $SO(10)$ theory~\cite{FritzMin}, for example, unifies the charges of
fermions and bosons separately. Analysing $SO(10)$ with respect to the
corresponding subgroups, the charges of fermions appear in fundamental
representations and bosons in adjoint representations~\footnote{The
space-time is in unifying theories $(3+1)$, consequently they have to relate 
handedness and charges ``by hand''~\cite{nh2017}, postulate the
existence of antiparticles, and the existence of scalar fields, as does the {\it
standard model.}}.\\
There are additional open questions answers of which the {\it spin-charge-family}
the theory offers.
\end{small}

The {\it spin-charge-family} theory has to answer the question common to all the 
Kaluza-Klein-like theories: How and why the space we observe has $d=(3+1)$ 
dimensions?  The proposed description of the internal spaces of fermion and boson 
fields might help.

\section{Some useful relations in Grassmann and Clifford algebras, needed also in App.~\ref{13+1representation} }
\label{A}

This appendix contains the helpful relations needed for the reader of this paper. For more detailed 
explanations and for proofs, the reader is kindly asked to read~\cite{nh2021RPPNP} 
and the \mbox{references therein.}

 
 For fermions, the operator of handedness $\Gamma^d$ is determined as follows:
  \begin{small}
\begin{eqnarray}
\label{Gamma}
 \Gamma^{(d)}= \prod_a (\sqrt{\eta^{aa}} \gamma^a)  \cdot \left \{ \begin{array}{l l}
 (i)^{\frac{d}{2}} \,, &\rm{ for\, d \,even}\,,\\
 (i)^{\frac{d-1}{2}}\,,&\rm{for \, d \,odd}\,,
  \end{array} \right.
 \end{eqnarray}
 \end{small}
%

The vacuum state for the Clifford odd ''basis vectors'', $|\psi_{oc}>$, is defined as
 \begin{small}
\begin{eqnarray}
\label{vaccliffodd1}
|\psi_{oc}>= \sum_{f=1}^{2^{\frac{d}{2}-1}}\,\hat{b}^{m}_{f}{}_{*_A}
\hat{b}^{m \dagger}_{f} \,|\,1\,>\,.
\end{eqnarray}
 \end{small}

Taking into account that the Clifford objects  $\gamma^a$ and $\tilde{\gamma}^a$ fulfil 
relations of Eq.~\ref{gammatildeantiher0}, one obtains beside the relations presented in 
Eq.~(\ref{usefulrel}) the following ones
%
\begin{small}
\begin{eqnarray}
\stackrel{ab}{(k)}\stackrel{ab}{(-k)}& =& \eta^{aa} \stackrel{ab}{[k]}\,,\quad 
\stackrel{ab}{(-k)}\stackrel{ab}{(k)} = \eta^{aa} \stackrel{ab}{[-k]}\,,\quad
\stackrel{ab}{(k)}\stackrel{ab}{[k]} =0\,,\quad 
\stackrel{ab}{(k)}\stackrel{ab}{[-k]} =
 \stackrel{ab}{(k)}\,,\quad 
 \nonumber\\
 \stackrel{ab}{(-k)}\stackrel{ab}{[k]} &=& \stackrel{ab}{(-k)}\,,\quad 
\stackrel{ab}{[k]}\stackrel{ab}{(k)}= \stackrel{ab}{(k)}\,,
\quad 
 \stackrel{ab}{[k]}\stackrel{ab}{(-k)} =0\,,\quad 
 \stackrel{ab}{[k]}\stackrel{ab}{[-k]} =0\,,\quad 
 \nonumber\\
\stackrel{ab}{\tilde{(k)}} \stackrel{ab}{(k)}&=& 0\,,\quad
\stackrel{ab}{\tilde{(k)}} \stackrel{ab}{(-k)}=-i \eta^{aa}\stackrel{ab}{[-k]}
\,,\quad
\stackrel{ab}{\widetilde{(-k)}} \stackrel{ab}{(k)}=-i \eta^{aa}\stackrel{ab}{[k]}
\,,\quad
\stackrel{ab}{\tilde{(k)}} \stackrel{ab}{[k]}= i  \stackrel{ab}{(k)}
\,, \quad \nonumber\\
%
\stackrel{ab}{\tilde{(k)}} \stackrel{ab}{[-k]}&=&0
\,,\quad
\stackrel{ab}{\widetilde{(-k)}} \stackrel{ab}{[k]}=0\,,
\quad
\stackrel{ab}{\widetilde{(-k)}} \stackrel{ab}{[-k]}=i  \stackrel{ab}{(-k)}\,,
\quad \stackrel{ab}{\tilde{[k]}} \stackrel{ab}{(k)}=\stackrel{ab}{(k)}\,,\nonumber\\
\stackrel{ab}{\tilde{[k]}} \stackrel{ab}{(-k)}&=&0\,, \quad
 \stackrel{ab}{\widetilde{[k]}} \stackrel{ab}{[k]}=0\,,\quad 
 \stackrel{ab}{\widetilde{[-k]}} \stackrel{ab}{[k]}=
\stackrel{ab}{[k]}\,,\quad
\stackrel{ab}{\tilde{[k]}}\stackrel{ab}{[-k]}=\stackrel{ab}{[-k]}\,,
\label{graficcliff0}
 \end{eqnarray}
\end{small}
The algebraic multiplication among $\stackrel{ab}{\tilde{(k)}}$ and
$\stackrel{ab}{\tilde{[k]}}$  goes as in the case of $\stackrel{ab}{(k)}$ and
$\stackrel{ab}{[k]}$
\begin{small}
\begin{eqnarray}
\stackrel{ab}{\tilde{(k)}}\stackrel{ab}{\tilde{[k]}}& =& 0\,,\quad
\stackrel{ab}{\tilde{[k]}}\stackrel{ab}{\tilde{(k)}}=  \stackrel{ab}{\tilde{(k)}}\,, 
\quad 
  \stackrel{ab}{\tilde{(k)}}\stackrel{ab}{\tilde{[-k]}} =  \stackrel{ab}{\tilde{(k)}}\,,
\quad \, \stackrel{ab}{\tilde{[k]}}\stackrel{ab}{\tilde{(-k)}} =0\,, \nonumber\\
  \stackrel{ab}{\widetilde{(-k)}}\stackrel{ab}{\tilde{(k)}}&=& \eta^{aa} \stackrel{ab}{[-k]}\,,\quad
 \stackrel{ab}{\widetilde{(-k)}} \stackrel{ab}{\widetilde{[-k]}}= 0\,,\quad 
%
\label{graficcliff1}
\end{eqnarray}
\end{small}
%
One can further find that
\begin{small}
\begin{eqnarray}
\label{graficfollow1}
S^{ac}\stackrel{ab}{(k)}\stackrel{cd}{(k)} &=& -\frac{i}{2} \eta^{aa} \eta^{cc} 
\stackrel{ab}{[-k]}\stackrel{cd}{[-k]}\,, \quad 
S^{ac}\stackrel{ab}{[k]}\stackrel{cd}{[k]} = 
\frac{i}{2} \stackrel{ab}{(-k)}\stackrel{cd}{(-k)}\,,\nonumber\\
S^{ac}\stackrel{ab}{(k)}\stackrel{cd}{[k]} &=& -\frac{i}{2} \eta^{aa}  
\stackrel{ab}{[-k]}\stackrel{cd}{(-k)}\,, \quad
S^{ac}\stackrel{ab}{[k]}\stackrel{cd}{(k)} = \frac{i}{2} \eta^{cc}  
\stackrel{ab}{(-k)}\stackrel{cd}{[-k]}\,.
\end{eqnarray}
\end{small}


%
\section{One family representation of Clifford odd ``basis vectors'' in $d=(13+1)$
}
\label{13+1representation}  

This appendix, is following App.~D of Ref.~\cite{n2023MDPI}, with a short comment
on the corresponding gauge vector and scalar fields and fermion and boson
representations in $d=(14+1)$-dimensional space included.

In even dimensional space $d=(13 +1)$~(\cite{n2022epjc}, App.~A), one irreducible
representation of the Clifford odd ``basis vectors'', analysed from the point of view of
the subgroups $SO(3,1)\times SO(4) $ (included in $SO(7,1)$) and $SO(7,1)\times
SO(6)$ (included in $SO(13,1)$, while $SO(6)$ breaks into $SU(3)\times U(1)$),
contains the Clifford odd ``basis vectors'' describing internal spaces of quarks and
leptons and antiquarks, and antileptons with the quantum numbers assumed by the
{\it standard model} before the electroweak break. Since $SO(4)$ contains two
$SU(2)$ groups, $Y=\tau^{23} + \tau^4$, one irreducible representation includes
the right-handed neutrinos and the left-handed antineutrinos, which are not in the
{\it standard model} scheme.\\

The Clifford even ``basis vectors'', analysed to the same subgroups,
offer the description of the internal spaces of the corresponding vector and scalar
fields, appearing in the {\it standard model} before the electroweak
break~\cite{n2021SQ,n2022epjc}; as explained in Subsect.~\ref{deven}.

For an overview of the properties of the vector and scalar gauge fields in the
{\it spin-charge-family} theory, the reader is invited to see Refs.~%
(\cite{nh2021RPPNP,nd2017} and the references therein). The vector gauge
fields, expressed as the superposition of spin connections and vielbeins, carrying the space
index $m=(0,1,2,3)$, manifest properties of the observed boson fields. The scalar
gauge fields, causing the electroweak break, carry the space index $s=(7,8)$ and
determine the symmetry of mass matrices of quarks and leptons.
~\cite{NHD,NH2006,nh2008}).

In this Table~\ref{Table so13+1.}, one can check the quantum numbers of the
Clifford odd ``basis vectors'' representing quarks and leptons {\it and antiquarks and
antileptons} if taking into account that all the nilpotents and projectors are eigenvectors
of one of the Cartan subalgebra members, ($S^{03}, S^{12}, S^{56}, \dots, $
$S^{13\,14}$), with the eigenvalues $\pm \frac{i}{2}$ for $\stackrel{ab}{(\pm i)}$
and $\stackrel{ab}{[\pm i]}$, and with the eigenvalues $\pm \frac{1}{2}$ for
$\stackrel{ab}{(\pm 1)}$ and $\stackrel{ab}{[\pm 1]}$.

Taking into account that the third component of the weak charge,
$\tau^{13}=\frac{1}{2} (S^{56}-S^{78})$, for the second $SU(2)$ charge,
$\tau^{23}=\frac{1}{2} (S^{56}+ S^{78})$, for the colour charge [$\tau^{33}=
\frac{1}{2} (S^{9\, 10}-S^{11\,12})$ and $\tau^{38}=
\frac{1}{2\sqrt{3}} (S^{9\, 10}+S^{11\,12} - 2 S^{13\,14})$], for the ``fermion
charge'' $\tau^4=-\frac{1}{3} (S^{9\, 10}+S^{11\,12} + S^{13\,14})$, for the hyper
charge $Y=\tau^{23} + \tau^4$, and electromagnetic charge $Q=Y + \tau^{13}$,
one reproduces all the quantum numbers of quarks, leptons, and {\it antiquarks, and
antileptons}. One notices that the $SO(7,1)$ part is the same for quarks and leptons
and the same for antiquarks and antileptons. Quarks distinguish from leptons only in
the colour and ``fermion'' quantum numbers and antiquarks distinguish from antileptons
only in the anti-colour and ``anti-fermion'' quantum numbers.

\vspace{2mm}

In odd dimensional space, $d=(14+1)$, the eigenstates of handedness are the superposition
of one irreducible representation of $SO(13,1)$, presented in Table~\ref{Table so13+1.},
and the one obtained if on each ``basis vector'' appearing in $SO(13,1)$ the operator
$S^{0 \, (14+1)}$ applies, Subsect.~\ref{dodd}, Ref.~\cite{n2023MDPI}.

Let me point out that in addition to the electroweak break of the {\it standard
model} the break at $\ge 10^{16}$ GeV is needed~(\cite{nh2021RPPNP}, and references
therein).
The condensate of the two right-handed neutrinos causes this break
(Ref.~\cite{nh2021RPPNP}, Table 6); it  interacts with all the scalar and vector gauge
fields, except the weak, $U(1), SU(3)$ and the gravitational field in $d=(3+1)$, leaving
these gauge fields massless up to the electroweak break, when the scalar fields, leaving
massless only the electromagnetic, colour and gravitational fields, cause masses of 
fermions and weak bosons.

The theory predicts two groups of four families: To the lower group of four families,
the three so far observed contribute. The theory predicts the symmetry of both groups
to be $SU(2)\times SU(2) \times U(1)$, Ref.~(\cite{nh2021RPPNP}, Sect. 7.3), which
enable to calculate mixing matrices of quarks and leptons for the accurately enough
measured $3\times 3$ sub-matrix of the $4\times 4$ unitary matrix. No sterile neutrinos
are needed, and no symmetry of the mass matrices must be guessed~\cite{gn2013}.

In the literature, one finds a lot of papers trying to reproduce mass matrices
and measured mixing matrices for quarks and leptons~\cite{Fritz,Frogatt,Jarlskog,%
Branco,Harari,Altarelli,FRI10}.

The stable of the upper four families predicted by the {\it spin-charge-family} theory is 
a candidate for the dark matter, as discussed in Refs.~\cite{gn2009,nh2021RPPNP}. In the
literature, there are several works suggesting candidates for the dark matter
and also for matter/antimatter asymmetry~\cite{Paul,Raby}.

%

\bottomcaption{\label{Table so13+1.}%
\begin{small}
The left-handed ($\Gamma^{(13,1)} = -1$, Eq.~(\ref{Gamma})) irreducible representation
of one family of spinors --- the product of the odd number of nilpotents and of projectors,
which are eigenvectors of the Cartan subalgebra of the $SO(13,1)$ 
group~\cite{n2014matterantimatter,nh02}, manifesting the subgroup $SO(7,1)$ of the
colour charged quarks and antiquarks and the colourless leptons and antileptons ---
is presented.
It contains the left-handed ($\Gamma^{(3,1)}=-1$) weak ($SU(2)_{I}$) charged
($\tau^{13}=\pm \frac{1}{2}$), 
and $SU(2)_{II}$ chargeless ($\tau^{23}=0$) 
quarks and leptons, and the right-handed
($\Gamma^{(3,1)}=1$) weak ($SU(2)_{I}$) chargeless and $SU(2)_{II}$ charged
($\tau^{23}=\pm \frac{1}{2}$) quarks and leptons, both with the spin $ S^{12}$ up
and down ($\pm \frac{1}{2}$, respectively).
Quarks distinguish from leptons only in the $SU(3) \times U(1)$ part: Quarks are triplets
of three colours ($c^i$ $= (\tau^{33}, \tau^{38})$ $ = [(\frac{1}{2},\frac{1}{2\sqrt{3}}),
(-\frac{1}{2},\frac{1}{2\sqrt{3}}), (0,-\frac{1}{\sqrt{3}}) $, 
carrying the "fermion charge" ($\tau^{4}=\frac{1}{6}$). 
The colourless leptons carry the "fermion charge" ($\tau^{4}=-\frac{1}{2}$).
The same multiplet contains also the left handed weak ($SU(2)_{I}$) chargeless and
$SU(2)_{II}$ charged antiquarks and antileptons and the right handed weak
($SU(2)_{I}$) charged and $SU(2)_{II}$ chargeless antiquarks and antileptons.
Antiquarks distinguish from antileptons again only in the $SU(3) \times U(1)$ part:
Antiquarks are anti-triplets carrying the "fermion charge" ($\tau^{4}=-\frac{1}{6}$).
The anti-colourless antileptons carry the "fermion charge" ($\tau^{4}=\frac{1}{2}$).
$Y=(\tau^{23} + \tau^{4})$ is the hyper charge, the electromagnetic charge
is $Q=(\tau^{13} + Y$).
%
\end{small}
}

\tablehead{\hline
i&$$&$|^a\psi_i>$&$\Gamma^{(3,1)}$&$ S^{12}$&
$\tau^{13}$&$\tau^{23}$&$\tau^{33}$&$\tau^{38}$&$\tau^{4}$&$Y$&$Q$\\
\hline
&& ${\rm (Anti)octet},\,\Gamma^{(7,1)} = (-1)\,1\,, \,\Gamma^{(6)} = (1)\,-1$&&&&&&&&& \\
&& ${\rm of \;(anti) quarks \;and \;(anti)leptons}$&&&&&&&&&\\
\hline\hline}
\tabletail{\hline \multicolumn{12}{r}{\emph{Continued on next page}}\\}
\tablelasttail{\hline}
\begin{tiny}
\begin{supertabular}{|r|c||c||c|c||c|c||c|c|c||r|r|}
1&$ u_{R}^{c1}$&$ \stackrel{03}{(+i)}\,\stackrel{12}{[+]}|
\stackrel{56}{[+]}\,\stackrel{78}{(+)}
||\stackrel{9 \;10}{(+)}\;\;\stackrel{11\;12}{[-]}\;\;\stackrel{13\;14}{[-]} $ &1&$\frac{1}{2}$&0&
$\frac{1}{2}$&$\frac{1}{2}$&$\frac{1}{2\,\sqrt{3}}$&$\frac{1}{6}$&$\frac{2}{3}$&$\frac{2}{3}$\\
\hline
2&$u_{R}^{c1}$&$\stackrel{03}{[-i]}\,\stackrel{12}{(-)}|\stackrel{56}{[+]}\,\stackrel{78}{(+)}
||\stackrel{9 \;10}{(+)}\;\;\stackrel{11\;12}{[-]}\;\;\stackrel{13\;14}{[-]}$&1&$-\frac{1}{2}$&0&
$\frac{1}{2}$&$\frac{1}{2}$&$\frac{1}{2\,\sqrt{3}}$&$\frac{1}{6}$&$\frac{2}{3}$&$\frac{2}{3}$\\
\hline
3&$d_{R}^{c1}$&$\stackrel{03}{(+i)}\,\stackrel{12}{[+]}|\stackrel{56}{(-)}\,\stackrel{78}{[-]}
||\stackrel{9 \;10}{(+)}\;\;\stackrel{11\;12}{[-]}\;\;\stackrel{13\;14}{[-]}$&1&$\frac{1}{2}$&0&
$-\frac{1}{2}$&$\frac{1}{2}$&$\frac{1}{2\,\sqrt{3}}$&$\frac{1}{6}$&$-\frac{1}{3}$&$-\frac{1}{3}$\\
\hline
4&$ d_{R}^{c1} $&$\stackrel{03}{[-i]}\,\stackrel{12}{(-)}|
\stackrel{56}{(-)}\,\stackrel{78}{[-]}
||\stackrel{9 \;10}{(+)}\;\;\stackrel{11\;12}{[-]}\;\;\stackrel{13\;14}{[-]} $&1&$-\frac{1}{2}$&0&
$-\frac{1}{2}$&$\frac{1}{2}$&$\frac{1}{2\,\sqrt{3}}$&$\frac{1}{6}$&$-\frac{1}{3}$&$-\frac{1}{3}$\\
\hline
5&$d_{L}^{c1}$&$\stackrel{03}{[-i]}\,\stackrel{12}{[+]}|\stackrel{56}{(-)}\,\stackrel{78}{(+)}
||\stackrel{9 \;10}{(+)}\;\;\stackrel{11\;12}{[-]}\;\;\stackrel{13\;14}{[-]}$&-1&$\frac{1}{2}$&
$-\frac{1}{2}$&0&$\frac{1}{2}$&$\frac{1}{2\,\sqrt{3}}$&$\frac{1}{6}$&$\frac{1}{6}$&$-\frac{1}{3}$\\
\hline
6&$d_{L}^{c1} $&$ - \stackrel{03}{(+i)}\,\stackrel{12}{(-)}|\stackrel{56}{(-)}\,\stackrel{78}{(+)}
||\stackrel{9 \;10}{(+)}\;\;\stackrel{11\;12}{[-]}\;\;\stackrel{13\;14}{[-]} $&-1&$-\frac{1}{2}$&
$-\frac{1}{2}$&0&$\frac{1}{2}$&$\frac{1}{2\,\sqrt{3}}$&$\frac{1}{6}$&$\frac{1}{6}$&$-\frac{1}{3}$\\
\hline
7&$ u_{L}^{c1}$&$ - \stackrel{03}{[-i]}\,\stackrel{12}{[+]}|\stackrel{56}{[+]}\,\stackrel{78}{[-]}
||\stackrel{9 \;10}{(+)}\;\;\stackrel{11\;12}{[-]}\;\;\stackrel{13\;14}{[-]}$ &-1&$\frac{1}{2}$&
$\frac{1}{2}$&0 &$\frac{1}{2}$&$\frac{1}{2\,\sqrt{3}}$&$\frac{1}{6}$&$\frac{1}{6}$&$\frac{2}{3}$\\
\hline
8&$u_{L}^{c1}$&$\stackrel{03}{(+i)}\,\stackrel{12}{(-)}|\stackrel{56}{[+]}\,\stackrel{78}{[-]}
||\stackrel{9 \;10}{(+)}\;\;\stackrel{11\;12}{[-]}\;\;\stackrel{13\;14}{[-]}$&-1&$-\frac{1}{2}$&
$\frac{1}{2}$&0&$\frac{1}{2}$&$\frac{1}{2\,\sqrt{3}}$&$\frac{1}{6}$&$\frac{1}{6}$&$\frac{2}{3}$\\
\hline\hline
\shrinkheight{0.25\textheight}
9&$ u_{R}^{c2}$&$ \stackrel{03}{(+i)}\,\stackrel{12}{[+]}|
\stackrel{56}{[+]}\,\stackrel{78}{(+)}
||\stackrel{9 \;10}{[-]}\;\;\stackrel{11\;12}{(+)}\;\;\stackrel{13\;14}{[-]} $ &1&$\frac{1}{2}$&0&
$\frac{1}{2}$&$-\frac{1}{2}$&$\frac{1}{2\,\sqrt{3}}$&$\frac{1}{6}$&$\frac{2}{3}$&$\frac{2}{3}$\\
\hline
10&$u_{R}^{c2}$&$\stackrel{03}{[-i]}\,\stackrel{12}{(-)}|\stackrel{56}{[+]}\,\stackrel{78}{(+)}
||\stackrel{9 \;10}{[-]}\;\;\stackrel{11\;12}{(+)}\;\;\stackrel{13\;14}{[-]}$&1&$-\frac{1}{2}$&0&
$\frac{1}{2}$&$-\frac{1}{2}$&$\frac{1}{2\,\sqrt{3}}$&$\frac{1}{6}$&$\frac{2}{3}$&$\frac{2}{3}$\\
\hline
11&$d_{R}^{c2}$&$\stackrel{03}{(+i)}\,\stackrel{12}{[+]}|\stackrel{56}{(-)}\,\stackrel{78}{[-]}
||\stackrel{9 \;10}{[-]}\;\;\stackrel{11\;12}{(+)}\;\;\stackrel{13\;14}{[-]}$
&1&$\frac{1}{2}$&0&
$-\frac{1}{2}$&$ - \frac{1}{2}$&$\frac{1}{2\,\sqrt{3}}$&$\frac{1}{6}$&$-\frac{1}{3}$&$-\frac{1}{3}$\\
\hline
12&$ d_{R}^{c2} $&$\stackrel{03}{[-i]}\,\stackrel{12}{(-)}|
\stackrel{56}{(-)}\,\stackrel{78}{[-]}
||\stackrel{9 \;10}{[-]}\;\;\stackrel{11\;12}{(+)}\;\;\stackrel{13\;14}{[-]} $
&1&$-\frac{1}{2}$&0&
$-\frac{1}{2}$&$-\frac{1}{2}$&$\frac{1}{2\,\sqrt{3}}$&$\frac{1}{6}$&$-\frac{1}{3}$&$-\frac{1}{3}$\\
\hline
13&$d_{L}^{c2}$&$\stackrel{03}{[-i]}\,\stackrel{12}{[+]}|\stackrel{56}{(-)}\,\stackrel{78}{(+)}
||\stackrel{9 \;10}{[-]}\;\;\stackrel{11\;12}{(+)}\;\;\stackrel{13\;14}{[-]}$
&-1&$\frac{1}{2}$&
$-\frac{1}{2}$&0&$-\frac{1}{2}$&$\frac{1}{2\,\sqrt{3}}$&$\frac{1}{6}$&$\frac{1}{6}$&$-\frac{1}{3}$\\
\hline
14&$d_{L}^{c2} $&$ - \stackrel{03}{(+i)}\,\stackrel{12}{(-)}|\stackrel{56}{(-)}\,\stackrel{78}{(+)}
||\stackrel{9 \;10}{[-]}\;\;\stackrel{11\;12}{(+)}\;\;\stackrel{13\;14}{[-]} $&-1&$-\frac{1}{2}$&
$-\frac{1}{2}$&0&$-\frac{1}{2}$&$\frac{1}{2\,\sqrt{3}}$&$\frac{1}{6}$&$\frac{1}{6}$&$-\frac{1}{3}$\\
\hline
15&$ u_{L}^{c2}$&$ - \stackrel{03}{[-i]}\,\stackrel{12}{[+]}|\stackrel{56}{[+]}\,\stackrel{78}{[-]}
||\stackrel{9 \;10}{[-]}\;\;\stackrel{11\;12}{(+)}\;\;\stackrel{13\;14}{[-]}$ &-1&$\frac{1}{2}$&
$\frac{1}{2}$&0 &$-\frac{1}{2}$&$\frac{1}{2\,\sqrt{3}}$&$\frac{1}{6}$&$\frac{1}{6}$&$\frac{2}{3}$\\
\hline
16&$u_{L}^{c2}$&$\stackrel{03}{(+i)}\,\stackrel{12}{(-)}|\stackrel{56}{[+]}\,\stackrel{78}{[-]}
||\stackrel{9 \;10}{[-]}\;\;\stackrel{11\;12}{(+)}\;\;\stackrel{13\;14}{[-]}$&-1&$-\frac{1}{2}$&
$\frac{1}{2}$&0&$-\frac{1}{2}$&$\frac{1}{2\,\sqrt{3}}$&$\frac{1}{6}$&$\frac{1}{6}$&$\frac{2}{3}$\\
\hline\hline
17&$ u_{R}^{c3}$&$ \stackrel{03}{(+i)}\,\stackrel{12}{[+]}|
\stackrel{56}{[+]}\,\stackrel{78}{(+)}
||\stackrel{9 \;10}{[-]}\;\;\stackrel{11\;12}{[-]}\;\;\stackrel{13\;14}{(+)} $ &1&$\frac{1}{2}$&0&
$\frac{1}{2}$&$0$&$-\frac{1}{\sqrt{3}}$&$\frac{1}{6}$&$\frac{2}{3}$&$\frac{2}{3}$\\
\hline
18&$u_{R}^{c3}$&$\stackrel{03}{[-i]}\,\stackrel{12}{(-)}|\stackrel{56}{[+]}\,\stackrel{78}{(+)}
||\stackrel{9 \;10}{[-]}\;\;\stackrel{11\;12}{[-]}\;\;\stackrel{13\;14}{(+)}$&1&$-\frac{1}{2}$&0&
$\frac{1}{2}$&$0$&$-\frac{1}{\sqrt{3}}$&$\frac{1}{6}$&$\frac{2}{3}$&$\frac{2}{3}$\\
\hline
19&$d_{R}^{c3}$&$\stackrel{03}{(+i)}\,\stackrel{12}{[+]}|\stackrel{56}{(-)}\,\stackrel{78}{[-]}
||\stackrel{9 \;10}{[-]}\;\;\stackrel{11\;12}{[-]}\;\;\stackrel{13\;14}{(+)}$&1&$\frac{1}{2}$&0&
$-\frac{1}{2}$&$0$&$-\frac{1}{\sqrt{3}}$&$\frac{1}{6}$&$-\frac{1}{3}$&$-\frac{1}{3}$\\
\hline
20&$ d_{R}^{c3} $&$\stackrel{03}{[-i]}\,\stackrel{12}{(-)}|
\stackrel{56}{(-)}\,\stackrel{78}{[-]}
||\stackrel{9 \;10}{[-]}\;\;\stackrel{11\;12}{[-]}\;\;\stackrel{13\;14}{(+)} $&1&$-\frac{1}{2}$&0&
$-\frac{1}{2}$&$0$&$-\frac{1}{\sqrt{3}}$&$\frac{1}{6}$&$-\frac{1}{3}$&$-\frac{1}{3}$\\
\hline
21&$d_{L}^{c3}$&$\stackrel{03}{[-i]}\,\stackrel{12}{[+]}|\stackrel{56}{(-)}\,\stackrel{78}{(+)}
||\stackrel{9 \;10}{[-]}\;\;\stackrel{11\;12}{[-]}\;\;\stackrel{13\;14}{(+)}$&-1&$\frac{1}{2}$&
$-\frac{1}{2}$&0&$0$&$-\frac{1}{\sqrt{3}}$&$\frac{1}{6}$&$\frac{1}{6}$&$-\frac{1}{3}$\\
\hline
22&$d_{L}^{c3} $&$ - \stackrel{03}{(+i)}\,\stackrel{12}{(-)}|\stackrel{56}{(-)}\,\stackrel{78}{(+)}
||\stackrel{9 \;10}{[-]}\;\;\stackrel{11\;12}{[-]}\;\;\stackrel{13\;14}{(+)} $&-1&$-\frac{1}{2}$&
$-\frac{1}{2}$&0&$0$&$-\frac{1}{\sqrt{3}}$&$\frac{1}{6}$&$\frac{1}{6}$&$-\frac{1}{3}$\\
\hline
23&$ u_{L}^{c3}$&$ - \stackrel{03}{[-i]}\,\stackrel{12}{[+]}|\stackrel{56}{[+]}\,\stackrel{78}{[-]}
||\stackrel{9 \;10}{[-]}\;\;\stackrel{11\;12}{[-]}\;\;\stackrel{13\;14}{(+)}$ &-1&$\frac{1}{2}$&
$\frac{1}{2}$&0 &$0$&$-\frac{1}{\sqrt{3}}$&$\frac{1}{6}$&$\frac{1}{6}$&$\frac{2}{3}$\\
\hline
24&$u_{L}^{c3}$&$\stackrel{03}{(+i)}\,\stackrel{12}{(-)}|\stackrel{56}{[+]}\,\stackrel{78}{[-]}
||\stackrel{9 \;10}{[-]}\;\;\stackrel{11\;12}{[-]}\;\;\stackrel{13\;14}{(+)}$&-1&$-\frac{1}{2}$&
$\frac{1}{2}$&0&$0$&$-\frac{1}{\sqrt{3}}$&$\frac{1}{6}$&$\frac{1}{6}$&$\frac{2}{3}$\\
\hline\hline
25&$ \nu_{R}$&$ \stackrel{03}{(+i)}\,\stackrel{12}{[+]}|
\stackrel{56}{[+]}\,\stackrel{78}{(+)}
||\stackrel{9 \;10}{(+)}\;\;\stackrel{11\;12}{(+)}\;\;\stackrel{13\;14}{(+)} $ &1&$\frac{1}{2}$&0&
$\frac{1}{2}$&$0$&$0$&$-\frac{1}{2}$&$0$&$0$\\
\hline
26&$\nu_{R}$&$\stackrel{03}{[-i]}\,\stackrel{12}{(-)}|\stackrel{56}{[+]}\,\stackrel{78}{(+)}
||\stackrel{9 \;10}{(+)}\;\;\stackrel{11\;12}{(+)}\;\;\stackrel{13\;14}{(+)}$&1&$-\frac{1}{2}$&0&
$\frac{1}{2}$ &$0$&$0$&$-\frac{1}{2}$&$0$&$0$\\
\hline
27&$e_{R}$&$\stackrel{03}{(+i)}\,\stackrel{12}{[+]}|\stackrel{56}{(-)}\,\stackrel{78}{[-]}
||\stackrel{9 \;10}{(+)}\;\;\stackrel{11\;12}{(+)}\;\;\stackrel{13\;14}{(+)}$&1&$\frac{1}{2}$&0&
$-\frac{1}{2}$&$0$&$0$&$-\frac{1}{2}$&$-1$&$-1$\\
\hline
28&$ e_{R} $&$\stackrel{03}{[-i]}\,\stackrel{12}{(-)}|
\stackrel{56}{(-)}\,\stackrel{78}{[-]}
||\stackrel{9 \;10}{(+)}\;\;\stackrel{11\;12}{(+)}\;\;\stackrel{13\;14}{(+)} $&1&$-\frac{1}{2}$&0&
$-\frac{1}{2}$&$0$&$0$&$-\frac{1}{2}$&$-1$&$-1$\\
\hline
29&$e_{L}$&$\stackrel{03}{[-i]}\,\stackrel{12}{[+]}|\stackrel{56}{(-)}\,\stackrel{78}{(+)}
||\stackrel{9 \;10}{(+)}\;\;\stackrel{11\;12}{(+)}\;\;\stackrel{13\;14}{(+)}$&-1&$\frac{1}{2}$&
$-\frac{1}{2}$&0&$0$&$0$&$-\frac{1}{2}$&$-\frac{1}{2}$&$-1$\\
\hline
30&$e_{L} $&$ - \stackrel{03}{(+i)}\,\stackrel{12}{(-)}|\stackrel{56}{(-)}\,\stackrel{78}{(+)}
||\stackrel{9 \;10}{(+)}\;\;\stackrel{11\;12}{(+)}\;\;\stackrel{13\;14}{(+)} $&-1&$-\frac{1}{2}$&
$-\frac{1}{2}$&0&$0$&$0$&$-\frac{1}{2}$&$-\frac{1}{2}$&$-1$\\
\hline
31&$ \nu_{L}$&$ - \stackrel{03}{[-i]}\,\stackrel{12}{[+]}|\stackrel{56}{[+]}\,\stackrel{78}{[-]}
||\stackrel{9 \;10}{(+)}\;\;\stackrel{11\;12}{(+)}\;\;\stackrel{13\;14}{(+)}$ &-1&$\frac{1}{2}$&
$\frac{1}{2}$&0 &$0$&$0$&$-\frac{1}{2}$&$-\frac{1}{2}$&$0$\\
\hline
32&$\nu_{L}$&$\stackrel{03}{(+i)}\,\stackrel{12}{(-)}|\stackrel{56}{[+]}\,\stackrel{78}{[-]}
||\stackrel{9 \;10}{(+)}\;\;\stackrel{11\;12}{(+)}\;\;\stackrel{13\;14}{(+)}$&-1&$-\frac{1}{2}$&
$\frac{1}{2}$&0&$0$&$0$&$-\frac{1}{2}$&$-\frac{1}{2}$&$0$\\
\hline\hline
33&$ \bar{d}_{L}^{\bar{c1}}$&$ \stackrel{03}{[-i]}\,\stackrel{12}{[+]}|
\stackrel{56}{[+]}\,\stackrel{78}{(+)}
||\stackrel{9 \;10}{[-]}\;\;\stackrel{11\;12}{(+)}\;\;\stackrel{13\;14}{(+)} $ &-1&$\frac{1}{2}$&0&
$\frac{1}{2}$&$-\frac{1}{2}$&$-\frac{1}{2\,\sqrt{3}}$&$-\frac{1}{6}$&$\frac{1}{3}$&$\frac{1}{3}$\\
\hline
34&$\bar{d}_{L}^{\bar{c1}}$&$\stackrel{03}{(+i)}\,\stackrel{12}{(-)}|\stackrel{56}{[+]}\,\stackrel{78}{(+)}
||\stackrel{9 \;10}{[-]}\;\;\stackrel{11\;12}{(+)}\;\;\stackrel{13\;14}{(+)}$&-1&$-\frac{1}{2}$&0&
$\frac{1}{2}$&$-\frac{1}{2}$&$-\frac{1}{2\,\sqrt{3}}$&$-\frac{1}{6}$&$\frac{1}{3}$&$\frac{1}{3}$\\
\hline
35&$\bar{u}_{L}^{\bar{c1}}$&$ - \stackrel{03}{[-i]}\,\stackrel{12}{[+]}|\stackrel{56}{(-)}\,\stackrel{78}{[-]}
||\stackrel{9 \;10}{[-]}\;\;\stackrel{11\;12}{(+)}\;\;\stackrel{13\;14}{(+)}$&-1&$\frac{1}{2}$&0&
$-\frac{1}{2}$&$-\frac{1}{2}$&$-\frac{1}{2\,\sqrt{3}}$&$-\frac{1}{6}$&$-\frac{2}{3}$&$-\frac{2}{3}$\\
\hline
36&$ \bar{u}_{L}^{\bar{c1}} $&$ - \stackrel{03}{(+i)}\,\stackrel{12}{(-)}|
\stackrel{56}{(-)}\,\stackrel{78}{[-]}
||\stackrel{9 \;10}{[-]}\;\;\stackrel{11\;12}{(+)}\;\;\stackrel{13\;14}{(+)} $&-1&$-\frac{1}{2}$&0&
$-\frac{1}{2}$&$-\frac{1}{2}$&$-\frac{1}{2\,\sqrt{3}}$&$-\frac{1}{6}$&$-\frac{2}{3}$&$-\frac{2}{3}$\\
\hline
37&$\bar{d}_{R}^{\bar{c1}}$&$\stackrel{03}{(+i)}\,\stackrel{12}{[+]}|\stackrel{56}{[+]}\,\stackrel{78}{[-]}
||\stackrel{9 \;10}{[-]}\;\;\stackrel{11\;12}{(+)}\;\;\stackrel{13\;14}{(+)}$&1&$\frac{1}{2}$&
$\frac{1}{2}$&0&$-\frac{1}{2}$&$-\frac{1}{2\,\sqrt{3}}$&$-\frac{1}{6}$&$-\frac{1}{6}$&$\frac{1}{3}$\\
\hline
38&$\bar{d}_{R}^{\bar{c1}} $&$ - \stackrel{03}{[-i]}\,\stackrel{12}{(-)}|\stackrel{56}{[+]}\,\stackrel{78}{[-]}
||\stackrel{9 \;10}{[-]}\;\;\stackrel{11\;12}{(+)}\;\;\stackrel{13\;14}{(+)} $&1&$-\frac{1}{2}$&
$\frac{1}{2}$&0&$-\frac{1}{2}$&$-\frac{1}{2\,\sqrt{3}}$&$-\frac{1}{6}$&$-\frac{1}{6}$&$\frac{1}{3}$\\
\hline
39&$ \bar{u}_{R}^{\bar{c1}}$&$\stackrel{03}{(+i)}\,\stackrel{12}{[+]}|\stackrel{56}{(-)}\,\stackrel{78}{(+)}
||\stackrel{9 \;10}{[-]}\;\;\stackrel{11\;12}{(+)}\;\;\stackrel{13\;14}{(+)}$ &1&$\frac{1}{2}$&
$-\frac{1}{2}$&0 &$-\frac{1}{2}$&$-\frac{1}{2\,\sqrt{3}}$&$-\frac{1}{6}$&$-\frac{1}{6}$&$-\frac{2}{3}$\\
\hline
40&$\bar{u}_{R}^{\bar{c1}}$&$\stackrel{03}{[-i]}\,\stackrel{12}{(-)}|\stackrel{56}{(-)}\,\stackrel{78}{(+)}
||\stackrel{9 \;10}{[-]}\;\;\stackrel{11\;12}{(+)}\;\;\stackrel{13\;14}{(+)}$
&1&$-\frac{1}{2}$&
$-\frac{1}{2}$&0&$-\frac{1}{2}$&$-\frac{1}{2\,\sqrt{3}}$&$-\frac{1}{6}$&$-\frac{1}{6}$&$-\frac{2}{3}$\\
\hline\hline
41&$ \bar{d}_{L}^{\bar{c2}}$&$ \stackrel{03}{[-i]}\,\stackrel{12}{[+]}|
\stackrel{56}{[+]}\,\stackrel{78}{(+)}
||\stackrel{9 \;10}{(+)}\;\;\stackrel{11\;12}{[-]}\;\;\stackrel{13\;14}{(+)} $
&-1&$\frac{1}{2}$&0&
$\frac{1}{2}$&$\frac{1}{2}$&$-\frac{1}{2\,\sqrt{3}}$&$-\frac{1}{6}$&$\frac{1}{3}$&$\frac{1}{3}$\\
\hline
42&$\bar{d}_{L}^{\bar{c2}}$&$\stackrel{03}{(+i)}\,\stackrel{12}{(-)}|\stackrel{56}{[+]}\,\stackrel{78}{(+)}
||\stackrel{9 \;10}{(+)}\;\;\stackrel{11\;12}{[-]}\;\;\stackrel{13\;14}{(+)}$
&-1&$-\frac{1}{2}$&0&
$\frac{1}{2}$&$\frac{1}{2}$&$-\frac{1}{2\,\sqrt{3}}$&$-\frac{1}{6}$&$\frac{1}{3}$&$\frac{1}{3}$\\
\hline
43&$\bar{u}_{L}^{\bar{c2}}$&$ - \stackrel{03}{[-i]}\,\stackrel{12}{[+]}|\stackrel{56}{(-)}\,\stackrel{78}{[-]}
||\stackrel{9 \;10}{(+)}\;\;\stackrel{11\;12}{[-]}\;\;\stackrel{13\;14}{(+)}$
&-1&$\frac{1}{2}$&0&
$-\frac{1}{2}$&$\frac{1}{2}$&$-\frac{1}{2\,\sqrt{3}}$&$-\frac{1}{6}$&$-\frac{2}{3}$&$-\frac{2}{3}$\\
\hline
44&$ \bar{u}_{L}^{\bar{c2}} $&$ - \stackrel{03}{(+i)}\,\stackrel{12}{(-)}|
\stackrel{56}{(-)}\,\stackrel{78}{[-]}
||\stackrel{9 \;10}{(+)}\;\;\stackrel{11\;12}{[-]}\;\;\stackrel{13\;14}{(+)} $
&-1&$-\frac{1}{2}$&0&
$-\frac{1}{2}$&$\frac{1}{2}$&$-\frac{1}{2\,\sqrt{3}}$&$-\frac{1}{6}$&$-\frac{2}{3}$&$-\frac{2}{3}$\\
\hline
45&$\bar{d}_{R}^{\bar{c2}}$&$\stackrel{03}{(+i)}\,\stackrel{12}{[+]}|\stackrel{56}{[+]}\,\stackrel{78}{[-]}
||\stackrel{9 \;10}{(+)}\;\;\stackrel{11\;12}{[-]}\;\;\stackrel{13\;14}{(+)}$
&1&$\frac{1}{2}$&
$\frac{1}{2}$&0&$\frac{1}{2}$&$-\frac{1}{2\,\sqrt{3}}$&$-\frac{1}{6}$&$-\frac{1}{6}$&$\frac{1}{3}$\\
\hline
46&$\bar{d}_{R}^{\bar{c2}} $&$ - \stackrel{03}{[-i]}\,\stackrel{12}{(-)}|\stackrel{56}{[+]}\,\stackrel{78}{[-]}
||\stackrel{9 \;10}{(+)}\;\;\stackrel{11\;12}{[-]}\;\;\stackrel{13\;14}{(+)} $
&1&$-\frac{1}{2}$&
$\frac{1}{2}$&0&$\frac{1}{2}$&$-\frac{1}{2\,\sqrt{3}}$&$-\frac{1}{6}$&$-\frac{1}{6}$&$\frac{1}{3}$\\
\hline
47&$ \bar{u}_{R}^{\bar{c2}}$&$\stackrel{03}{(+i)}\,\stackrel{12}{[+]}|\stackrel{56}{(-)}\,\stackrel{78}{(+)}
||\stackrel{9 \;10}{(+)}\;\;\stackrel{11\;12}{[-]}\;\;\stackrel{13\;14}{(+)}$
 &1&$\frac{1}{2}$&
$-\frac{1}{2}$&0 &$\frac{1}{2}$&$-\frac{1}{2\,\sqrt{3}}$&$-\frac{1}{6}$&$-\frac{1}{6}$&$-\frac{2}{3}$\\
\hline
48&$\bar{u}_{R}^{\bar{c2}}$&$\stackrel{03}{[-i]}\,\stackrel{12}{(-)}|\stackrel{56}{(-)}\,\stackrel{78}{(+)}
||\stackrel{9 \;10}{(+)}\;\;\stackrel{11\;12}{[-]}\;\;\stackrel{13\;14}{(+)}$
&1&$-\frac{1}{2}$&
$-\frac{1}{2}$&0&$\frac{1}{2}$&$-\frac{1}{2\,\sqrt{3}}$&$-\frac{1}{6}$&$-\frac{1}{6}$&$-\frac{2}{3}$\\
\hline\hline
49&$ \bar{d}_{L}^{\bar{c3}}$&$ \stackrel{03}{[-i]}\,\stackrel{12}{[+]}|
\stackrel{56}{[+]}\,\stackrel{78}{(+)}
||\stackrel{9 \;10}{(+)}\;\;\stackrel{11\;12}{(+)}\;\;\stackrel{13\;14}{[-]} $ &-1&$\frac{1}{2}$&0&
$\frac{1}{2}$&$0$&$\frac{1}{\sqrt{3}}$&$-\frac{1}{6}$&$\frac{1}{3}$&$\frac{1}{3}$\\
\hline
50&$\bar{d}_{L}^{\bar{c3}}$&$\stackrel{03}{(+i)}\,\stackrel{12}{(-)}|\stackrel{56}{[+]}\,\stackrel{78}{(+)}
||\stackrel{9 \;10}{(+)}\;\;\stackrel{11\;12}{(+)}\;\;\stackrel{13\;14}{[-]} $&-1&$-\frac{1}{2}$&0&
$\frac{1}{2}$&$0$&$\frac{1}{\sqrt{3}}$&$-\frac{1}{6}$&$\frac{1}{3}$&$\frac{1}{3}$\\
\hline
51&$\bar{u}_{L}^{\bar{c3}}$&$ - \stackrel{03}{[-i]}\,\stackrel{12}{[+]}|\stackrel{56}{(-)}\,\stackrel{78}{[-]}
||\stackrel{9 \;10}{(+)}\;\;\stackrel{11\;12}{(+)}\;\;\stackrel{13\;14}{[-]} $&-1&$\frac{1}{2}$&0&
$-\frac{1}{2}$&$0$&$\frac{1}{\sqrt{3}}$&$-\frac{1}{6}$&$-\frac{2}{3}$&$-\frac{2}{3}$\\
\hline
52&$ \bar{u}_{L}^{\bar{c3}} $&$ - \stackrel{03}{(+i)}\,\stackrel{12}{(-)}|
\stackrel{56}{(-)}\,\stackrel{78}{[-]}
||\stackrel{9 \;10}{(+)}\;\;\stackrel{11\;12}{(+)}\;\;\stackrel{13\;14}{[-]}  $&-1&$-\frac{1}{2}$&0&
$-\frac{1}{2}$&$0$&$\frac{1}{\sqrt{3}}$&$-\frac{1}{6}$&$-\frac{2}{3}$&$-\frac{2}{3}$\\
\hline
53&$\bar{d}_{R}^{\bar{c3}}$&$\stackrel{03}{(+i)}\,\stackrel{12}{[+]}|\stackrel{56}{[+]}\,\stackrel{78}{[-]}
||\stackrel{9 \;10}{(+)}\;\;\stackrel{11\;12}{(+)}\;\;\stackrel{13\;14}{[-]} $&1&$\frac{1}{2}$&
$\frac{1}{2}$&0&$0$&$\frac{1}{\sqrt{3}}$&$-\frac{1}{6}$&$-\frac{1}{6}$&$\frac{1}{3}$\\
\hline
54&$\bar{d}_{R}^{\bar{c3}} $&$ - \stackrel{03}{[-i]}\,\stackrel{12}{(-)}|\stackrel{56}{[+]}\,\stackrel{78}{[-]}
||\stackrel{9 \;10}{(+)}\;\;\stackrel{11\;12}{(+)}\;\;\stackrel{13\;14}{[-]} $&1&$-\frac{1}{2}$&
$\frac{1}{2}$&0&$0$&$\frac{1}{\sqrt{3}}$&$-\frac{1}{6}$&$-\frac{1}{6}$&$\frac{1}{3}$\\
\hline
55&$ \bar{u}_{R}^{\bar{c3}}$&$\stackrel{03}{(+i)}\,\stackrel{12}{[+]}|\stackrel{56}{(-)}\,\stackrel{78}{(+)}
||\stackrel{9 \;10}{(+)}\;\;\stackrel{11\;12}{(+)}\;\;\stackrel{13\;14}{[-]} $ &1&$\frac{1}{2}$&
$-\frac{1}{2}$&0 &$0$&$\frac{1}{\sqrt{3}}$&$-\frac{1}{6}$&$-\frac{1}{6}$&$-\frac{2}{3}$\\
\hline
56&$\bar{u}_{R}^{\bar{c3}}$&$\stackrel{03}{[-i]}\,\stackrel{12}{(-)}|\stackrel{56}{(-)}\,\stackrel{78}{(+)}
||\stackrel{9 \;10}{(+)}\;\;\stackrel{11\;12}{(+)}\;\;\stackrel{13\;14}{[-]} $&1&$-\frac{1}{2}$&
$-\frac{1}{2}$&0&$0$&$\frac{1}{\sqrt{3}}$&$-\frac{1}{6}$&$-\frac{1}{6}$&$-\frac{2}{3}$\\
\hline\hline
57&$ \bar{e}_{L}$&$ \stackrel{03}{[-i]}\,\stackrel{12}{[+]}|
\stackrel{56}{[+]}\,\stackrel{78}{(+)}
||\stackrel{9 \;10}{[-]}\;\;\stackrel{11\;12}{[-]}\;\;\stackrel{13\;14}{[-]} $ &-1&$\frac{1}{2}$&0&
$\frac{1}{2}$&$0$&$0$&$\frac{1}{2}$&$1$&$1$\\
\hline
58&$\bar{e}_{L}$&$\stackrel{03}{(+i)}\,\stackrel{12}{(-)}|\stackrel{56}{[+]}\,\stackrel{78}{(+)}
||\stackrel{9 \;10}{[-]}\;\;\stackrel{11\;12}{[-]}\;\;\stackrel{13\;14}{[-]}$&-1&$-\frac{1}{2}$&0&
$\frac{1}{2}$ &$0$&$0$&$\frac{1}{2}$&$1$&$1$\\
\hline
59&$\bar{\nu}_{L}$&$ - \stackrel{03}{[-i]}\,\stackrel{12}{[+]}|\stackrel{56}{(-)}\,\stackrel{78}{[-]}
||\stackrel{9 \;10}{[-]}\;\;\stackrel{11\;12}{[-]}\;\;\stackrel{13\;14}{[-]}$&-1&$\frac{1}{2}$&0&
$-\frac{1}{2}$&$0$&$0$&$\frac{1}{2}$&$0$&$0$\\
\hline
60&$ \bar{\nu}_{L} $&$ - \stackrel{03}{(+i)}\,\stackrel{12}{(-)}|
\stackrel{56}{(-)}\,\stackrel{78}{[-]}
||\stackrel{9 \;10}{[-]}\;\;\stackrel{11\;12}{[-]}\;\;\stackrel{13\;14}{[-]} $&-1&$-\frac{1}{2}$&0&
$-\frac{1}{2}$&$0$&$0$&$\frac{1}{2}$&$0$&$0$\\
\hline
61&$\bar{\nu}_{R}$&$\stackrel{03}{(+i)}\,\stackrel{12}{[+]}|\stackrel{56}{(-)}\,\stackrel{78}{(+)}
||\stackrel{9 \;10}{[-]}\;\;\stackrel{11\;12}{[-]}\;\;\stackrel{13\;14}{[-]}$&1&$\frac{1}{2}$&
$-\frac{1}{2}$&0&$0$&$0$&$\frac{1}{2}$&$\frac{1}{2}$&$0$\\
\hline
62&$\bar{\nu}_{R} $&$ - \stackrel{03}{[-i]}\,\stackrel{12}{(-)}|\stackrel{56}{(-)}\,\stackrel{78}{(+)}
||\stackrel{9 \;10}{[-]}\;\;\stackrel{11\;12}{[-]}\;\;\stackrel{13\;14}{[-]} $&1&$-\frac{1}{2}$&
$-\frac{1}{2}$&0&$0$&$0$&$\frac{1}{2}$&$\frac{1}{2}$&$0$\\
\hline
63&$ \bar{e}_{R}$&$\stackrel{03}{(+i)}\,\stackrel{12}{[+]}|\stackrel{56}{[+]}\,\stackrel{78}{[-]}
||\stackrel{9 \;10}{[-]}\;\;\stackrel{11\;12}{[-]}\;\;\stackrel{13\;14}{[-]}$ &1&$\frac{1}{2}$&
$\frac{1}{2}$&0 &$0$&$0$&$\frac{1}{2}$&$\frac{1}{2}$&$1$\\
\hline
64&$\bar{e}_{R}$&$\stackrel{03}{[-i]}\,\stackrel{12}{(-)}|\stackrel{56}{[+]}\,\stackrel{78}{[-]}
||\stackrel{9 \;10}{[-]}\;\;\stackrel{11\;12}{[-]}\;\;\stackrel{13\;14}{[-]}$&1&$-\frac{1}{2}$&
$\frac{1}{2}$&0&$0$&$0$&$\frac{1}{2}$&$\frac{1}{2}$&$1$\\
\hline
\end{supertabular}
\end{tiny}

\vspace {3mm}


%

%
\section*{Acknowledgment} 

The author thanks Department of Physics, FMF, University of Ljubljana, Society of Mathematicians, Physicists and Astronomers of Slovenia,  for supporting the research on the {\it spin-charge-family} theory by offering the room and computer facilities and Matja\v z Breskvar of Beyond Semiconductor for donations, in particular for the annual workshops entitled "What comes beyond the standard models".

\vspace{3mm}


\end{document}